\documentclass[prl,aps,superscriptaddress,nofootinbib,twocolumn,notitlepage,floatfix,10pt]{revtex4-2}
\usepackage{amsmath,mathtools,amsthm,amssymb,pifont}
\usepackage[utf8]{inputenc}
\usepackage[american]{babel}
\usepackage{graphicx,xcolor,bbold,titlesec}
\usepackage{braket}
\usepackage{MnSymbol}

\usepackage[colorlinks,
citecolor=red,
linkcolor=red,
urlcolor=red]{hyperref}
\usepackage{tikz,ifthen}
\usepackage{bbold}
\usepackage{tikz-network}
\usetikzlibrary{patterns,decorations.pathreplacing,calligraphy}
\usepackage{orcidlink}
\usepackage{color}
\usepackage{MnSymbol}
\usepackage{adjustbox}
\usepackage{float}
\usepackage{bibunits}

\usepackage{listings}
\usepackage{xcolor}
\usepackage{orcidlink}

\makeatletter

\let\savedrevtexaffiliation\affiliation
\let\savedrevtexemail\email
\let\savedrevtexthanks\thanks

\newcommand{\resetrevtexfrontmatter}{%
  \frontmatter@init
  \let\title\frontmatter@title
  \let\author\frontmatter@author
  \let\affiliation\savedrevtexaffiliation
  \let\email\savedrevtexemail
  \let\thanks\savedrevtexthanks
  \let\maketitle\frontmatter@maketitle
}

\makeatother

\newcommand{\Tr}{\mathrm{Tr}}

\begin{document}
\title{Random Gaussian Augmented Matrix Product States}

\author{Stavya Puri~\orcidlink{0009-0001-4889-6235}}
\email{stavya.puri@bsc.es}
\affiliation{Barcelona Supercomputing Center Plaça Eusebi G\"uell, 1-3 08034, Barcelona, Spain}
\affiliation{Departament de Física Quàntica i Astrofísica, Facultat de Física, Universitat de Barcelona (UB), Martí i Franquès, 1, 08028 Barcelona, Spain.}

\author{Poetri Sonya Tarabunga~\orcidlink{0000-0001-8079-9040}}
%\email{poetri.tarabunga@tum.de}
\affiliation{Technical University of Munich, TUM School of Natural Sciences, Physics Department, 85748 Garching, Germany}
\affiliation{Munich Center for Quantum Science and Technology (MCQST), Schellingstraße 4, 80799 M{\"u}nchen, Germany}

\author{Piotr Sierant~\orcidlink{0000-0001-9219-7274}}
\email{piotr.sierant@bsc.es}
\affiliation{Barcelona Supercomputing Center Plaça Eusebi G\"uell, 1-3 08034, Barcelona, Spain}

\setlength{\abovedisplayskip}{6pt}
\setlength{\belowdisplayskip}{6pt}
\setlength{\abovedisplayshortskip}{4pt}
\setlength{\belowdisplayshortskip}{4pt}

\begin{abstract}
Strong entanglement limits the reach of tensor-network descriptions of quantum many-body systems. We study Gaussian-augmented matrix product states (GAMPS), formed by applying a free-fermionic (Gaussian) unitary to a matrix product state. To explore the capabilities of this family, we consider \emph{random GAMPS}, a minimally structured ensemble obtained by independently sampling random matrix product states and Gaussian unitaries. Exploiting the algebraic structure of Gaussian unitaries, we develop a replica tensor-network method that computes averages over the random GAMPS ensemble for hundreds of qubits. 
Random GAMPS exhibit nearly maximal volume-law entanglement already at modest bond dimension, while other quantum resource quantifiers rapidly approach Haar-random state values as the bond dimension increases. The entanglement scaling suggests an exponential reduction in the bond dimension needed to approach the volume-law entropy compared with unaugmented matrix product states. We find a state two-design behavior of random GAMPS at fixed accuracy with a bond dimension independent of system size. Our results reveal how Gaussian augmentation expands the expressive power of tensor networks.
\end{abstract}

\maketitle

Quantum many-body physics underpins our understanding of phases of matter~\cite{Anderson72More,polkovnikov2011colloquium} and the development of quantum technologies~\cite{Georgescu14Simulation,Acin18Roadmap}. Progress in both areas relies on tractable descriptions of many-body states, which tensor-network methods provide by exploiting entanglement structure~\cite{Verstraete08mps,Orus14MPS,Cirac21rmp}.
In one dimension, matrix product states (MPS) support ground-state calculations with the density-matrix renormalization group~\cite{White1992,White1993,Schollwock2011mps} and nonequilibrium simulations~\cite{Vidal04tebd,WhiteFeiguin04tdmrg,Daley04tdmrg,Haegeman16}. For open chains, MPS bond dimension $\chi$ bounds entanglement entropy as $S\leq\log_2\chi$~\cite{Schuch08}. The linear entanglement growth typical of global quenches~\cite{CalabreseCardy05,Kim13ballistic} thus demands exponentially growing $\chi$, imposing an \emph{entanglement barrier}~\cite{Paeckel19}.

Disentangling schemes seek to ease the entanglement constraints of tensor network approaches~\cite{Schuch08,Cirac21rmp} by combining a less entangled tensor-network state with an auxiliary unitary. This strategy underlies the multiscale entanglement renormalization ansatz (MERA)~\cite{Vidal2007EntanglementRenormalization,Evenbly2009AlgorithmsEntanglementRenormalization}, augmented tree tensor networks and MPS~\cite{Felser21augmented,Qian23disentanglers}, and disentangling-based state preparation~\cite{Mansuroglu2026PreparationCircuits, Szoldra2026BeliefPropagationDisentanglers}. Clifford-augmented matrix product states (CAMPS) employ Clifford circuits as the auxiliary layer~\cite{Qian24prl, qian2024cliffordcircuitsaugmentedtimedependent, Collura2024hybridMPO, masotllima2024stabilizertensornetworks, fux2024disentanglingunitarydynamicsclassically, huang2024cliffordcircuitsaugmentedmatrix, MasotLlima2026LimitsCliffordDisentangling, Fioroni26camps,frau2024stabilizer}. This layer can generate extensive entanglement while preserving the nonstabilizerness~\cite{Veitch2014theresourcetheory,leone2022stabilizerrenyientropy}, or magic resources, of the MPS input, with Pauli expectation values remaining efficiently computable. Random MPS~\cite{Garnerone10rmps, Garnerone10a_rmps, Haferkamp21rmps, Lami25anti, magni2025anti} supply substantial nonstabilizerness even at modest bond dimensions, and their augmentation by random Clifford unitaries yields ensembles whose statistics approach those of Haar-random states as $\chi$ increases~\cite{Lami25cmps}.

\begin{figure}
\centering
\includegraphics[width=\linewidth]{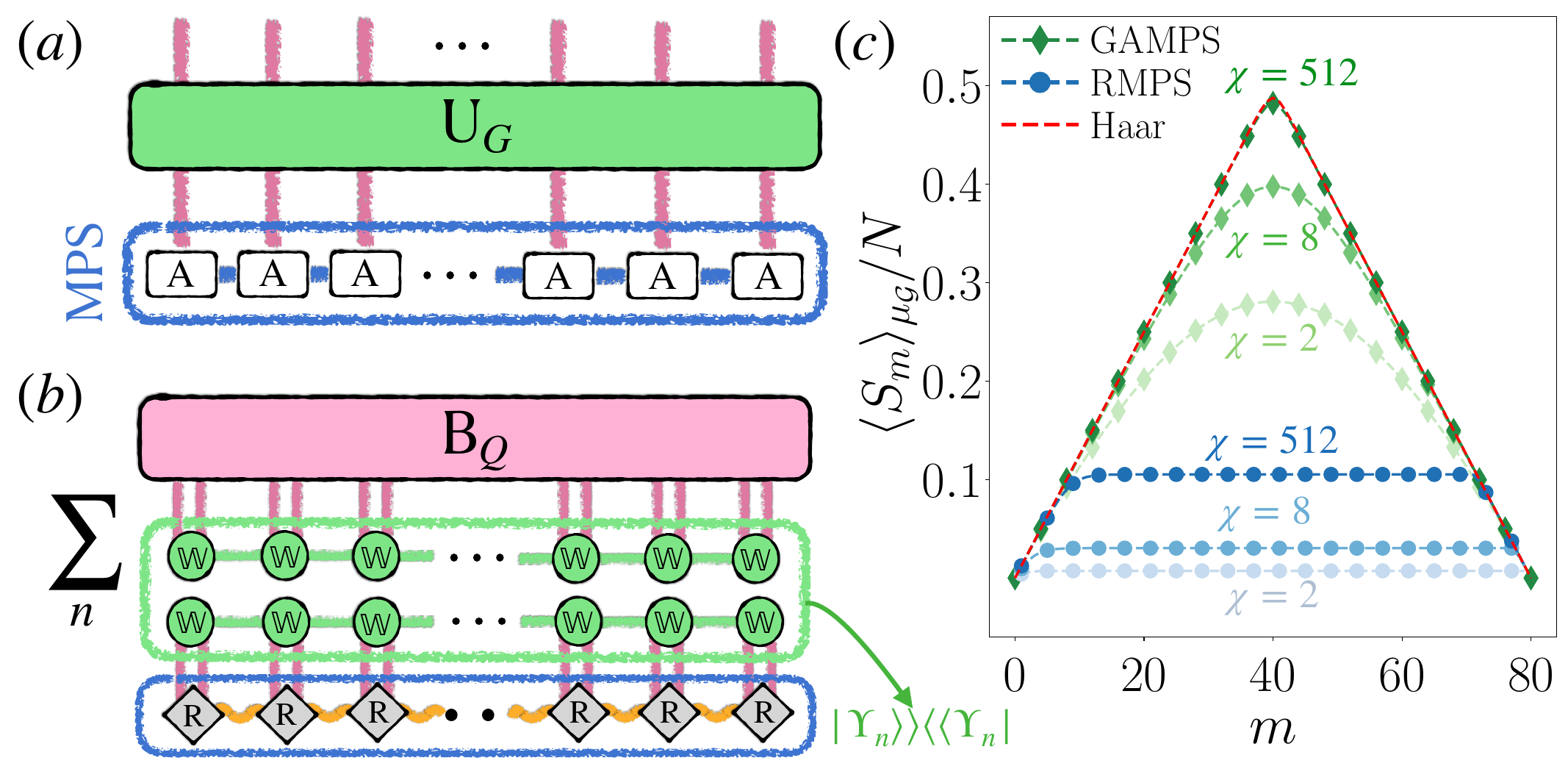}
\caption{\((a)\) GAMPS construction by applying a fermionic Gaussian unitary \(U_G\) to an MPS.
\((b)\)  Evaluation of \(\langle Q\rangle\) by contracting three layers: the averaged RMPS tensors, the fermionic commutant operators \(\Gamma_n\), and the boundary operator \(B_Q\).
  \((c)\) Page curve for annealed R\'enyi-2 entropy density \(\langle S_m\rangle/N\) versus subsystem size \(m\) for GAMPS and RMPS at \(N=80\) and varying \(\chi\). GAMPS exhibit near-Haar entanglement
  at modest \(\chi\).} 
\label{fig1}
\end{figure}
Fermionic Gaussian unitaries (FGUs) generate the matchgate circuits~\cite{Valiant01Matchgate,
Knill01fermionic, Valiant02, Terhal02Matchgate, Jozsa08Matchgate, Bravyi05flo, Surace22FGS} to offer a complementary route to disentangling~\cite{morralyepes2026disentangling,Langer26}. They can generate extensive entanglement~\cite{Lydzba20quadratic, Bianchi2021fermionicPagecurve} and nonstabilizerness~\cite{Collura24fermionicgaussian, Turkeshi26pauli} while their action on Gaussian input states remains efficiently simulable using covariance matrix formalism. As auxiliary transformations, FGUs can reduce the entanglement that an MPS must encode, while the MPS captures the remaining non-Gaussian correlations. This principle underlies orbital and mode optimization~\cite{Krumnow16,Krumnow21}, Gaussian circuit constructions~\cite{Fishman15,Wu25fermDisent}, and matchgate-augmented DMRG~\cite{Huang25augmenting}. 
These applications motivate a broader question: \textit{how much does Gaussian augmentation expand the expressive power of MPS?}
% \textit{how does Gaussian augmentation change the entanglement and statistical properties of MPS?}

In this Letter, we address this question by studying Gaussian-augmented matrix product states (GAMPS),
\begin{align}
  |\Psi_{\mathcal G}\rangle = U_G|\Phi_\chi\rangle,
  \end{align}
where $U_G$ is an fermionic Gaussian unitary and $|\Phi_\chi\rangle$ is an MPS of bond dimension $\chi$. Since an FGU acts linearly on Majorana operators, $U_G^\dagger\gamma_\mu\gamma_\nu U_G$ is a sum of $O(N^2)$ bilinears, so two-point correlators of $\ket{\Psi_{\mathcal G}}$ remain efficiently computable from polynomially many MPS expectation values. We probe the typical properties of this family by independently sampling a random FGU $U_G$ and $|\Phi_\chi\rangle$ from the ensemble of RMPS, see Fig.~\ref{fig1}(a). We develop a replica tensor-network framework based on fermionic commutant theory~\cite{Wan23Matchgateshadows, Sierant26matchgatecommutant, Braccia2026matchgatecommutant, Lastres2026geometry} that combines analytical ensemble integration with numerical evaluation, allowing us to the study the GAMPS ensemble in large systems comprising hundreds of qubits. 
Our results demonstrate that GAMPS ensemble features nearly maximal volume-law entanglement entropy, while other quantum resource quantifiers~\cite{chitambar2019quantumresourcetheories} approach values typical of Haar-random states as $\chi$ increases. 
We further find evidence for approximate state two-design behavior~\cite{Ambainis07tdesigns, Cotler23emergent, Mele2024introductiontoHaar}, showing that measurements on two identical copies of a GAMPS yield, on average, nearly the same statistics as for Haar-random states.

\paragraph{Random matrix product states.} We consider a system of $N$ qubits with
%the global 
Hilbert space 
%of
dimension \(D=d^N\) for \(d=2\) as local qubit dimension. 
Each pure state \(|\psi\rangle = \sum_{\mathbf{i}}c_{\mathbf{i}}|\mathbf{i}\rangle\) admits an MPS representation~\cite{Schollwock2011mps},
  \(
  |\psi\rangle=\sum_{\mathbf{i},\boldsymbol{\lambda}}
  A^{i_1}_{\lambda_1}
  A^{i_2}_{(\lambda_1,\lambda_2)}
  \cdots
  A^{i_N}_{\lambda_{N-1}}
  |\mathbf{i}\rangle,
  \)
where \(\mathbf{i}=(i_1,\ldots,i_N)\) collects the physical indices \(i_j\in\{0,1\}\), with \(|\mathbf{i}\rangle=|i_1\cdots i_N\rangle\), and \(\boldsymbol{\lambda}=(\lambda_1,\ldots, \lambda_{N-1})\) collects the bond indices \(\lambda_j\in\{0,\ldots,\chi-1\}\). A bulk tensor \(A^{i_j}_{(\lambda_{j-1},\lambda_j)}\) is graphically represented by \(\adjustbox{valign=c}{\includegraphics[height=2em]{ 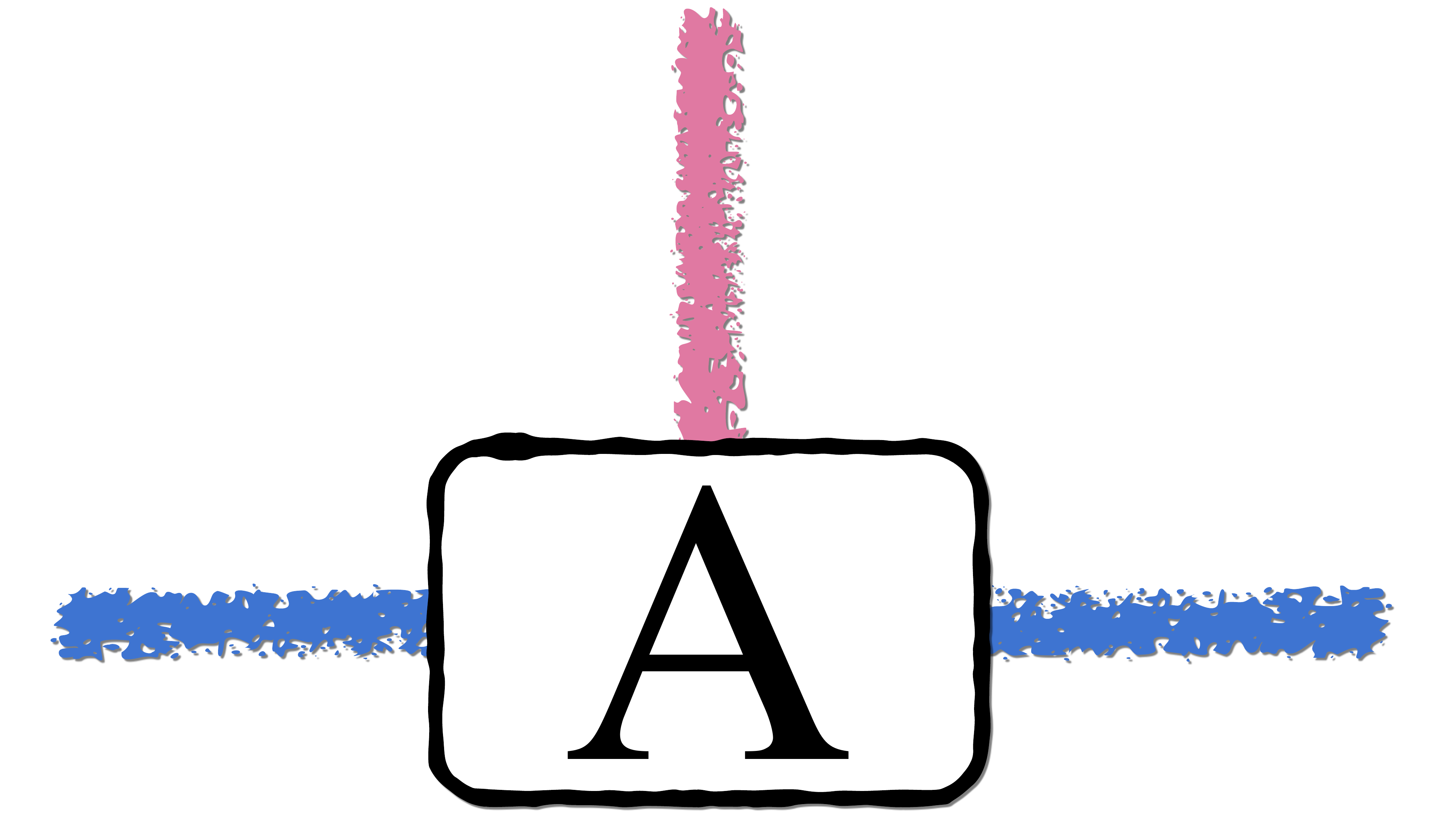}}\), see Fig.~\ref{fig1}(a).

Limited entanglement can enable accurate MPS approximations with modest bond dimension $\chi$~\cite{Vidal03mps, Verstraete08mps, Vidal07tebd,Collura2024TNforQC}. The resulting MPS has R\'enyi-2 entropy $S=-\log_2\Tr(\rho_m^2)\leq\log_2\chi$ across the cut separating the first $m$ qubits from the rest~\cite{Renyi1961,NielsenChuang2010,Calabrese04}.

To investigate typical properties of MPS at fixed \(\chi\), we consider the ensemble of random matrix product states (RMPS)~\cite{Garnerone10rmps, Garnerone10a_rmps, Haferkamp21rmps}. Expressing each local tensor as
\(A^{i_j}_{(\lambda_{j-1},\lambda_{j})}=U^{(\lambda_{j-1},\lambda_{j})}_{i_j,0}\) recasts the MPS as a chain of \(d\chi\times d\chi\) unitary matrices acting on the reference state \(|0\rangle\)~\cite{Haag2023TypicalTNcorrlength, Schon2005SeqGeneration}. The RMPS ensemble is obtained by independently sampling $U$ at each site \(j\) from the Haar measure on the unitary group $\mathcal{U}(d\chi)$.
Accordingly, an RMPS density matrix \(\rho_{\Phi_\chi} = |\Phi_{\chi}\rangle\langle\Phi_\chi|\) is graphically represented using \(\rho_0 =|0\rangle\langle0|\) as shown below, 
\begin{align}
\rho_{\Phi_\chi} 
= \adjustbox{valign=c}{\includegraphics[height=4.5em]{ 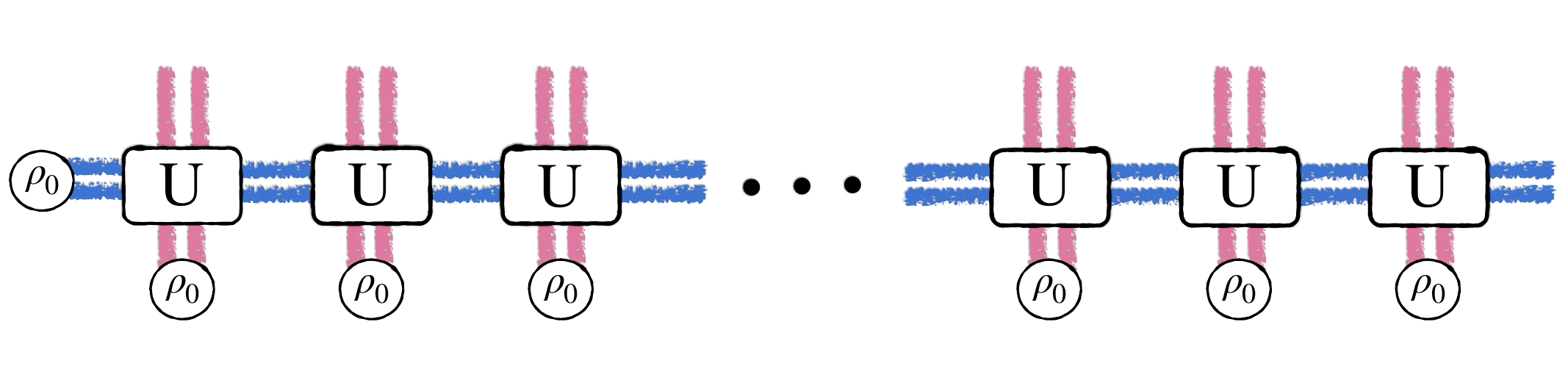}}
\label{eq:density_Unitary_chain}
\end{align}
We depict both the bra and ket physical indices on top of each tensor for pictorial convenience.
\paragraph{Fermionic gaussian Unitaries.} For $N$ fermionic modes, FGUs form a subgroup $\mathcal{U}_G(D)\subset\mathcal{U}(D)$ generated by quadratic Hamiltonians,
\begin{align}
U_G = \exp \bigg[\frac 14 \sum_{\mu,\nu=1}^{2N}H_{\mu \nu}\gamma_\mu\gamma_\nu\bigg]
\end{align}
where $H$ is a real antisymmetric $2N\times2N$ matrix and $\{\gamma_\mu\}_{\mu=1}^{2N}$ are Majorana operators. Their qubit representation follows from the Jordan-Wigner transformation of Pauli strings, given by \(\gamma_{2k-1} = ( \prod_{p=1}^{k-1} Z_p )X_k\) and \(\gamma_{2k} = ( \prod_{p=1}^{k-1} Z_p )Y_k\) \cite{Jordan1928, Surace22FGS,Mbeng24isingchain}, under which FGUs correspond to matchgate unitaries \cite{Valiant01Matchgate}. These FGUs induce an orthogonal rotation within the Majorana operator set, \(U_G^\dagger\gamma_\eta U_G = \sum_{\mu=1}^{2N}G_{\eta \mu}\gamma_\mu\), where \(G \in \mathrm{SO}(2N)\) has the form~\(G=\exp(H)\)\cite{Note_G}.
%However, in this work 
We additionally include parity reflections generated by $X_N$ in $\mathcal U_G(D)$, extending the corresponding transformations to $G\in\mathrm{O}(2N)$. Fermionic Gaussian states form the orbit $\ket{\Psi_G}=U_G\ket{0}$ and admit efficient classical simulation through their covariance matrices \cite{Terhal02Matchgate,Negele18Wick}.

\paragraph{Quantum resource quantifiers.} 
We characterize entanglement through the subsystem purity $\mathcal P_m(\ket{\psi})=\Tr(\rho_m^2)$, where $\rho_m$ is the reduced density matrix of the first $m$ qubits~\cite{Page93}.
To probe anticoncentration~\cite{Dalzell22anti}, we use the second-order inverse participation ratio (IPR), $I_2(\ket{\psi})=\sum_{\mathbf i\in\mathcal B}|\langle\mathbf i|\psi\rangle|^4$, where $\mathcal B$ labels the computational basis~\cite{Kramer93localization,Luitz14IPR,Liu25ipralgo}. Fermionic non-Gaussianity is quantified by the fermionic antiflatness (FAF), $\mathcal F_1(\ket{\psi})=N-\frac12\Tr(M^TM)$~\cite{Sierant26FAF,Haug2026FAFwitness}, where $M_{\mu\nu}=-\frac{i}{2}\langle\psi|[\gamma_\mu,\gamma_\nu]|\psi\rangle$ is the covariance matrix. The averaged purity defines the annealed R\'enyi-2 entropy, $\langle S_m\rangle_{\mu_{\mathcal G}} =-\log_2\langle\mathcal P_m\rangle_{\mu_{\mathcal G}}$.

To characterize random GAMPS, we compute $\langle Q\rangle_{\mu_{\mathcal G}} =\mathbb E_{\mu_{\mathcal G}}[Q(\ket{\Psi_{\mathcal G}})]$ for $Q\in\{\mathcal P_m,I_2,\mathcal F_1\}$ where $\mu_{\mathcal G}$ denotes the Haar measure on the ensemble of GAMPS. The quadratic dependence of these quantifiers on the density matrix  $\rho_{\mathcal G}=\ket{\Psi_{\mathcal G}}\bra{\Psi_{\mathcal G}}$ allows their averages to be extracted from a common two-copy density operator~\cite{Gross07Unitarydesign, Mele2024introductiontoHaar}:
\begin{align}
  \langle Q\rangle_{\mu_{\mathcal G}}
  =\Tr\!\left[
  B_Q\,\mathbb E_{\mu_{\mathcal G}}
  \!\left[\rho_{\mathcal G}^{\otimes2}\right]
  \right].
\label{eq:quantifier_general}
\end{align}
The boundary operator $B_Q$ selects the quantifier in the tensor-network contraction~\cite{Turkeshi24HSdelocalization}, see Fig.~\ref{fig1}(b). Explicitly,  $B_{\mathcal P_m}=\mathbb F^{\otimes m}\otimes\mathbb I^{\otimes(N-m)}$ for purity, $B_{I_2}=\sum_{\mathbf i\in\mathcal B}(\ket{\mathbf i}\bra{\mathbf i})^{\otimes2}$ for IPR, and $B_{\mathcal F_1}=N\mathbb I^{\otimes N}+\sum_{1\leq\mu<\nu\leq2N}(\gamma_\mu\gamma_\nu)^{\otimes2}$ for FAF. Here $\mathbb F$ swaps the two copies of a single qubit, and $\mathbb I$ is the identity on this pair.

\paragraph{Replica method for random GAMPS.}
Averaging over the random GAMPS ensemble amounts to independent averages over the FGUs and RMPS. Denoting their respective measures by $\mu_G$ and $ \mu_\chi$, Eq.~\eqref{eq:quantifier_general} becomes
\begin{align}
  \langle Q\rangle_{\mu_{\mathcal G}}
  =\Tr\!\left[
  B_Q\,\mathbb E_{\mu_G}\!\left[
  U_G^{\otimes2}
  \mathbb E_{\mu_\chi}\!\left[\rho_{\Phi_\chi}^{\otimes2}\right]
  (U_G^\dagger)^{\otimes2}
  \right]
  \right],
  \label{eq:quantifier_rmps_fgus}
\end{align}
where $\rho_{\Phi_\chi}=\ket{\Phi_\chi}\bra{\Phi_\chi}$.
The unitary network structure of Eq.~\eqref{eq:density_Unitary_chain} allows the RMPS average to be performed locally, since the Haar-random unitaries at different sites are independent.
Each local average is described by the two-copy Haar-twirling channel $\mathcal T_{\mathrm{Haar}}^{(2)}(X) =\mathbb E_{U\sim\mu_{\mathrm H}}[U^{\otimes2}X(U^\dagger)^{\otimes2}]$, where $X$ is a two-copy operator and $\mu_{\mathrm H}$ is Haar measure on $\mathcal U(d\chi)$.
Schur-Weyl duality expands this channel in the permutation basis \(\{P_\sigma^{(d\chi)}\}_{\sigma\in S_2}\) with $\sigma\in S_2$~\cite{Weyl1939SchurWeyl, Collins03SchurWeyl, Collins2006HaarTwirl}, with coefficients determined by the Weingarten matrix $W^{(d\chi)}$~\cite{Kostenberger21Weingarten}.
Inserting this decomposition into the sequential RMPS network~\cite{Schon2005SeqGeneration} yields a matrix product operator (MPO) for this averaged RMPS as \(\varrho^{(2)}_{\mu_\chi} = \mathbb E_{\mu_\chi}[\rho_{\Phi_\chi}^{\otimes 2}]\) having bond dimension two [Fig.~\ref{fig1}(b)] and the local tensor as~\cite{Haag2023TypicalTNcorrlength}
 \begin{align}
    \adjustbox{valign=c}{\includegraphics[height=3em]{ 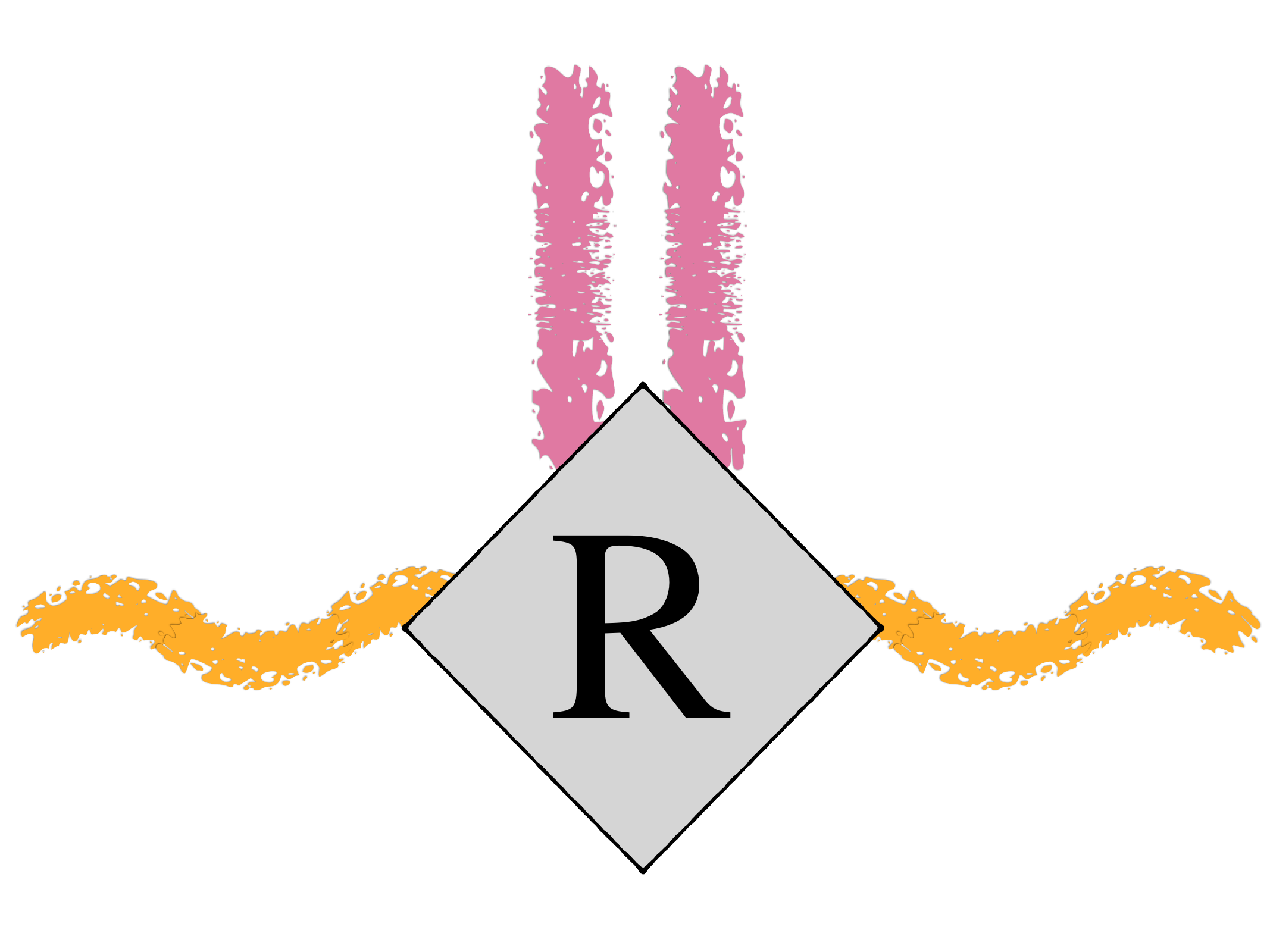}} = \sum_{\eta} W_{\sigma,\eta}^{(d\chi)}P_\eta^{(2)} G_{\eta,\tau}^{(\chi)} = \adjustbox{valign=c}{\includegraphics[height=3.5em]{ 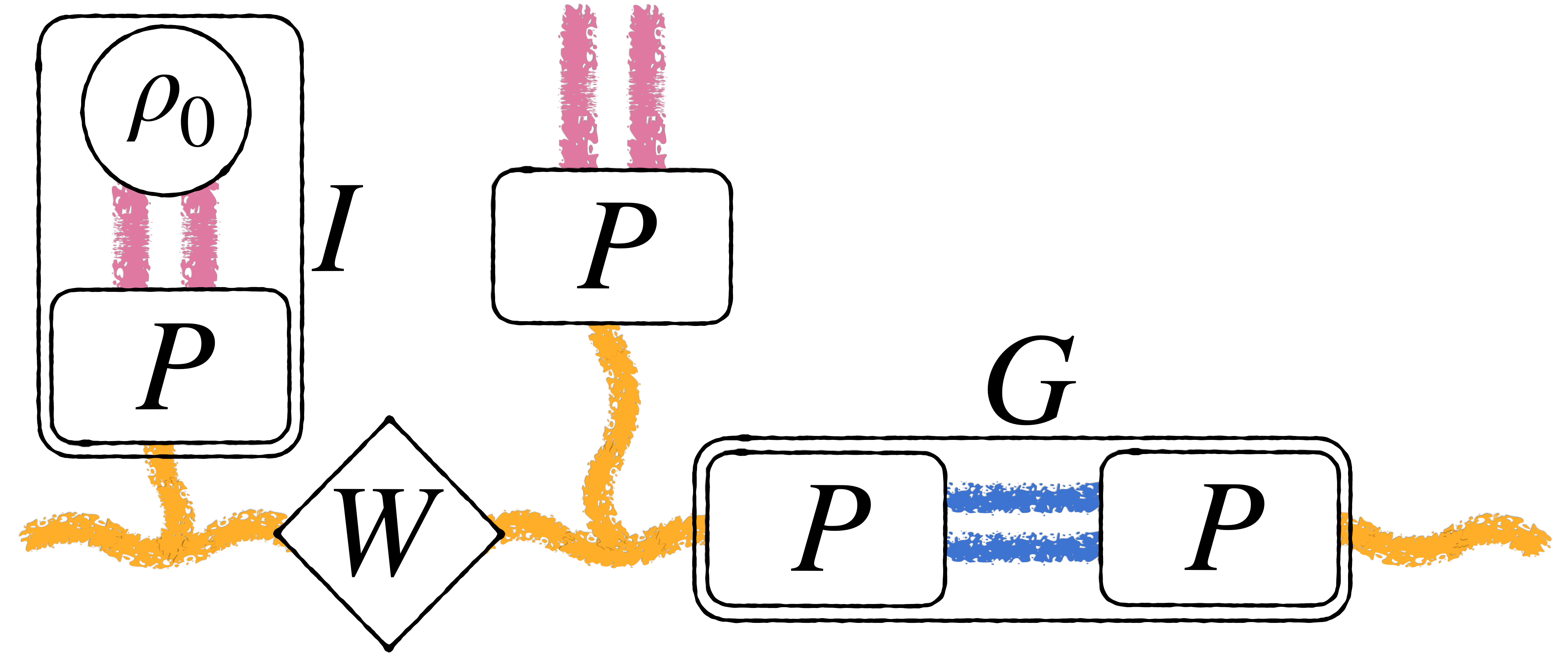}}.
    \label{eq:MPO_singleterm}
\end{align}
Here $G^{(\chi)}$ is the Gram matrix of the permutation operators on the virtual bond space. 

We next evaluate the Gaussian average. The two-copy twirling channel
$\mathcal T_G^{(2)}(X)=\mathbb E_{\mu_G}[U_G^{\otimes2}X(U_G^\dagger)^{\otimes2}]$ projects onto the operators invariant under Gaussian conjugation, which form the two-copy fermionic commutant~\cite{Wan23Matchgateshadows,Sierant26matchgatecommutant}.
For the Gaussian group considered here, including parity reflections, this space has an orthonormal pairing-tensor basis~\cite{Wan23Matchgateshadows, Sierant26matchgatecommutant}, giving
\begin{align}
  \mathcal T_G^{(2)}(\rho^{\otimes2})
  =\sum_{n=0}^{2N}\alpha_n(\rho)\Upsilon_n^{(2)},
  \label{eq:PT_expansion}
\end{align}
where $\alpha_n(\rho)=\Tr[\Upsilon_n^{(2)}\rho^{\otimes2}]$. Each pairing tensor $\Upsilon_n^{(2)}$ combines Majorana strings of the same length across the two copies: $\Upsilon_n^{(2)}=C_n\sum_{|S|=n}\gamma_S\otimes\gamma_S$. Here $\gamma_S=\gamma_{\mu_1}\cdots\gamma_{\mu_n}$ for $S=\{\mu_1<\cdots<\mu_n\}\subset[2N]$~\cite{Note_2Nset}, and $C_n=2^{-N}\binom{2N}{n}^{-1/2}$ ensures orthonormality.
Substituting the expansion~\eqref{eq:PT_expansion} into Eq.~\eqref{eq:quantifier_rmps_fgus} yields
\begin{align}
  \langle Q\rangle_{\mu_{\mathcal G}}
  =\sum_{n=0}^{2N}\beta_n(Q)\,
  \alpha_n(\varrho^{(2)}_{\mu_\chi}),
  \label{eq:quantifier_rmps_fgus_ptbasis}
\end{align}
where $\beta_n(Q)=\Tr[B_Q\Upsilon_n^{(2)}]$. 
The coefficients $\alpha_n(\varrho^{(2)}_{\mu_\chi})$ are evaluated by contracting the averaged RMPS with the pairing tensors, while the boundary coefficients $\beta_n(Q)$ are determined by the resource measure $Q$ [Fig.~\ref{fig1}(b)].
These coefficients admit closed-form expressions, for instance, for FAF, $\beta_0(\mathcal F_1)=N2^N$ and $\beta_2(\mathcal F_1)=C_2^{-1}$, with all other coefficients vanishing.
The remaining boundary coefficients are derived in the End Matter.

\begin{figure}[!t]
    \centering
    \includegraphics[width=1\linewidth]{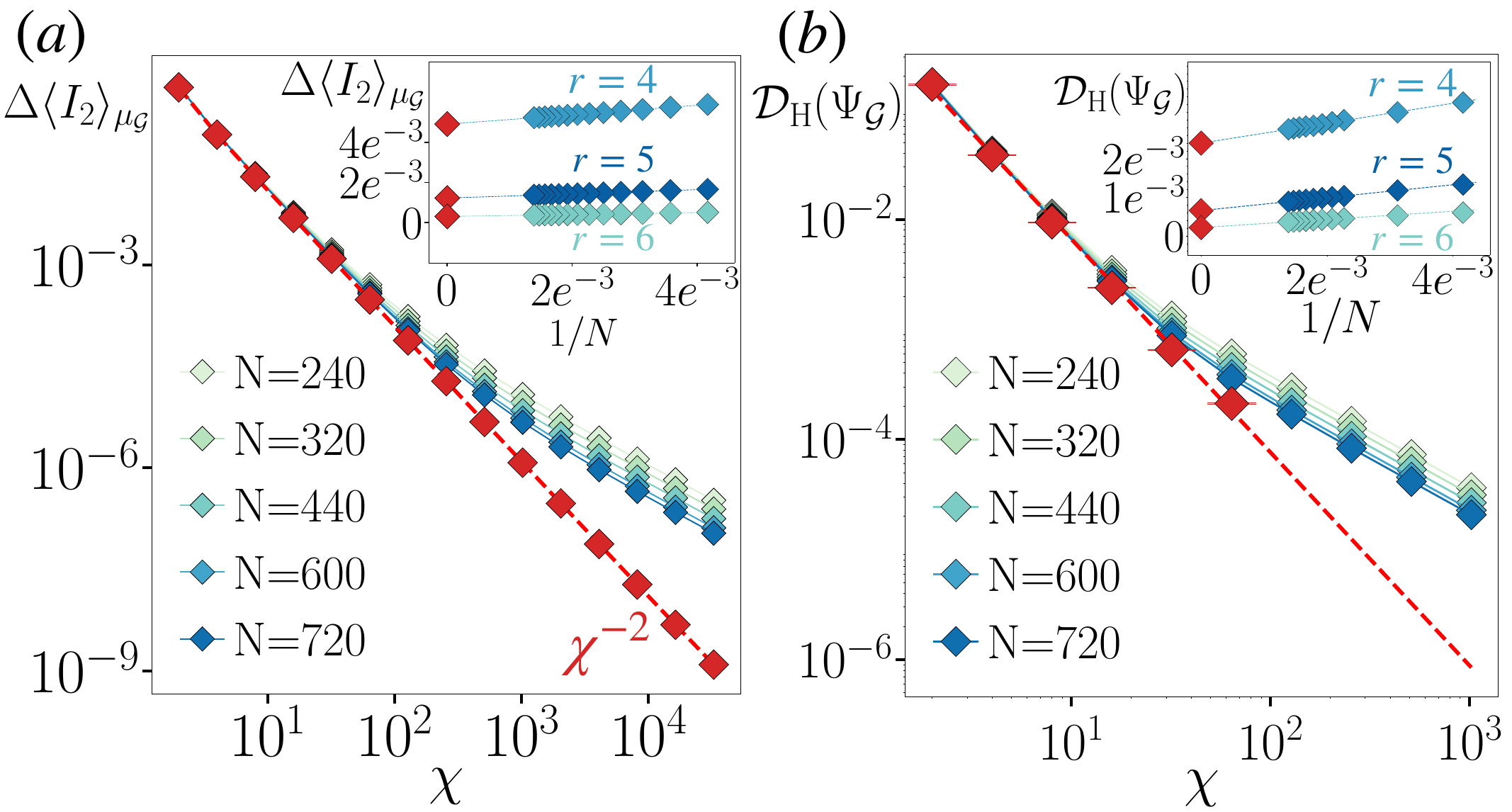}
    \caption{Anticoncentration and \(\epsilon-\)approximate \(2-\)design properties of GAMPS for \(N\) qubits. \((a)\) The deviation of the ensemble-averaged second-order inverse participation ratio from the Haar value, \(\Delta\langle I_2\rangle_{\mu_\mathcal G}\), exhibits the asymptotic scaling \(\chi^{-2}\), with the inset depicting its thermodynamic-limit extrapolation. \((b)\) The Trace distance \(\mathcal D_\mathrm H({\Psi_{\mathcal G}})\) also depicts a similar power-law asymptotic scaling of \(\chi^{-2}\) corresponding to the extrapolation in \(N \rightarrow \infty\) limit (see inset).}
    \label{fig2}
\end{figure}

\paragraph{Matrix product operator representation of pairing tensors.}
The key computational step is to represent the pairing tensor \(\Upsilon_n^{(2)}\) as a compact MPO. A finite-state-machine construction~\cite{Schollwock2011mps} generates the local tensors $\mathbb W_n^{[j]}$, which, together with boundary vectors $v_L$ and $v_R$, yield the pairing tensor
representation shown in Fig.~\ref{fig1}(b):
\begin{align}
    \Upsilon_n^{(2)}
    =C_n\adjustbox{valign=c}{\includegraphics[height=2.4em]{ 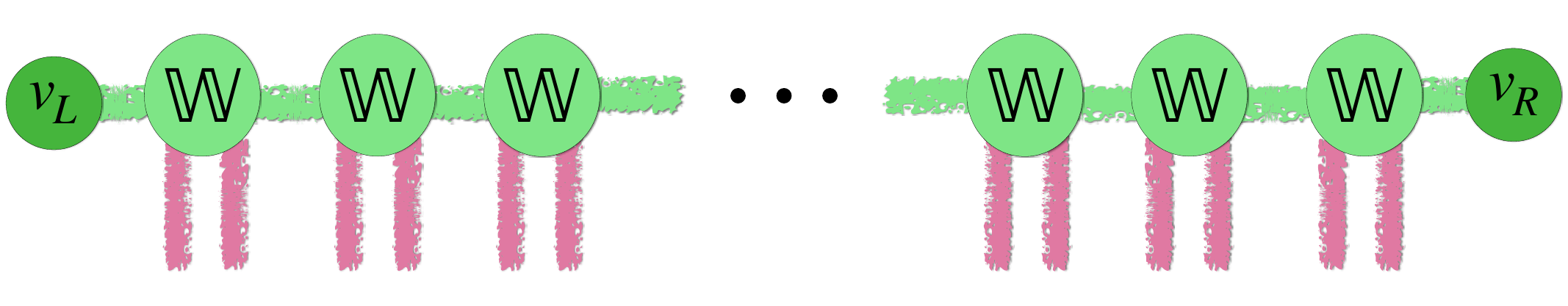}}.
    \label{eq:Upsilon_MPO_form}
\end{align}
The construction of $\mathbb W_n^{[j]}$ and the corresponding boundary vectors is detailed in the End Matter. Importantly, this construction renders the complete two-copy Gaussian twirl efficiently computable for an arbitrary MPS, at a cost polynomial in $N$ and $\chi$, and with it the monotones defined from the two-copy commutant~\cite{tarabunga2026fermionic}. This is done through contraction of the constructed MPO having bond dimension \(n+1\) with the MPS enabling the efficient evaluation of the coefficient \(\alpha_n\) ~\cite{itensor-r0.3,Pfeifer14optimalTN}. Applying it to the averaged RMPS of bond dimension $2$ yields a transfer matrix of size $2(n+1)\times2(n+1)$. 
% Combining the pairing tensor MPO representation of bond dimension $n+1$ with the averaged RMPS of bond dimension $2$ yields a transfer matrix of size $2(n+1)\times2(n+1)$.
Since both sets of bulk tensors are site independent, the averaged coefficients $ \alpha_n(\varrho^{(2)}_{\mu_\chi})$ are obtained from powers of this transfer matrix contracted with the corresponding boundary vectors~\cite{Fannes1992TransferMat,Orus14MPS}.

\paragraph{Fermionic non-Gaussianity of GAMPS.} 
Fermionic non-Gaussianity is a quantum resource characterizing departures from fermionic Gaussian states~\cite{Sierant26FAF, tarabunga2026nongaussmsrs, Leone26ferm, Turkeshi26williamson,tarabunga2026fermionic}. Since FAF is invariant under Gaussian unitaries, GAMPS and their RMPS inputs have identical FAF values. The $n=0$ and $n=2$ contributions yield $\langle\mathcal F_1\rangle_{\mu_{\mathcal G}} =N+ \alpha_2(\varrho^{(2)}_{\mu_\chi})/C_2 =N+\mathcal L T^{N-r-1}\mathcal R$, where $\mathcal L$ and $\mathcal R$ are boundary vectors and $T=\adjustbox{valign=c}{\includegraphics[height=2em]{ 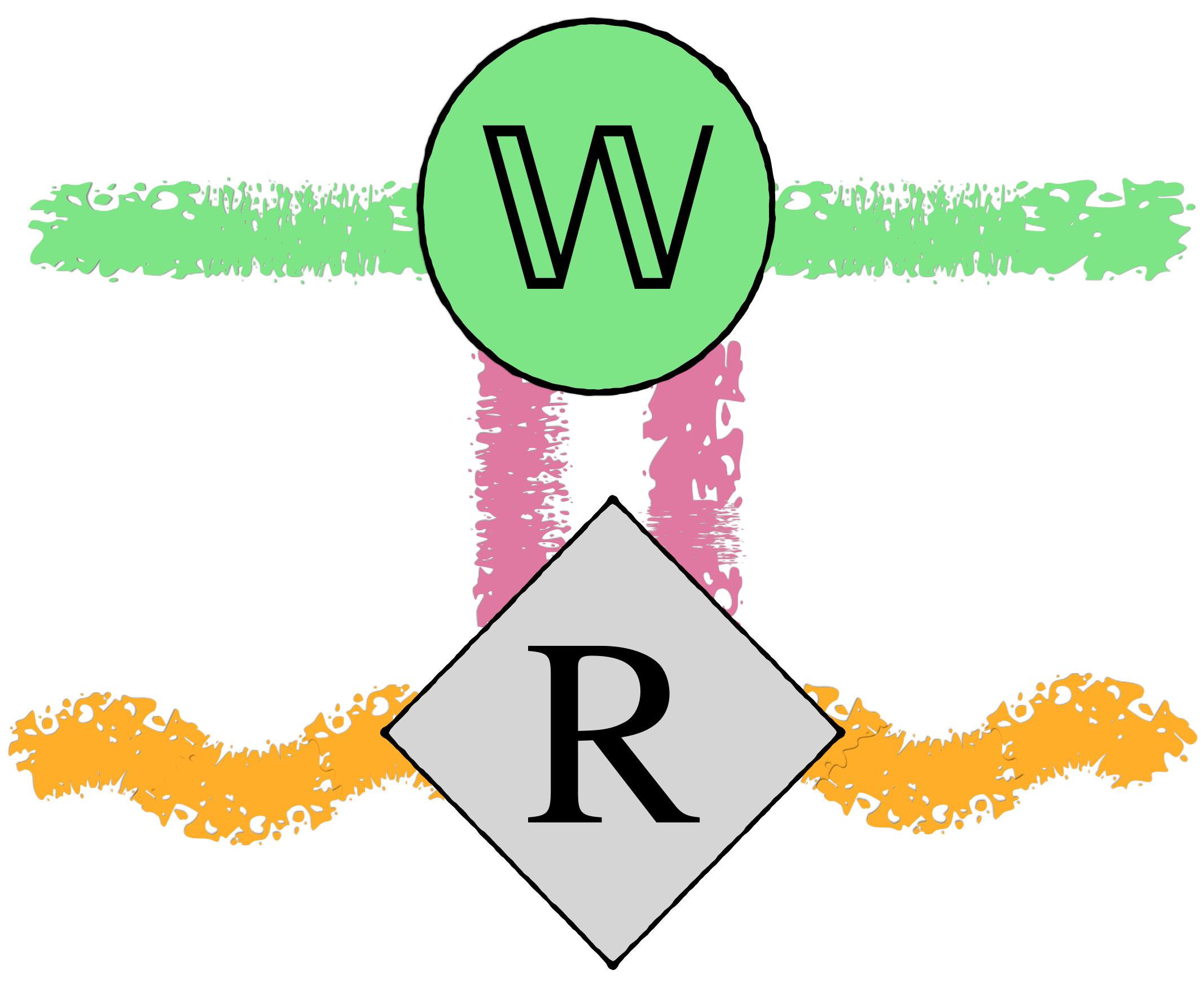}}$ is the $6\times6$ transfer matrix obtained from the MPO construction. The deviation from the Haar value~\cite{Sierant26FAF}, recovered at $r=N-1$, is
  \begin{equation}
  \Delta\langle\mathcal F_1\rangle_{\mu_{\mathcal G}}
  =\langle\mathcal F_1\rangle_{\mu_H}
  -\langle\mathcal F_1\rangle_{\mu_{\mathcal G}}
  =-\mathcal L T^{N-r-1}\mathcal R
+ O(2^{-N}).
  \label{eq:del_faf}
\end{equation}
For $\chi>1$, $T$ is not diagonalizable, so we evaluate its powers using the Jordan decomposition $T^{N-r-1}=VJ^{N-r-1}V^{-1}$~\cite{Horn1985Jordan_decomp}. The Jordan form $J$ contains a $2\times2$ block with eigenvalue $1$. Its powers generate a term proportional to $N-r-1$, producing the contribution linear in $N$. The remaining eigenvalues, $0$, $\lambda_2(\chi)$, and $\lambda_3(\chi)$, have magnitude below $1/2$, so their contributions decay exponentially with $N$ at fixed $\chi$.
For $N\gg r=\log_2\chi$ and large $\chi$, evaluating the nondecaying contribution yields
\begin{equation}
    \frac{ \Delta \langle \mathcal F_1 \rangle_{\mu_{\mathcal G}}}{N} = \frac{5}{N\chi} + \frac{5}{2\chi^2} + O\bigg(\frac{\log(\chi)^2}{N\chi^2}\bigg) + O \bigg(\frac{1}{\chi^4}\bigg).
\end{equation}
Taking $N\to\infty$ at fixed $\chi$, followed by large $\chi$, gives a deficit proportional to $\chi^{-2}$, while the leading finite-size correction scales as $1/(N\chi)$. This analytical result for GAMPS explains the numerical findings for RMPS~\cite{Sierant26FAF}.

\paragraph{Inverse participation ratio.}
We compare the anticoncentration of GAMPS with Haar-random states through the relative IPR deviation
$\Delta\langle I_2\rangle_{\mu_{\mathcal G}}
  =\langle I_2\rangle_{\mu_{\mathcal G}}/
  \langle I_2\rangle_{\mu_H}-1$.
The thermodynamic extrapolation in Fig.~\ref{fig2}(a) supports a deviation scaling as $\chi^{-2}$.
For unaugmented RMPS, the leading deviation scales as $N/\chi$ in the near-Haar regime~\cite{Lami25anti}. This difference demonstrates that Gaussian augmentation removes the system-size dependence of the bond dimension required for Haar-like anticoncentration, using only $\chi=O(\epsilon^{-1/2})$ for a given relative anticoncentration error $\epsilon$.
% This difference demonstrates that Gaussian augmentation drives the GAMPS ensemble more rapidly with increasing $\chi$ toward Haar-like anticoncentration.
The leading contribution to $\Delta\langle I_2\rangle_{\mu_{\mathcal G}}
$ in Eq.~\eqref{eq:quantifier_rmps_fgus_ptbasis} comes from the $n=2$ sector and is proportional to the FAF deficit. The full IPR deviation shares the same $\chi^{-2}$ scaling, linking the deficit in fermionic non-Gaussianity to departures from Haar-like anticoncentration.

\begin{figure}
    \centering
    \includegraphics[width=1\linewidth]{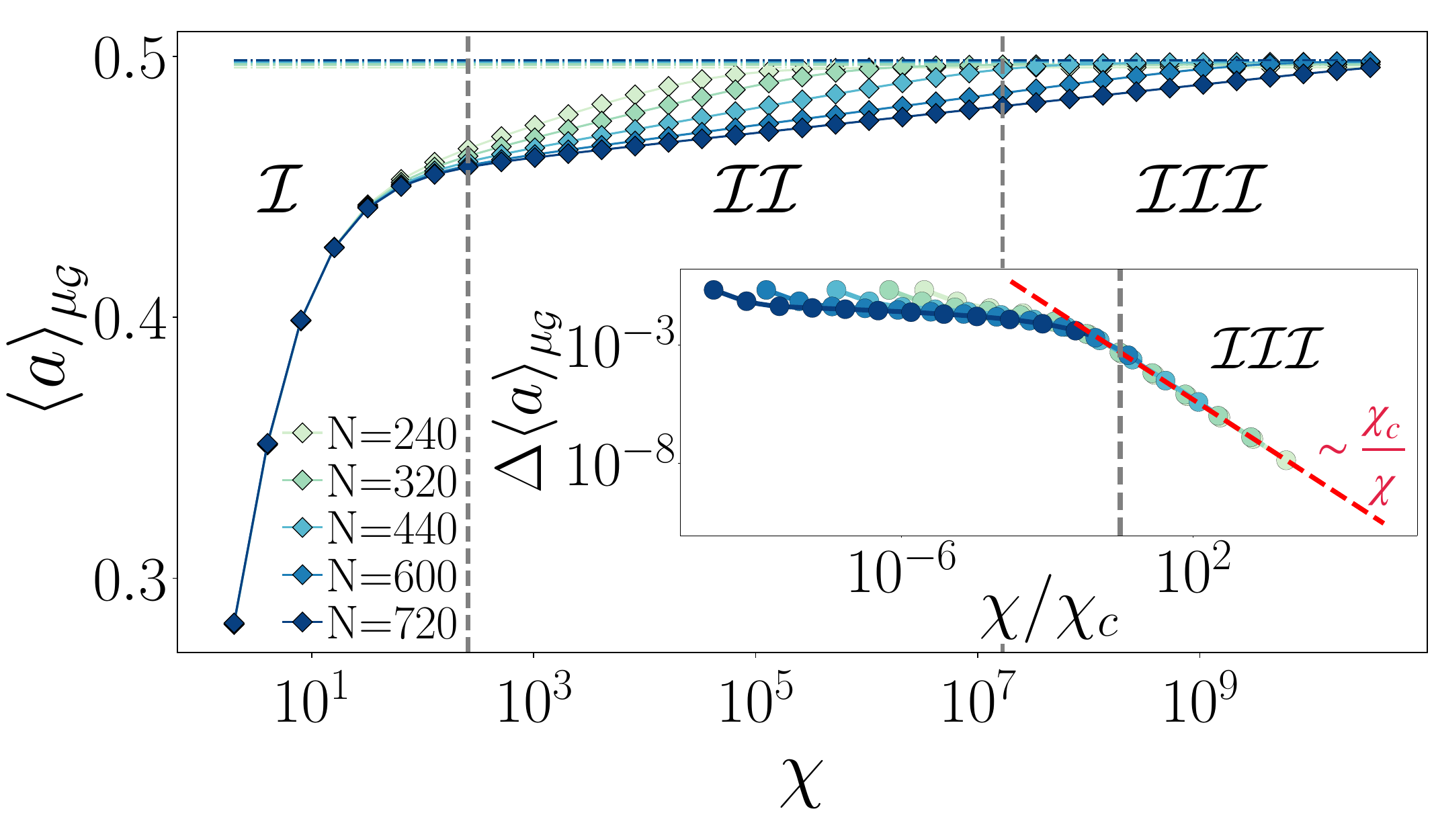}
    \caption{The half-chain R\'enyi-2 entanglement entropy density \(\langle a\rangle_{\mu_{\mathcal G}}\) exhibits different behaviors in the three regimes determined by the bond dimension \(\chi\). The crossover bond dimension \(\chi_c\) rescales the deviation \(\Delta\langle a\rangle_{\mu_{\mathcal G}}=a_{\mathrm{Haar}}-\langle a\rangle_{\mu_{\mathcal G}}\), collapsing the data to \(\chi_c/\chi\) and identifying regime \(\mathcal{III}\), where the entropy is within \(10^{-3}\) of the Haar value (see inset).}
    \label{fig3}
\end{figure}

\paragraph{State design properties.}
An $\epsilon$-approximate state two-design reproduces the Haar second moment to within trace distance $\epsilon$~\cite{Ambainis07tdesigns, Cotler23emergent, Mele2024introductiontoHaar}.
For GAMPS, we evaluate this distance as
 % $\mathcal D_{\mathrm H}
 %  =\frac12\|\Omega_{\mathcal G}^{(2)}-\Omega_{\mathrm H}^{(2)}\|_1$,
  $\mathcal D_{\mathrm H}
  =\frac12\|\mathcal T_{\mathcal G}^{(2)}-\mathcal T_{\mathrm{Haar}}^{(2)}\|_1$,
where
  % $\Omega_{\mathcal G}^{(2)}
  $\mathcal T_{\mathcal G}^{(2)}
  =\mathcal T_G^{(2)}(\varrho^{(2)}_{\mu_\chi})$
is the ensemble second-moment operator and $\mathcal T_{\mathrm{Haar}}^{(2)}$ its Haar counterpart.
To evaluate the trace norm, we use an alternative basis of the two-replica fermionic commutant consisting of mutually orthogonal projectors $\{\mathcal E_p\}_{p=0}^{2N}$~\cite{Sierant26matchgatecommutant, Braccia2026matchgatecommutant,tarabunga2026fermionic}.
These satisfy $\mathcal E_p\mathcal E_q=\delta_{pq}\mathcal E_p$ and have ranks $\Tr\mathcal E_p=\binom{2N}{p}$.
The pairing tensors are related to this projector basis by
$
  \Upsilon_n^{(2)}
  =C_n\sum_{p=0}^{2N}K_n(p;2N)\mathcal E_p,
 $
where $K_n(p;2N)$ are binary Krawtchouk polynomials~\cite{Krawtchouk1929}.
In this basis, the GAMPS second-moment operator has eigenvalues
  $\kappa_p^{\mathcal G}
  =\sum_{n=0}^{2N}C_nK_n(p;2N)
   \alpha_n(\varrho^{(2)}_{\mu_\chi})$.
The trace distance therefore reduces to
\begin{align}
  \mathcal D_{\mathrm H}
  =\frac12\sum_{p=0}^{2N}\binom{2N}{p}
  \left|\kappa_p^{\mathcal G}-\kappa_p^{\mathrm{Haar}}\right|, 
\end{align}
where $\kappa_p^{\mathrm{Haar}}=2/[2^N(2^N+1)]$ for $(p-N)\bmod 4\in\{0,1\}$ and vanishes otherwise; see \cite{supmat} for details.
Figure~\ref{fig2}(b) shows that, at finite $N$, the trace distance $\mathcal D_{\mathrm H}$ crosses over from $\chi^{-2}$ decay at small bond dimensions to a slower, approximately $\chi^{-1}$ decay at larger $\chi$. To determine its thermodynamic behavior, we extrapolate $\mathcal D_{\mathrm H}$ as a function of $1/N$ to $N\to \infty$ at each fixed $\chi$ [inset of Fig.~\ref{fig2}(b)]. The crossover disappears in the extrapolated data, which recover $\mathcal D_{\mathrm H}\sim\chi^{-2}$ scaling. 
GAMPS therefore form an $\epsilon$-approximate state two-design with $\epsilon=\mathcal{O}(\chi^{-2})$. 

\paragraph{Entanglement entropy regimes.} The incorporation of entanglement-generating FGUs significantly enlarges the manifold of Hilbert space accessible to bond-dimension-constrained RMPS. At fixed bond dimension, RMPS obey an area law, whereas typical fermionic Gaussian states exhibit volume-law entanglement with an entropy density below the Haar value~\cite{Bianchi2021fermionicPagecurve, Pathak2021Fermionicpagecurve, quinn2026typicalentanglementsuperpositions}. Figure~\ref{fig1}(c) shows that GAMPS achieve nearly maximal volume-law entanglement, closely approaching the Haar Page curve~\cite{Lubkin1978purity,Page93,Siddhartha96Avgentropy} already at modest $\chi$.
We quantify this behavior through the annealed half-chain R\'enyi-2 entanglement entropy density
$
    \langle a\rangle_{\mu_\mathcal G}  =  \langle S_{N/2} \rangle_{\mu_\mathcal G}/N.$
We define $\chi_c$ as the crossover bond dimension at which the relative deviation from the Haar entropy falls below $(S_{\mathrm{Haar}}-S)/S_{\mathrm{Haar}}\leq10^{-3}$. Fig.~\ref{fig3} reveals three distinct scaling regimes of \(\langle a\rangle_{\mu_\mathcal G}\) under increasing \(\chi\).

For \(\chi \ll \chi_c\) (regime $\mathcal{I}$), GAMPS rapidly develop volume-law entanglement with the entanglement entropy density remarkably close to the Page value (\(a_{\rm Page} = 1/2\)). This indicates that the entanglement-generating FGU layer drives the ensemble to near-maximal volume-law entanglement even at small \(\chi\). In the intermediate regime $\mathcal{}II$, for \(\chi \lesssim \chi_c\), numerical fits reveal that the additional entanglement growth follows the characteristic logarithmic dependence of RMPS \cite{Garnerone10a_rmps,Lami25cmps}, \(\langle a\rangle_{\mu_{\mathcal G}} \approx a_0 + \frac{b \log\chi}{N} + \frac{c}{N}\), where $a_0 \approx 0.46$. The increase of bond dimension therefore contributes a sub-leading logarithmic correction to the volume-law background $a_0N$ generated by GAMPS in the regime $\mathcal{I}$. Finally, for \(\chi\gtrsim\chi_c\), the entropy converges to the Haar value with a power-law scaling of \(\langle a\rangle_{\mu_{\mathcal G}} = a_{\mathrm{Haar}}\left(1 - \frac{\chi_c}{\chi}\right)\). Numerically we obtain \(\chi_c\sim 2^{0.04N}\) demonstrating an exponential improvement in attaining a Haar-like entanglement over RMPS, which requires \(\chi\sim2^{0.5N}\). A key observation is that the extensive entanglement generated for \(\chi \ll \chi_c\) establishes a near-Haar background, such that even the logarithmic RMPS-like corrections of the \(\chi \lesssim\chi_c\) are sufficient to drive a rapid convergence to the Haar value.

\paragraph{Conclusions.}
We studied Gaussian-augmented matrix product states (GAMPS), obtained by applying fermionic Gaussian unitaries to matrix product states. Our replica tensor-network method enables the evaluation of ensemble-averaged properties of random GAMPS for hundreds of qubits. Numerical extrapolations indicate that the inverse participation ratio, the trace distance from the Haar second-moment operator, and the state frame potential approach their Haar values to fixed accuracy at bond dimensions independent of system size, with corrections that decay polynomially in $\chi$ and more rapidly than for RMPS. Our closed-form expression for the average fermionic antiflatness quantifies the approach of non-Gaussianity to values typical of Haar-random states as $\chi$ increases. Gaussian invariance ensures that this result applies equally to GAMPS and their RMPS inputs. Numerical results for nonstabilizerness show a similar trend (End Matter).

Gaussian augmentation also substantially increases the entanglement accessible at fixed bond dimension. Random GAMPS exhibit volume-law entanglement with a nearly maximal coefficient already at modest $\chi$. The observed crossover scaling shows that to reach the same relative deviation from the Page curve entropy, GAMPS require an exponentially smaller bond dimension than unaugmented MPS.
Numerical results further indicate self-averaging  of the entropy density (End Matter). In~\cite{supmat}, we show that analogous entanglement and resource trends persist when fermionic parity is conserved, with the relevant Haar comparison then restricted to the corresponding parity sector. Overall, our results link two quantum resources, entanglement and fermionic non-Gaussianity, by showing that an MPS of bounded entanglement acquires Haar-like character under unitaries that cannot generate non-Gaussianity.

Our results motivate several directions. Extending the replica construction to higher moments may determine whether random GAMPS form approximate state $k$-designs for $k>2$.  A further direction is to investigate whether supplementing the Gaussian layer with non-Gaussian gates can reduce the MPS bond dimension needed to reproduce Haar statistics, building on studies of doped Clifford and matchgate circuits~\cite{Haferkamp23nonclifford, Fabian26dopedmatchgate,Leone26Noncliffcost}. GAMPS may also provide an ansatz for Gaussian-augmented DMRG~\cite{Krumnow16,Huang25augmenting,Wu25fermDisent} and a setting for learning non-Gaussian states with matchgate shadows~\cite{Wan23Matchgateshadows,Zhao21, Bittel24optimal,Mele25learning}. The fixed-parity GAMPS may facilitate applications to fermionic quantum simulation and matchgate-based hardware~\cite{Kivlichan18lineardepth, Google20hartreefock,Helsen22matchgatebench}. Determining the classical simulation cost of numerical algorithms employing GAMPS will clarify how their increased expressive power can translate into practical methods for quantum many-body physics.

\paragraph{Acknowledgments.}
P.S. acknowledges a fellowship within the ``Generaci\'on D'' initiative, Red.es, Ministerio para la Transformaci\'on Digital y de la Funci\'on P\'ublica, for talent attraction (C005/24-ED CV1), funded by the European Union NextGenerationEU funds, through PRTR. P.S.T. acknowledges funding from the European Union (ERC, DynaQuant, No. 101169765).

\bibliography{bib.bib}

\begin{thebibliography}{6}%
\makeatletter
\providecommand \@ifxundefined [1]{%
 \@ifx{#1\undefined}
}%
\providecommand \@ifnum [1]{%
 \ifnum #1\expandafter \@firstoftwo
 \else \expandafter \@secondoftwo
 \fi
}%
\providecommand \@ifx [1]{%
 \ifx #1\expandafter \@firstoftwo
 \else \expandafter \@secondoftwo
 \fi
}%
\providecommand \natexlab [1]{#1}%
\providecommand \enquote  [1]{``#1''}%
\providecommand \bibnamefont  [1]{#1}%
\providecommand \bibfnamefont [1]{#1}%
\providecommand \citenamefont [1]{#1}%
\providecommand \href@noop [0]{\@secondoftwo}%
\providecommand \href [0]{\begingroup \@sanitize@url \@href}%
\providecommand \@href[1]{\@@startlink{#1}\@@href}%
\providecommand \@@href[1]{\endgroup#1\@@endlink}%
\providecommand \@sanitize@url [0]{\catcode `\\12\catcode `\$12\catcode `\&12\catcode `\#12\catcode `\^12\catcode `\_12\catcode `\%12\relax}%
\providecommand \@@startlink[1]{}%
\providecommand \@@endlink[0]{}%
\providecommand \url  [0]{\begingroup\@sanitize@url \@url }%
\providecommand \@url [1]{\endgroup\@href {#1}{\urlprefix }}%
\providecommand \urlprefix  [0]{URL }%
\providecommand \Eprint [0]{\href }%
\providecommand \doibase [0]{https://doi.org/}%
\providecommand \selectlanguage [0]{\@gobble}%
\providecommand \bibinfo  [0]{\@secondoftwo}%
\providecommand \bibfield  [0]{\@secondoftwo}%
\providecommand \translation [1]{[#1]}%
\providecommand \BibitemOpen [0]{}%
\providecommand \bibitemStop [0]{}%
\providecommand \bibitemNoStop [0]{.\EOS\space}%
\providecommand \EOS [0]{\spacefactor3000\relax}%
\providecommand \BibitemShut  [1]{\csname bibitem#1\endcsname}%
\let\auto@bib@innerbib\@empty
%</preamble>
\bibitem [{\citenamefont {Haag}\ \emph {et~al.}(2023)\citenamefont {Haag}, \citenamefont {Baccari},\ and\ \citenamefont {Styliaris}}]{Haag2023TypicalTNcorrlength}%
  \BibitemOpen
  \bibfield  {author} {\bibinfo {author} {\bibfnamefont {D.}~\bibnamefont {Haag}}, \bibinfo {author} {\bibfnamefont {F.}~\bibnamefont {Baccari}},\ and\ \bibinfo {author} {\bibfnamefont {G.}~\bibnamefont {Styliaris}},\ }\href {https://doi.org/10.1103/PRXQuantum.4.030330} {\bibfield  {journal} {\bibinfo  {journal} {PRX Quantum}\ }\textbf {\bibinfo {volume} {4}},\ \bibinfo {pages} {030330} (\bibinfo {year} {2023})}\BibitemShut {NoStop}%
\bibitem [{\citenamefont {Sierant}\ \emph {et~al.}(2026)\citenamefont {Sierant}, \citenamefont {Turkeshi},\ and\ \citenamefont {Tarabunga}}]{Sierant26matchgatecommutant}%
  \BibitemOpen
  \bibfield  {author} {\bibinfo {author} {\bibfnamefont {P.}~\bibnamefont {Sierant}}, \bibinfo {author} {\bibfnamefont {X.}~\bibnamefont {Turkeshi}},\ and\ \bibinfo {author} {\bibfnamefont {P.~S.}\ \bibnamefont {Tarabunga}},\ }\href {https://arxiv.org/abs/2603.12392} {\bibinfo {title} {Theory of the matchgate commutant}} (\bibinfo {year} {2026}),\ \Eprint {https://arxiv.org/abs/2603.12392} {arXiv:2603.12392 [quant-ph]} \BibitemShut {NoStop}%
\bibitem [{\citenamefont {Sch\"on}\ \emph {et~al.}(2005)\citenamefont {Sch\"on}, \citenamefont {Solano}, \citenamefont {Verstraete}, \citenamefont {Cirac},\ and\ \citenamefont {Wolf}}]{Schon2005SeqGeneration}%
  \BibitemOpen
  \bibfield  {author} {\bibinfo {author} {\bibfnamefont {C.}~\bibnamefont {Sch\"on}}, \bibinfo {author} {\bibfnamefont {E.}~\bibnamefont {Solano}}, \bibinfo {author} {\bibfnamefont {F.}~\bibnamefont {Verstraete}}, \bibinfo {author} {\bibfnamefont {J.~I.}\ \bibnamefont {Cirac}},\ and\ \bibinfo {author} {\bibfnamefont {M.~M.}\ \bibnamefont {Wolf}},\ }\href {https://doi.org/10.1103/PhysRevLett.95.110503} {\bibfield  {journal} {\bibinfo  {journal} {Phys. Rev. Lett.}\ }\textbf {\bibinfo {volume} {95}},\ \bibinfo {pages} {110503} (\bibinfo {year} {2005})}\BibitemShut {NoStop}%
\bibitem [{\citenamefont {Braccia}\ \emph {et~al.}(2026)\citenamefont {Braccia}, \citenamefont {Diaz}, \citenamefont {Larocca}, \citenamefont {Cerezo},\ and\ \citenamefont {García-Martín}}]{Braccia2026matchgatecommutant}%
  \BibitemOpen
  \bibfield  {author} {\bibinfo {author} {\bibfnamefont {P.}~\bibnamefont {Braccia}}, \bibinfo {author} {\bibfnamefont {N.~L.}\ \bibnamefont {Diaz}}, \bibinfo {author} {\bibfnamefont {M.}~\bibnamefont {Larocca}}, \bibinfo {author} {\bibfnamefont {M.}~\bibnamefont {Cerezo}},\ and\ \bibinfo {author} {\bibfnamefont {D.}~\bibnamefont {García-Martín}},\ }\href {https://arxiv.org/abs/2603.19210} {\bibinfo {title} {The commutant of fermionic gaussian unitaries}} (\bibinfo {year} {2026}),\ \Eprint {https://arxiv.org/abs/2603.19210} {arXiv:2603.19210 [quant-ph]} \BibitemShut {NoStop}%
\bibitem [{\citenamefont {Krawtchouk}(1929)}]{Krawtchouk1929}%
  \BibitemOpen
  \bibfield  {author} {\bibinfo {author} {\bibfnamefont {M.}~\bibnamefont {Krawtchouk}},\ }\href@noop {} {\bibfield  {journal} {\bibinfo  {journal} {Comptes Rendus de l'Acad{\'e}mie des Sciences}\ }\textbf {\bibinfo {volume} {189}},\ \bibinfo {pages} {620} (\bibinfo {year} {1929})}\BibitemShut {NoStop}%
\bibitem [{\citenamefont {Tarabunga}(2026)}]{tarabunga2026fermionic}%
  \BibitemOpen
  \bibfield  {author} {\bibinfo {author} {\bibfnamefont {P.~S.}\ \bibnamefont {Tarabunga}},\ }\href@noop {} {\bibinfo {title} {Fermionic non-{Gaussianity} via {Bell} sampling: monotones and efficient quantum algorithms}} (\bibinfo {year} {2026}),\ \Eprint {https://arxiv.org/abs/2606.05066} {arXiv:2606.05066} \BibitemShut {NoStop}%
\end{thebibliography}%

\onecolumngrid % Switch to single column
\section*{End Matter} 
\twocolumngrid % Switch back to double column

\setcounter{equation}{0}
\renewcommand{\theequation}{A\arabic{equation}}

\paragraph{MPO construction for Pairing tensors.} To simplify the ensemble averages in Eq.~\eqref{eq:quantifier_rmps_fgus_ptbasis}, we construct a compact MPO representation of the pairing-tensors (PT) operators $\Upsilon_n^{(2)}$ with bond dimension $n+1$. This reduces the evaluation of \(\alpha_n(\varrho^{(2)}_{\mu_\chi})\) to low dimensional tensor network contractions. Introducing the two-replica bridge generators \(\Gamma_{\mu_i}=\gamma_{\mu_i}\otimes\gamma_{\mu_i}\), we write
\begin{align}
    \Upsilon_n^{(2)} = C_n\sum_{1 \le \mu_1 < \mu_2 \cdots < \mu_n \le 2N} \Gamma_{\mu_1} \Gamma_{\mu_2} \cdots \Gamma_{\mu_{n}}.
\end{align}
We then invoke finite state machine framework to generate the local MPO tensor \(\mathbb W_n^{[j]}\).  For convenience, we introduce the two-replica Pauli operators, \(\{\mathcal I=I\otimes I,\,\mathcal X=X\otimes X,\,\mathcal Y=Y\otimes Y,\,\mathcal Z=Z\otimes Z\} \in \mathbb C^2\otimes \mathbb C^2\). 
Under the Jordan--Wigner transformation, each bridge generator $\Gamma_{\mu_i}$ maps to a Pauli string consisting of $\mathcal Z_j$ on sites $j<\lceil\mu_i/2\rceil$, followed by $\mathcal X_k$ ($\mathcal Y_k$) for odd (even) $\mu_i$ at $k=\lceil\mu_i/2\rceil$, with $\mathcal I$ on all remaining sites.
This Pauli-string structure determines the states of the finite state machine and the transition rules.

The finite state machine requires $n+1$ states $\{\mathbf{S}_0,\mathbf{S}_1,\ldots,\mathbf{S}_n\}$ that record the number of ordered bridge generators incorporated while traversing the Pauli string from left to right generated upto the current site \(j\). After processing the site \(j\), the state $\mathbf{S}_k$ (\(0\le k\le n\)) indicates that the terminal $\mathcal{X}$ or $\mathcal{Y}$ operators associated with $\Gamma_{\mu_1},\ldots,\Gamma_{\mu_k}$ have been inserted, whereas those associated with $\Gamma_{\mu_{k+1}},\ldots,\Gamma_{\mu_n}$ have not yet been reached. Equivalently this condition reads 
\begin{align}
j\ge\lceil\mu_i/2\rceil \quad \forall\,i\le k,
  \qquad
  j<\lceil\mu_i/2\rceil \quad \forall\,k<i\le n.
\end{align} Thus, \(\mathbf{S_0}\) denotes that none of the \(\Gamma_{\mu_i}\) have been reached, whereas \(\mathbf{S_n}\) corresponds to all \(n\) bridge generators having been incorporated in the generated Pauli string.

The generation of the two-replica Pauli strings along the chain is governed by the transition matrix of the finite state machine.  The row and column indices of $\mathbb W_n^{[j]}$ label the counts before and after processing site $j$, respectively. Its entry $(\mathbb W_n^{[j]})_{k,\ell}$ encodes the transition $\mathbf S_k \to\mathbf S_\ell$. Since each site hosts at most two selected terminal operators, the allowed transitions are \(\mathbf{S_k}\!\rightarrow\!\mathbf{S_k}, \;\mathbf{S_k}\!\rightarrow\!\mathbf{S_{k+1}} \;\text{ and } \; \mathbf{S_k}\!\rightarrow\!\mathbf{S_{k+2}},\) corresponding to the insertion of zero, one, or two terminal operators satisfying 
$j=\lceil\mu_i/2\rceil$, respectively. The transition $\mathbf S_k\to\mathbf S_{k+2}$ requires the
terminal operators of $\Gamma_{\mu_{k+1}}$ and $\Gamma_{\mu_{k+2}}$ to occur at the current site $j$, that is,
$\lceil\mu_{k+1}/2\rceil=\lceil\mu_{k+2}/2\rceil=j$. Consequently, the local transition operators utilized in constructing the MPO \(\mathbb W_n^{[j]}\) are
\begin{align}
T_{\mathrm{init}} &=
\begin{cases}
\mathcal I, & n \text{ even},\\
\mathcal Z, & n \text{ odd},
\end{cases}
&
T_{\mathrm{local}} &=
\begin{cases}
-\mathcal Z, & n \text{ even},\\
-\mathcal I, & n \text{ odd},
\end{cases}
\nonumber\\[1mm]
T_{\mathrm{start}} &=
\begin{cases}
-(\mathcal X+\mathcal Y), & n \text{ even},\\
\mathcal X+\mathcal Y, & n \text{ odd},
\end{cases}
&
\tilde{T}_x &= \mathcal Z T_x,
\end{align}
where \(x\in\{\mathrm{init},\mathrm{local}\}\). For even $p=0,2,4,\ldots$, the nonzero elements of $\mathbb{W}_n^{[j]}$ are
\begin{align}
\big(\mathbb W^{[j]}_{(n)}\big)_{p,p}=T_{\mathrm{init}},\quad
\big(\mathbb W^{[j]}_{(n)}\big)_{p,p+1}=T_{\mathrm{start}},\quad
\nonumber\\
\big(\mathbb W^{[j]}_{(n)}\big)_{p+1,p+1}=\tilde T_{\mathrm{init}},\quad
\big(\mathbb W^{[j]}_{(n)}\big)_{p+1,p+2}=-T_{\mathrm{start}},\nonumber\\
\big(\mathbb W^{[j]}_{(n)}\big)_{p,p+2}=T_{\mathrm{local}},\quad
\big(\mathbb W^{[j]}_{(n)}\big)_{p+1,p+3}=\tilde T_{\mathrm{local}},\nonumber
\end{align}
With boundary vectors
$v_L=(1,0,\ldots,0)$ and $v_R=(0,\ldots,0,1)^{\mathsf T}$, the resulting MPO is
\begin{align}
    \Upsilon_n^{(2)} &= C_n \; v_L \left(\prod_{j=1}^{N} \mathbb W^{[j]}_{n} \right)v_R.
    \label{eq:MPO_Ug}
\end{align}
Its graphical representation, used in Eq.~\eqref{eq:Upsilon_MPO_form}, is \(v_L \left(\prod_{j=1}^{N} \mathbb W^{[j]}_{(n)}  \right)v_R = \adjustbox{valign=c}{\includegraphics[height=2em]{ MPO_PTbasis.pdf}}\). Each $\mathbb W_n^{[j]}$ is an $(n+1)\times(n+1)$ array of $4\times4$ operators acting on the two replica qubits at site $j$. The product contracts auxiliary indices, with tensor products understood between physical operators on different sites.
    
\paragraph{Details of Resource quantifiers.} We derive the coefficients $\beta_n(Q)$ appearing in Eq.~\eqref{eq:quantifier_rmps_fgus_ptbasis}. For the IPR, the inclusion of the reflection operator \(X_N\in\mathcal U_G(D)\) reduces to \(\beta_n(I_{2})=2^N\Tr\big[\rho_0^{\otimes2}\Upsilon_n^{(2)}\big]\).
The odd coefficients vanish because of the parity of the vacuum state, while the even sectors give
\begin{align}
    \beta_{2\ell}(I_{2}) = (-1)^{\ell} \,\binom{2N}{2\ell}^{-1/2}\binom{N}{\ell}
\end{align}

The calculation for the coefficient of subsystem purity is a bit more involved. Eq.~(\ref{eq:quantifier_rmps_fgus_ptbasis}) suggests that \(\beta_n(\mathcal P_m) = C_n \sum_{|S|=n} \Tr[F^{\otimes m} \, \mathbb I^{\otimes N-m} (\gamma_S \otimes \gamma_S)]\).
The SWAP operator restricted to the \(m\)-qubit subsystem, obeys the relation \(\Tr[F^{\otimes m} \, \mathbb I^{\otimes N-m} (\gamma_S \otimes \gamma_S)] = \Tr_m[(\gamma_{S}^{m})^2] \, \Tr_{N-m}^2[\gamma_{S}^{N-m}]\). Here, \(\gamma_{S}^{m}\) and \(\gamma_{S}^{N-m}\) denote the components of the Majorana string \(\gamma_S\) supported on the \(m\)-qubit subsystem and its complement, respectively. 
Any nontrivial Pauli operator on the complement is traceless that is, \(\Tr_{N-m}^2[\gamma_{S}^{N-m}] = 0\). So, only Majorana strings supported entirely within the $m$-qubit subsystem contribute. Using the Majorana anti-commutation relations, \((\gamma_S)^2 = (-1)^{n(n-1)/2} \gamma_{\mu_1}^2 \gamma_{\mu_2}^2 \ldots  \gamma_{\mu_{|S|}}^2\), the Majorana strings fully supported on \(m-\)qubit bipartition yield \(\Tr_{N-m}(\gamma^{N-m}_{S})=2^{N-m}\) and \(\Tr_{m}((\gamma^{m}_{S})^2) = (-1)^{\lfloor n/2\rfloor}2^m\). Consequently, summing over the $\binom{2m}{n}$ contributing strings gives
\begin{align}
    \beta_{n}(\mathcal P_m) = C_n \binom{2m}{n} (-1)^{\lfloor n/2 \rfloor}\,2^{2N-m}.
\end{align}

\begin{figure}
    \centering
    \includegraphics[width=1\linewidth]{ 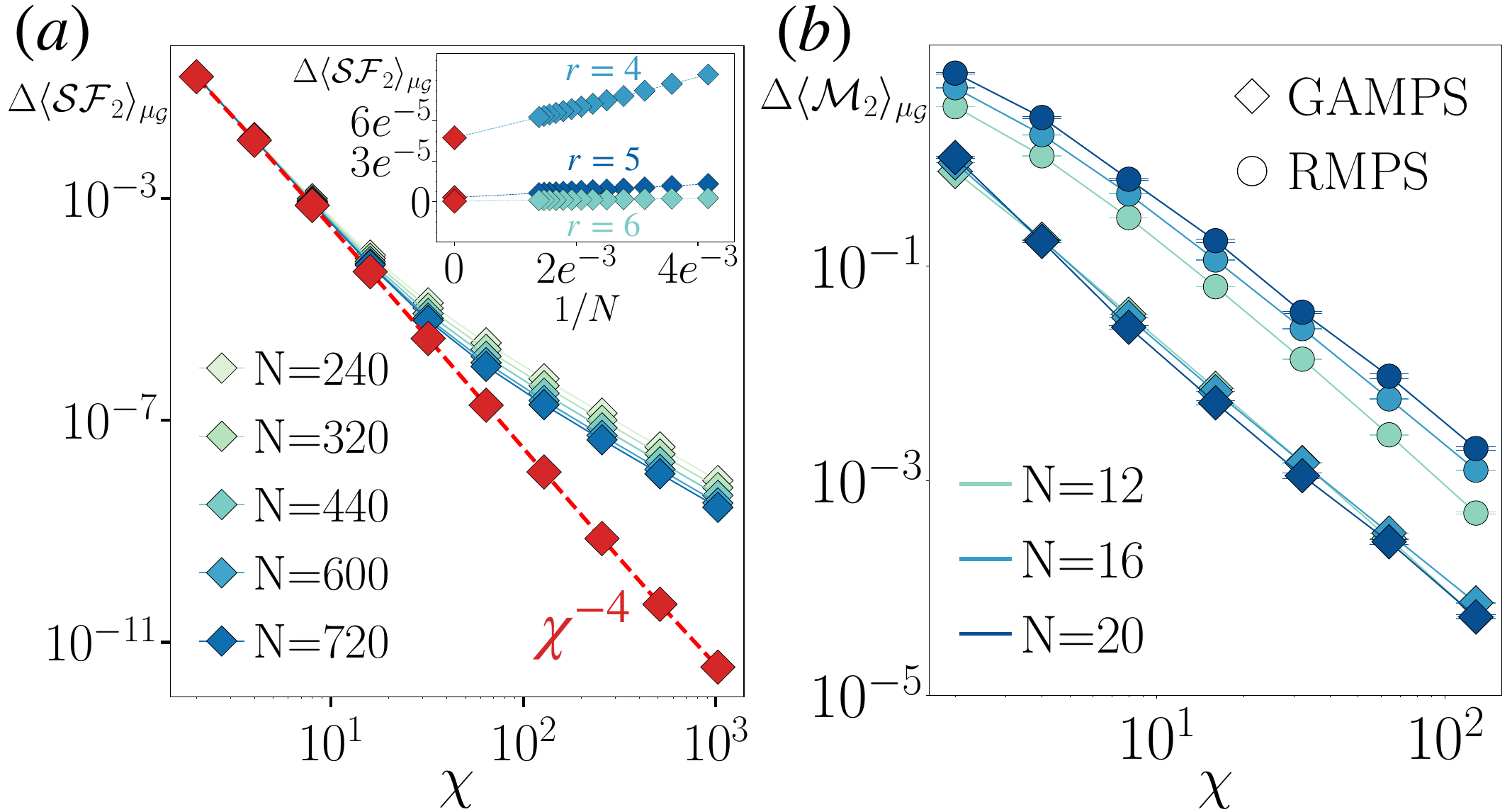}
    \caption{Second order state frame potential and the second order Stabilizer Renyi entropy for \(N\) qubits. \((a)\) The deviation \(\Delta \langle \mathcal{SF}_{2} \rangle_{\mu_{\mathcal G}}\) as a function of the bond dimension exhibits an asymptotic power-law \(\chi^{-4}\) decay upon extrapolation to the thermodynamic limit (inset). \((b)\) The second-order SRE of GAMPS remains consistently below that of RMPS by an approximately constant offset as the bond dimension \(\chi\) increases while depicting a scaling independent of the system size unlike the RMPS.}
    \label{fig:fig4}
\end{figure}

\paragraph{State frame potential.} We further characterize the state two-design properties of GAMPS through the second order state frame potential , \(\langle \mathcal{SF}_{2}\rangle_{\mu_\mathcal G} = \mathbb E_{\Psi_\mathcal G, \Psi'_\mathcal G \sim \mu_\mathcal G}\big(|\langle\Psi_\mathcal G|\Psi'_\mathcal G\rangle|^4\big) \) \cite{Gross07Unitarydesign,Mele2024introductiontoHaar}. Analogous to the previous discussion, the quadratic structure allows us to identify the boundary operator in Eq.~\eqref{eq:quantifier_general} as \(B_{\mathcal{SF}_{2}} = \mathbb E_{\Psi'_\mathcal G \sim \mu_\mathcal G}(\Psi^{' \, \otimes 2}_\mathcal G)\). Using the PT expansion in Eq.~\eqref{eq:quantifier_rmps_fgus_ptbasis}, we obtain 
\(\beta_n(\mathcal{SF}_{2}) = \sum_{n'}
 \alpha_{n'}(\varrho^{(2)}_{\mu_\chi})
\Tr\big[\Upsilon_{n}^{(2)}\Upsilon_{n'}^{(2)}\big]\) where orthonormality of the pairing tensors results in
\begin{align}
    \beta_n(\mathcal{SF}_{2}) =  \alpha_n(\varrho^{(2)}_{\mu_\chi})
\end{align}
We quantify the deviation from the Haar SFP through the relative state frame potential, \(\Delta \langle \mathcal{SF}_{2} \rangle_{\mu_{\mathcal G}} = \langle \mathcal{SF}_{2}\rangle_{\mu_\mathcal G} /\langle \mathcal{SF}_{2}\rangle_{\mathrm{Haar}} - 1\). This quantity bounds the trace distance as,
\begin{align}
    \mathcal D_{\mathrm H}(\Psi_\mathcal G) \leq \sqrt{\Delta \langle \mathcal{SF}_{2} \rangle_{\mu_{\mathcal G}}}.
\end{align}
so $\Delta\langle\mathcal{SF}_2\rangle_{\mu_{\mathcal G}}\leq\delta^2$ is sufficient for a $\delta$-approximate state two-design. Fig.~\ref{fig:fig4}(a) illustrates that the relative frame potential approaches the asymptotic scaling $\chi^{-4}$, converging faster than the $\chi^{-2}$ behavior of RMPS \cite{Lami25anti}. This result is consistent with the trace-distance scaling and yields an \(\epsilon\)-approximate state \(2\)-design with \(\epsilon\sim\mathcal{O}(\chi^{-2})\).

\begin{figure}[!t]
    \centering
    \includegraphics[width=0.65\linewidth]{ 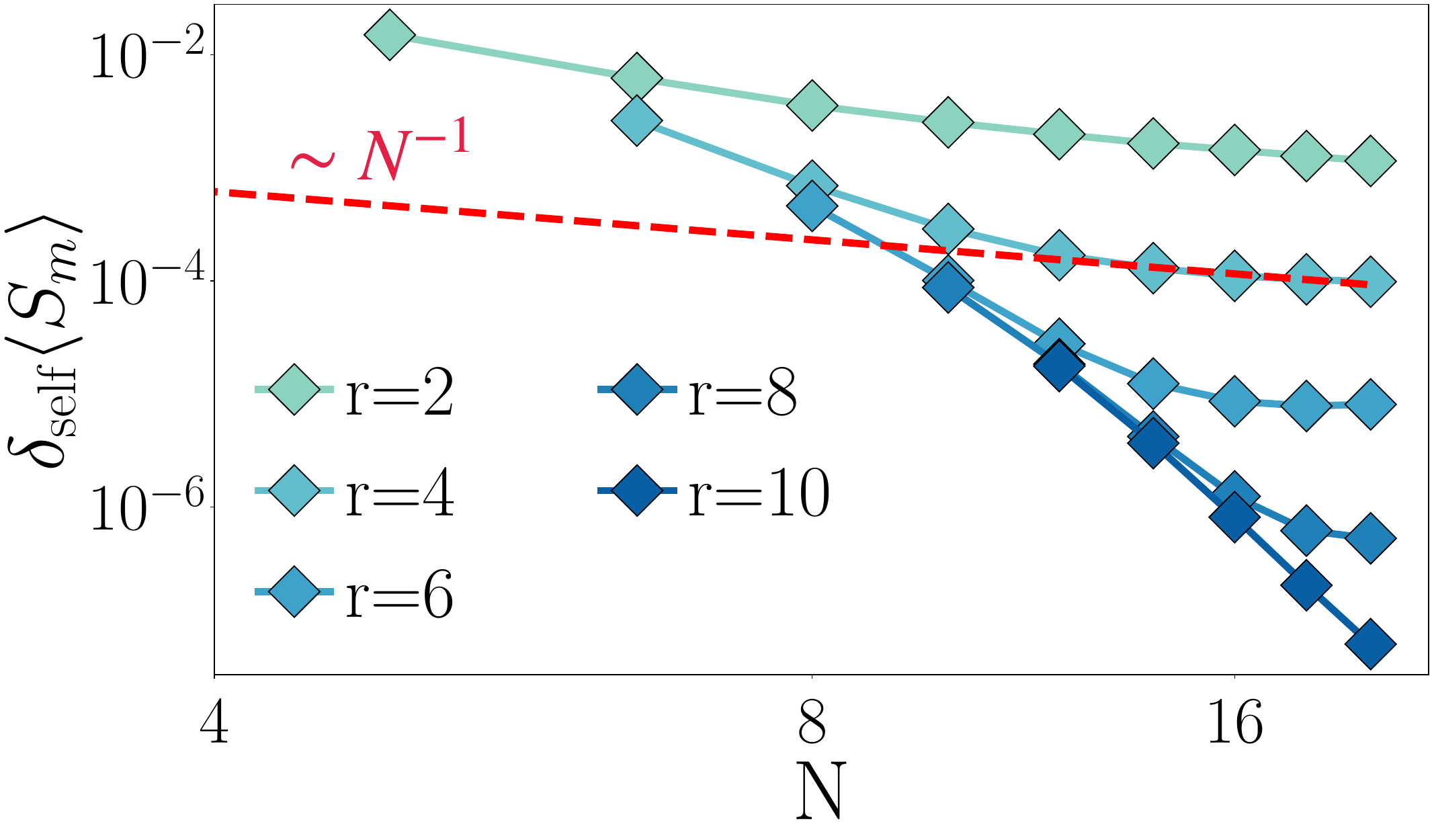}
    \caption{For GAMPS, the deviation between the quenched and annealed average R\'enyi-2 entropies decreases with system size, exhibiting an asymptotic \(N^{-1}\) power-law decay for different values of \(r=\log_2{\chi}\) indicative of self-averaging.}
    \label{fig:self_avg}
\end{figure}
\paragraph{Stabilizer R\'enyi entropy.}
We quantify nonstabilizerness, or magic, through the stabilizer R\'enyi entropy (SRE) \cite{leone2022stabilizerrenyientropy},
\begin{align}
\mathcal{M}q(|\psi\rangle) = \frac{1}{1-q}\log_2\bigg[\sum_{P \in P_N}\frac{\langle \psi | P |\psi \rangle^{2q}}{2^N}\bigg]
\end{align}
where $\mathcal P_N=\{I,X,Y,Z\}^{\otimes N}$ denotes the set of $N$-qubit Pauli strings, and \(q \geq 2\). The SRE vanishes for stabilizer states and increases as a state departs from the stabilizer manifold. Low-bond-dimension RMPS can already exhibit substantial non-stabilizerness, which may be further enhanced by fermionic Gaussian transformations in the qubit representation \cite{haug2023quantifying,tarabunga2024mps,frau2024nonstabilizernessversusentanglementmatrix,Lami25cmps,Collura24fermionicgaussian}. We examine whether this augmentation drives the SRE toward its Haar value faster than for RMPS.

We define the deviation from the Haar value as
\begin{align}
\Delta\langle\mathcal{M}_2\rangle_{\mu_{\mathcal G}}
=
\langle\mathcal{M}_2\rangle_{\mu_{\mathrm H}}
-
\langle\mathcal{M}_2\rangle_{\mu_{\mathcal G}}.
\end{align}
Fig.~\ref{fig:fig4}(b) shows this deviation as a function of $\chi$. Both ensembles approach the Haar value with an asymptotic power-law scaling $\chi^{-2}$. However, the GAMPS data remain systematically below the RMPS results, with the two ensembles exhibiting the same asymptotic scaling but differing by an approximately constant offset. In addition, GAMPS exhibit negligible dependence on $N$ unlike the RMPS SRE deviation. These results show that Gaussian augmentation drives the ensemble toward Haar-typical non-stabilizerness at smaller bond dimensions than RMPS while strongly suppressing its dependence on system size.

\paragraph{Self-averaging properties of GAMPS.} The replica tensor-network method directly yields the ensemble-averaged subsystem purity $\langle\mathcal{P}_m\rangle_{\mu_{\mathcal G}}$, but not the quenched R\'enyi-2 entropy \(\langle S_m\rangle_{\mu_{\mathcal G}}^{\mathrm{que}} = -\mathbb{E}_{\mu_{\mathcal G}}
\left[\log_2\mathcal{P}_m\right]\). Consequently, in our calculations so far, we use the efficiently accessible annealed entropy \(\langle S_m \rangle_{\mu_\mathcal G}^{\mathrm{ann}} = -\log_2(\langle P_m \rangle_{\mu_\mathcal G})\). To establish its equivalence to the quenched entropy, we quantify their relative difference through \(\delta_{\mathrm{self}} \langle S_m \rangle_{\mu_\mathcal G} = 1-\langle S_m \rangle_{\mu_\mathcal G}^{\mathrm{ann}}/\langle S_m \rangle_{\mu_\mathcal G}^{\mathrm{que}}\). Fig.~\ref{fig:self_avg} shows the asymptotic power-law scaling exhibited by
$\delta_{\mathrm{self}} \langle S_m \rangle_{\mu_\mathcal G} \sim N^{-1}$ resulting in,
\begin{align}
\delta_{\mathrm{self}} \langle S_m \rangle_{\mu_\mathcal G}
\xrightarrow[N\rightarrow\infty]{}
0.
\end{align}
Therefore, the annealed and quenched entropies coincide in the thermodynamic limit, implying that a typical GAMPS realization exhibits the ensemble-averaged entanglement entropy and establishing self-averaging.

\clearpage

\begin{bibunit}[apsrev4-2]

\begingroup

\resetrevtexfrontmatter

\title{Supplemental Material for:\\
``Random Gaussian Augmented Matrix Product States''}

\author{Stavya Puri\,\orcidlink{0009-0001-4889-6235}}
\email{stavya.puri@bsc.es}
\affiliation{Barcelona Supercomputing Center,
Plaça Eusebi Güell 1--3, 08034 Barcelona, Spain}
\affiliation{Departament de Física Quàntica i Astrofísica,
Facultat de Física, Universitat de Barcelona (UB),
Martí i Franquès 1, 08028 Barcelona, Spain}

\author{Poetri Sonya Tarabunga\,\orcidlink{0000-0001-8079-9040}}
\affiliation{Technical University of Munich,
TUM School of Natural Sciences, Physics Department,
85748 Garching, Germany}
\affiliation{Munich Center for Quantum Science and Technology (MCQST),
Schellingstraße 4, 80799 München, Germany}

\author{Piotr Sierant\,\orcidlink{0000-0001-9219-7274}}
\email{piotr.sierant@bsc.es}
\affiliation{Barcelona Supercomputing Center,
Plaça Eusebi Güell 1--3, 08034 Barcelona, Spain}

\maketitle

\endgroup
\onecolumngrid

\tableofcontents

\setcounter{equation}{0}
\renewcommand{\theequation}{S\arabic{equation}}

\section{Two-replica random matrix product states}
\label{sec:rmps}
Any pure state
\(
|\psi\rangle
=
\sum_{i_1,\ldots,i_N}
c_{i_1,\ldots,i_N}
|i_1\cdots i_N\rangle
\)
admits an exact matrix product state (MPS) representation parametrized by the bond dimension \(\chi\) depicted as
\begin{align}
    |\psi\rangle=\sum_{\mathbf{i}}\sum_{\lambda_1, \lambda_2, \cdots, \lambda_{N-1}} A^{i_1}_{\lambda_1} A^{i_2}_{(\lambda_1,\lambda_2)} \cdots A^{i_N}_{\lambda_{N-1}} |\mathbf{i}\rangle = \adjustbox{valign=c}{\includegraphics[height=4em]{ 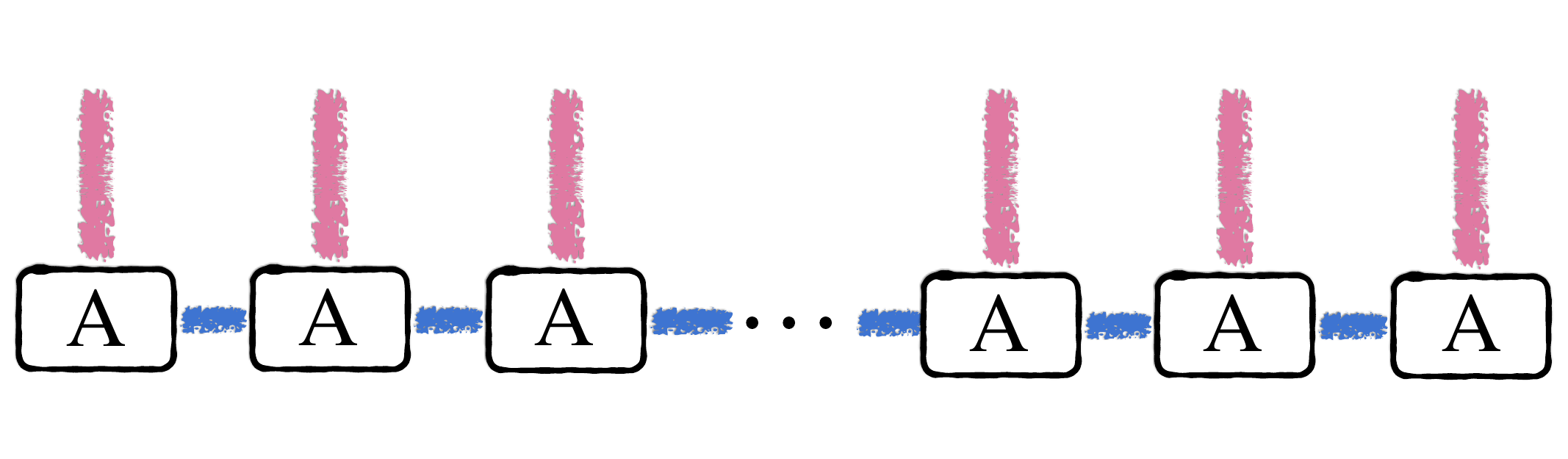}}
\end{align}
where
\(|\mathbf{i}\rangle=|i_1 i_2\cdots i_N\rangle\). The blue lines denote the virtual bond indices
\(\lambda_j\in\{0,\ldots,\chi-1\}\), while the pink lines denote the physical indices
\(i_j\) at site \(j\). To investigate typical states at fixed bond dimension \(\chi\), we consider an ensemble of random matrix-product states (RMPS). We parametrize each local MPS tensor by a unitary \(U_j\) acting on the physical and virtual degrees of freedom,
\begin{equation}
A^{i_j}_{\lambda_{j-1},\lambda_j} = \langle i_j,\lambda_j| U_j |\lambda_{j-1},0\rangle
\label{eq:rmps-local-tensor}
\end{equation}
with the boundary indices fixed by
\(\lambda_0=\lambda_N=1\). The RMPS density matrix inherits the sequential unitary structure of the underlying state,
\begin{align}
\rho_{\Phi_\chi}
=
\ket{\Phi_\chi}\!\bra{\Phi_\chi}
=
\adjustbox{valign=c}{
\includegraphics[height=5.5em]{ density_mps_unitary_chain.pdf}
}.
\label{eq:density_Unitary_chain}
\end{align}
The unitary network structure of the RMPS implies that taking the expectation of this RMPS is equivalent to independently taking Haar average over the local Unitary \(\mathcal U(d\chi) \in \mu_\mathrm{H}\) acting on state \(\rho_0 = |0\rangle\langle0|\). For graphical clarity, the physical indices of both the ket and bra layers are placed above each tensor. Since the unitaries $\{U_j\}$ are sampled independently, the full RMPS ensemble average factorizes into local Haar averages before the virtual indices are contracted. Graphically,
\begin{align}
    \mathbb E_{\mu_\chi}[\rho_{\Phi_\chi}] = \mathbb E\bigg[\adjustbox{valign=c}{\includegraphics[height=5em]{ density_mps_unitary_chain.pdf}}\bigg] = \mathbb E\bigg[\adjustbox{valign=c}{\includegraphics[height=3em]{ 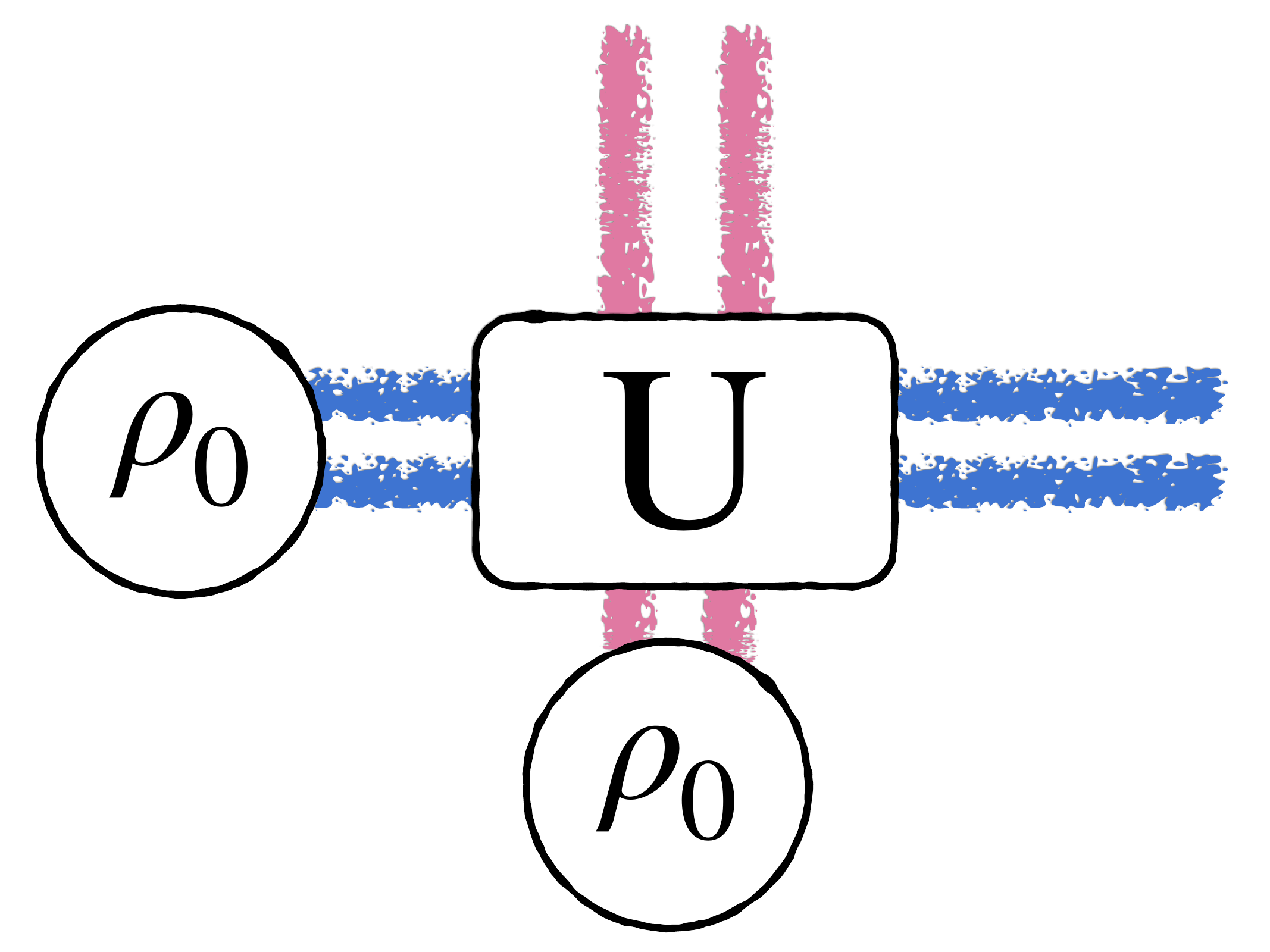}}\bigg]  \; \mathbb E\bigg[\adjustbox{valign=c}{\includegraphics[height=3em]{ 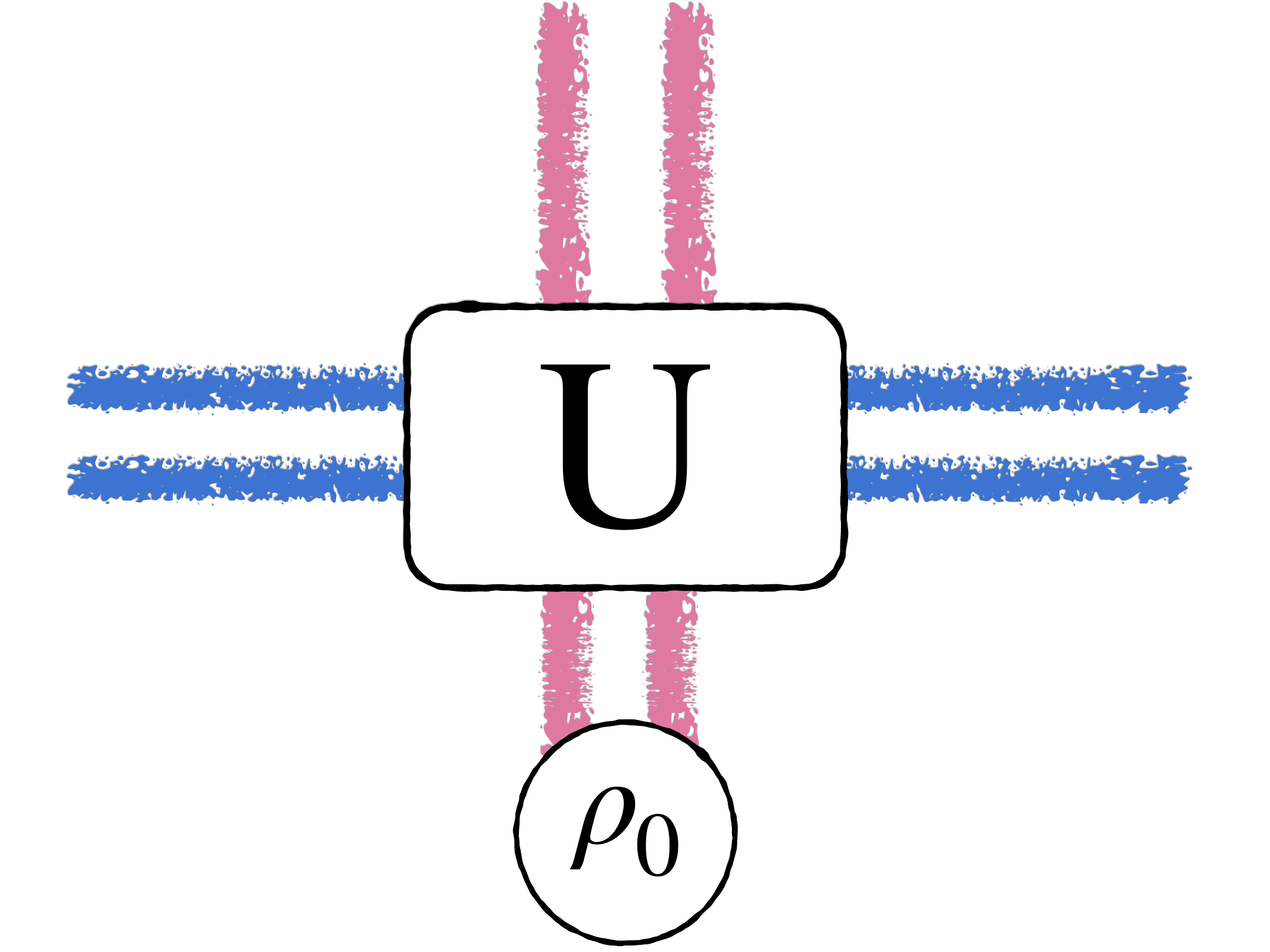}}\bigg] \; \cdots \; \mathbb E\bigg[\adjustbox{valign=c}{\includegraphics[height=3em]{ density_mps_unitary_op.pdf}}\bigg].
\end{align}
The resource quantifiers considered in the main text depend quadratically on the density matrix and therefore require the second ensemble moment \(\mathbb E_{\mu_\chi}\left[\rho_{\Phi_\chi}^{\otimes2}
\right]\). Replicating the unitary network before performing the ensemble average preserves the sitewise factorization,
\begin{align}
    \mathbb E\bigg[\bigg(\adjustbox{valign=c}{\includegraphics[height=5em]{ density_mps_unitary_chain.pdf}}\bigg)^{\otimes 2}\bigg] = 
    \mathbb E\bigg[\bigg(\adjustbox{valign=c}{\includegraphics[height=3em]{ density_mps_unitary_op2.pdf}}\bigg)^{\otimes 2}\bigg]  \;
    \mathbb E\bigg[\bigg(\adjustbox{valign=c}{\includegraphics[height=3em]{ density_mps_unitary_op.pdf}}\bigg)^{\otimes 2}\bigg]  \; \cdots \; 
    \mathbb E\bigg[\bigg(\adjustbox{valign=c}{\includegraphics[height=3em]{ density_mps_unitary_op.pdf}}\bigg)^{\otimes 2}\bigg].
\end{align}
Importantly, this expression is the average of $\rho_{\Phi_\chi}^{\otimes2}$ generated by the same RMPS realization in both replicas, rather than the tensor product of two independently averaged density matrices.

At each site, the replicated local average is the second-moment Haar-twirling channel acting on $\rho_0^{\otimes2}$,
\begin{align}
\mathbb E
\left[
\left(
\adjustbox{valign=c}{
\includegraphics[height=3em]{ density_mps_unitary_op.pdf}
}
\right)^{\otimes2}
\right]
&=
\mathbb E_{U\sim\mu_{\mathrm H}}
\left[
U^{\otimes2}
\rho_0^{\otimes2}
(U^\dagger)^{\otimes2}
\right]
\nonumber\\
&\equiv
\mathcal T_{\mathrm{Haar}}^{(2)}
\left(
\rho_0^{\otimes2}
\right).
\label{eq:local_second_moment_twirl}
\end{align}

Consequently, we evaluate the local Haar twirl using Schur-Weyl duality, which restricts the second-moment channel to the permutation algebra of $S_2$. For a local unitary acting on $\mathbb C^{D} = \mathbb C^d\otimes\mathbb C^\chi$, the twirled tensor takes the form
\begin{align}
    \mathcal{T}^{(2)}_{\mathrm{Haar}} (\rho_0^{\otimes 2}) = \sum_{\sigma,\tau \in S_2} P_{\sigma}^{(d \chi)} W_{\sigma, \tau}^{(d\chi)} \Tr_d[(P_{\tau}^{(d\chi)})^T(\rho_0^{\otimes 2})]
    \label{eq:local_haar_permutation}
\end{align}
where $d=2$ is the physical qubit dimension and the partial trace leaves the virtual indices open. The permutation operators factorize across the physical and virtual spaces according to \(P_{\sigma}^{d\chi} = \sum_{\theta,\eta} P_{\theta}^{\chi}P_{\eta}^{d}\delta_{\sigma,\theta}\delta_{\theta,\eta}\).
Using this decomposition, Eq.~\eqref{eq:local_haar_permutation} becomes
\begin{align}
    \mathcal{T}^{(2)}_{\mathrm{Haar}} (\rho_0^{\otimes 2}) = \sum_{\sigma,\tau,\theta,\eta \in S_2} P_{\theta}^{(d)}  P_{\eta}^{(\chi)} W_{\sigma, \tau}^{(d\chi)} (P_{\tau}^{(\chi)})^T \Tr_d\left[(P_{\tau}^{(d)})^T(\rho_0^{\otimes 2})\right]\delta_{\sigma,\theta} \delta_{\theta,\eta}
\label{eq:local_haar_factorized}
\end{align}
For two replicas, the permutation basis consists of the identity and swap operators. In the ordered basis $(\mathbb I,\mathbb F)$, the Gram and Weingarten matrices are
\begin{align}
G^{(D)}
&=
\begin{pmatrix}
D^2 & D\\
D & D^2
\end{pmatrix}, \qquad
W^{(D)}
=
\left(G^{(D)}\right)^{-1}
=
\frac{1}{D^2-1}
\begin{pmatrix}
1 & -D^{-1}\\
-D^{-1} & 1
\end{pmatrix}.
\label{eq:second_moment_weingarten}
\end{align}

When neighboring local tensors are contracted, the virtual permutation operators combine through \(G_{\sigma, \tau}^{\chi} = \Tr[(P_{\sigma}^{\chi})^T P_{\tau}^{\chi}]\). This contraction absorbs the virtual permutation and Weingarten reformulates the chain of tensors having on the permutation indices as the free dimension.
\begin{align}
R_{\sigma,\tau}^{(d)}
=
\sum_{\eta\in S_2}
W_{\sigma,\eta}^{(d\chi)}
P_\eta^{(d)}
G_{\eta,\tau}^{(\chi)}.
\label{eq:local_permutation_tensor}
\end{align}
Its equivalent operator and graphical representations are
\begin{align}
R_{\sigma,\tau}^{(d)}
\quad\Longleftrightarrow\quad
\adjustbox{valign=c}{
\includegraphics[height=3em]{ density_mps_permutation3_op.pdf}
}
=
\adjustbox{valign=c}{
\includegraphics[height=3.5em]{ density_mps_permutation4_op.pdf}
}.
\label{eq:MPO_singleterm}
\end{align}
where the yellow curved lines are permutation indices. The corresponding left and right boundary tensors are
\begin{align}
    \adjustbox{valign=c}{\includegraphics[height=3em]{ 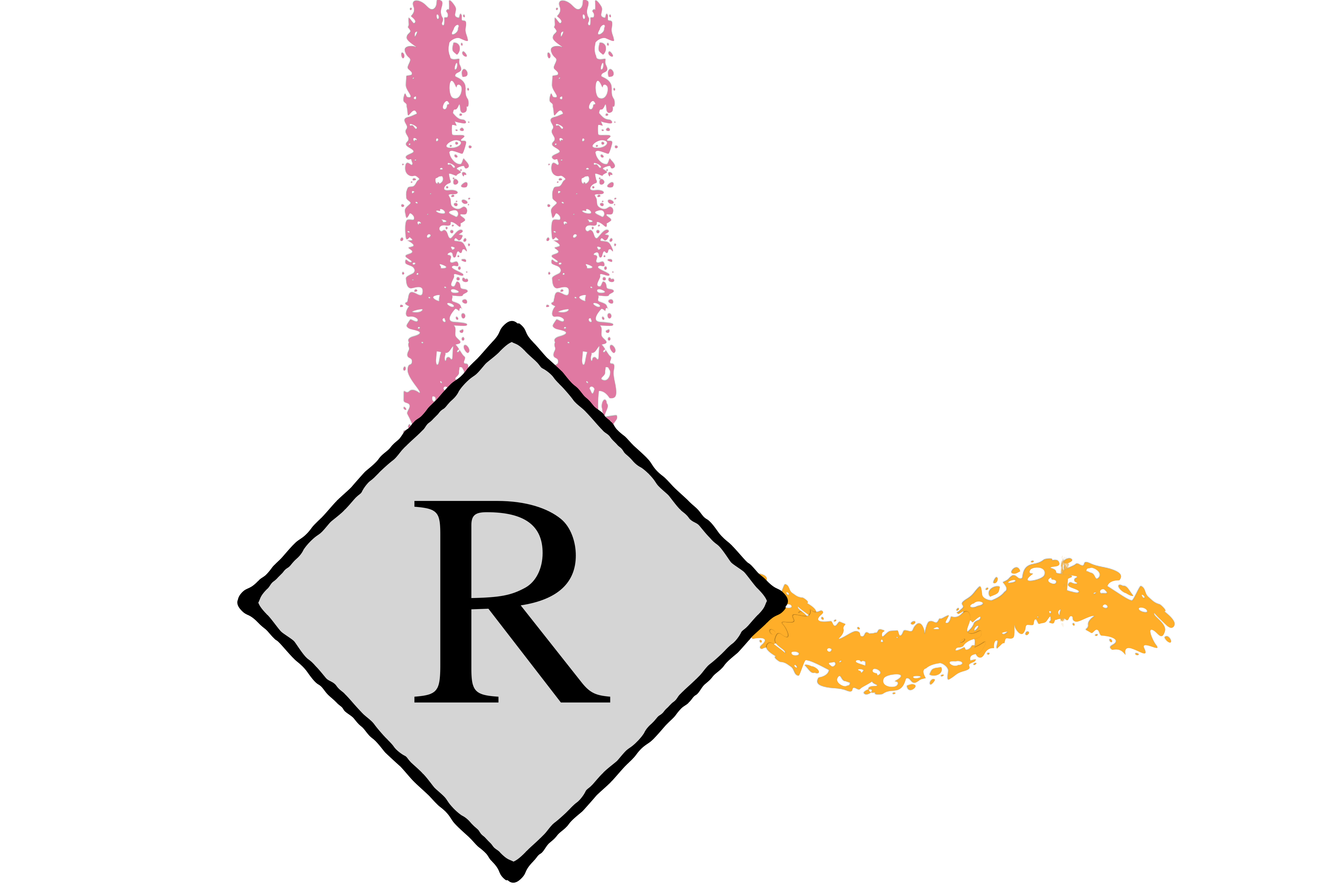}} = \adjustbox{valign=c}{\includegraphics[height=3.5em]{ 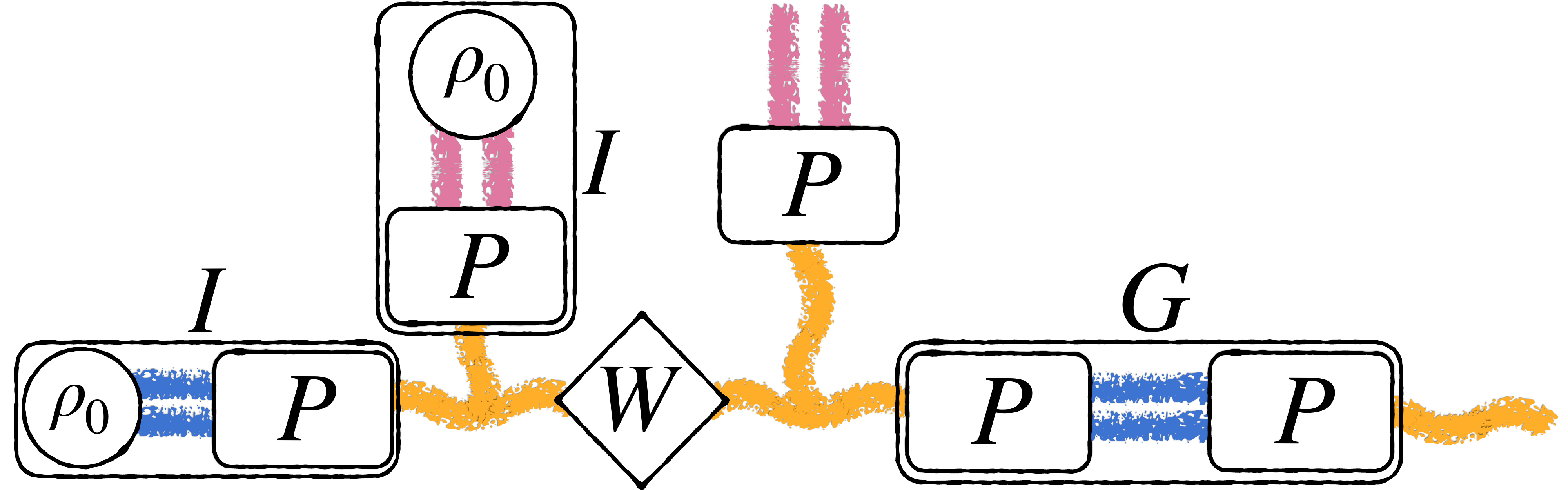}} \quad \text{and} \quad \adjustbox{valign=c}{\includegraphics[height=3em]{ 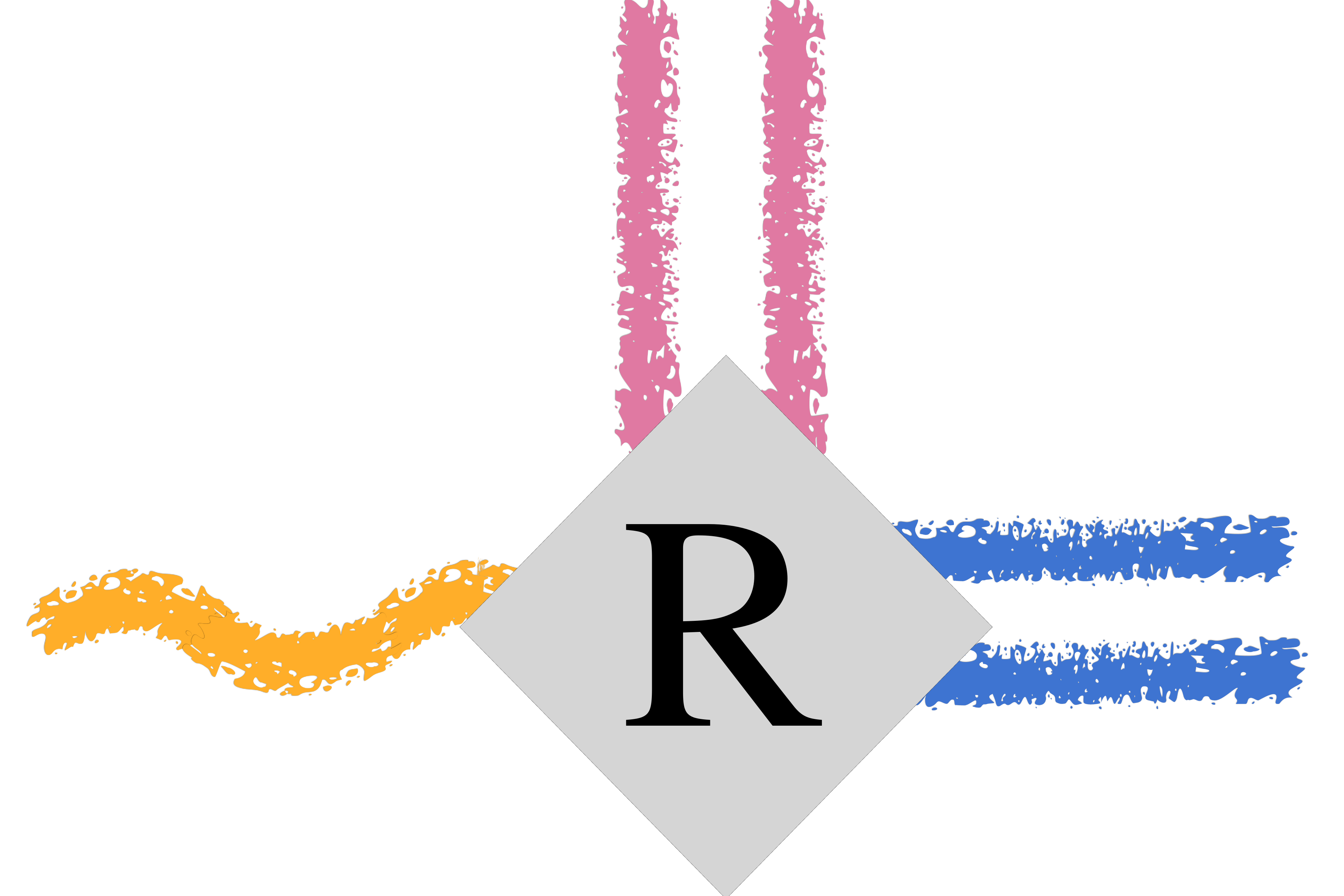}} = \adjustbox{valign=c}{\includegraphics[height=4em]{ 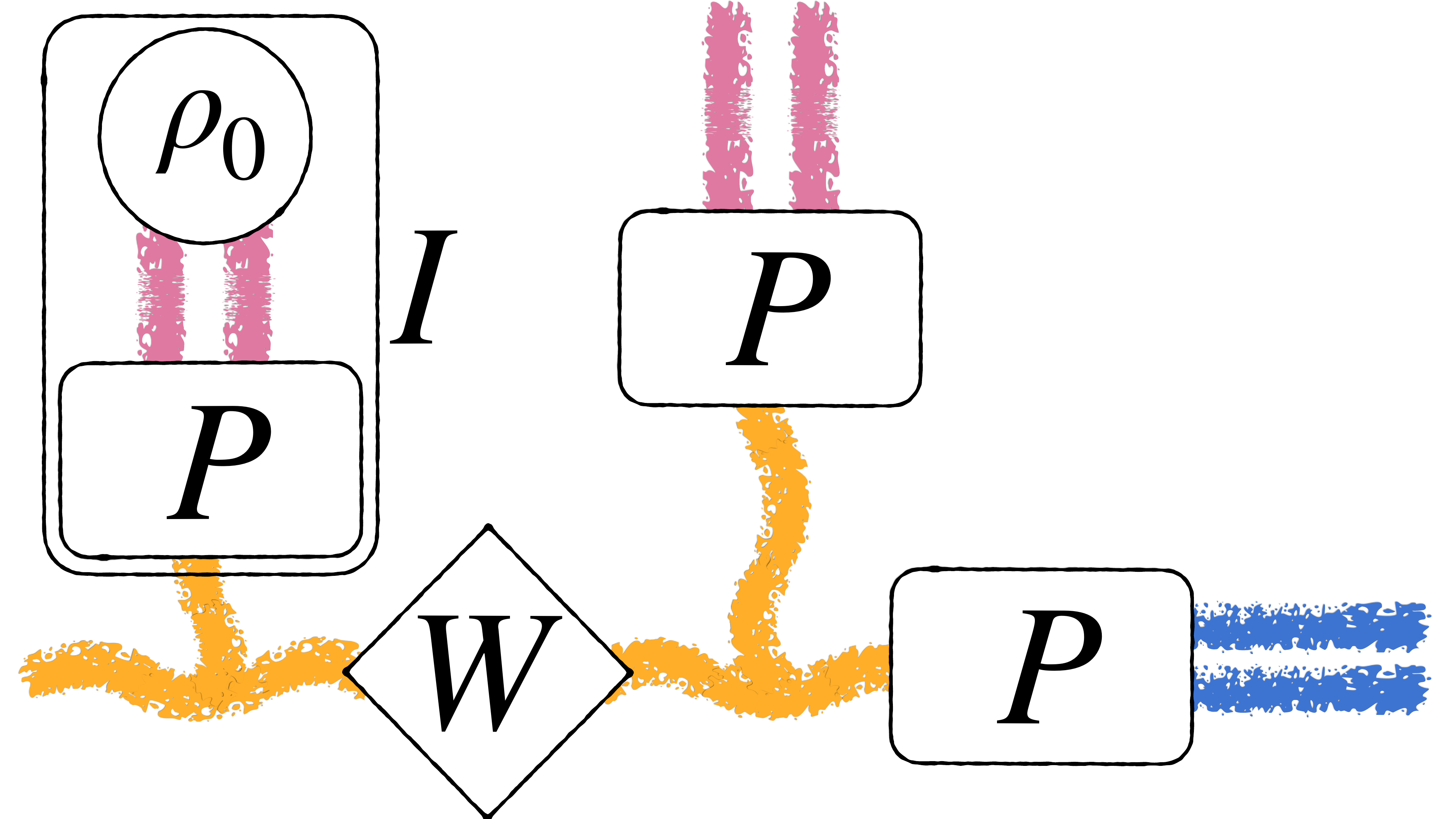}}
\end{align}
respectively. Combining the bulk and boundary tensors gives the MPO representation
\begin{align}
\mathbb E_{\mu_\chi}
\left[
\rho_{\Phi_\chi}^{\otimes2}
\right]
=
\adjustbox{valign=c}{
\includegraphics[height=3em]{ 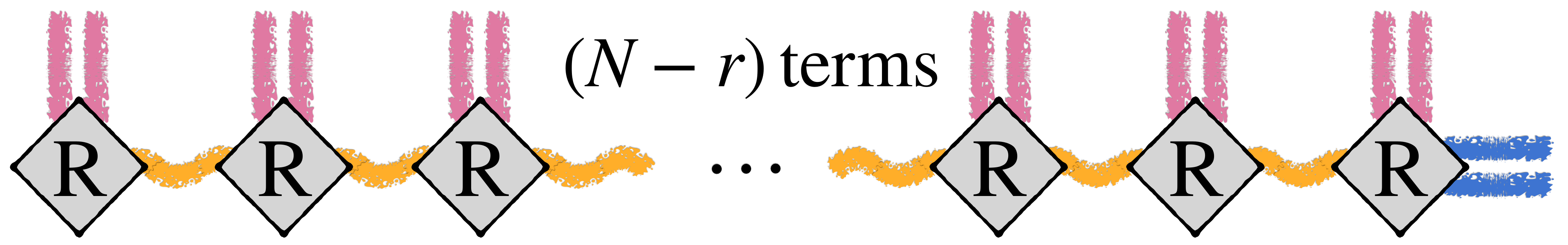}
}.
\label{eq:averaged_rmps_permutation_mpo}
\end{align}

The original replicated network retains virtual indices whose dimensions grow with $\chi$. After Haar averaging, these indices are replaced by permutation labels $\sigma\in S_2$, giving an MPO bond dimension $|S_2|=2$ that is independent of $\chi$. Together with the closed-form expressions for $W_{\sigma,\tau}$, $G_{\sigma,\tau}$, and $P_\sigma$, this representation substantially simplifies both analytical calculations and numerical contractions. This MPS network yields an MPO form for the replica averged RMPS as \(\varrho_{\mu_\chi}^{(2)} = \mathbb E_{\mu_\chi}[\rho^{\otimes 2}_{\Phi_\chi}]\)~\cite{Haag2023TypicalTNcorrlength}.

\section{MPO construction for pairing tensors}

The ensemble-averaged resource quantifiers derived in the main text take the form
\begin{align}
\langle Q\rangle_{\mu_{\mathcal G}}
  =\sum_{n=0}^{2N}\beta_n(Q)\,
  \alpha_n(\varrho^{(2)}_{\mu_\chi}),
\label{eq:quantifier_pt_expansion_supp}
\end{align}
where
\(\beta_n(Q) = \Tr\left[B_Q\Upsilon_n^{(2)}\right]\) and \(\alpha_n(\rho)=\Tr[\Upsilon_n^{(2)}\rho^{\otimes2}]\). The coefficients $\beta_n(Q)$ depend only on the boundary operator associated with the quantifier and admit closed form expressions. Meanwhile, $\alpha_n(\varrho^{(2)}_{\mu_\chi})$ contains the RMPS ensemble dependence. To evaluate the latter efficiently, we construct an MPO representation of the pairing-tensor operator $\Upsilon_n^{(2)}$ with bond dimension $n+1$. As derived in Sec.~\ref{sec:rmps}, the evaluation of $\alpha_n(\varrho^{(2)}_{\mu_\chi})$ then reduces tensor-network contraction between this MPO and the averaged RMPS at cost polynomial in \(N\) and \(\chi\).

Following the finite-state-machine construction introduced in the End Matter of the main text, we build the local MPO tensor $\mathbb W_n^{[j]}$ from the two-replica Pauli operators
\begin{align}
\mathcal I=I\otimes I,
\qquad
\mathcal X=X\otimes X,
\qquad
\mathcal Y=Y\otimes Y,
\qquad
\mathcal Z=Z\otimes Z.
\end{align}
Consequently, the required transition operators are
\begin{align}
T_{\mathrm{init}} &=
\begin{cases}
\mathcal I, & n \text{ even},\\
\mathcal Z, & n \text{ odd},
\end{cases}
&
T_{\mathrm{local}} &=
\begin{cases}
-\mathcal Z, & n \text{ even},\\
-\mathcal I, & n \text{ odd},
\end{cases}
\nonumber\\[1mm]
T_{\mathrm{start}} &=
\begin{cases}
-(\mathcal X+\mathcal Y), & n \text{ even},\\
\mathcal X+\mathcal Y, & n \text{ odd},
\end{cases}
&
\tilde{T}_x &= \mathcal Z T_x,
\end{align}
where \(x\in\{\mathrm{init},\mathrm{local}\}\). For even $p=0,2,4,\ldots$, the nonzero entries of the $(n+1)\times(n+1)$ local MPO tensor are
\begin{align}
\big(\mathbb W^{[j]}_{n}\big)_{p,p}=T_{\mathrm{init}},\quad
\big(\mathbb W^{[j]}_{n}\big)_{p,p+1}=T_{\mathrm{start}},\quad
\nonumber\\
\big(\mathbb W^{[j]}_{n}\big)_{p+1,p+1}=\tilde T_{\mathrm{init}},\quad
\big(\mathbb W^{[j]}_{n}\big)_{p+1,p+2}=-T_{\mathrm{start}},\nonumber\\
\big(\mathbb W^{[j]}_{n}\big)_{p,p+2}=T_{\mathrm{local}},\quad
\big(\mathbb W^{[j]}_{n}\big)_{p+1,p+3}=\tilde T_{\mathrm{local}},\nonumber
\end{align}
\label{eq:fsm_local_mpo}
whenever the corresponding indices lie in $\{0,\ldots,n\}$, with all remaining entries equal to zero.

Introducing the boundary vectors \(v_L=(1,0,\ldots,0) \) and \(v_R=(0,\ldots,0,1)^{\mathsf T},\) the pairing tensor admits the compact MPO representation
\begin{align}
    \Upsilon_n^{(2)} &= C_n \; v_L \left(\prod_{j=1}^{N} \mathbb W^{[j]}_{n} \right)v_R = C_n \adjustbox{valign=c}{\includegraphics[height=2.4em]{ MPO_PTbasis.pdf}}.
    \label{eq:Upsilon_MPO_form}
\end{align}
where \(C_n = 2^{-N}\binom{2N}{n}^{-1/2}\) is the normalization constant. The boundary vectors select $\mathbf S_0$ before the first site and $\mathbf S_n$ after the last site. Each $\mathbb W_n^{[j]}$ is an $(n+1)\times(n+1)$ array of $4\times4$ operators acting on the two replica qubits at site $j$.
The product contracts auxiliary indices, with tensor products
understood between physical operators on different sites.

\subsection{\(\mathit{n=2}\) example}
To elucidate the MPO construction, we first consider the representative example of the two-replica Majorana correlator 
\begin{align}
    \Upsilon_2^{(2)} = C_2 \sum_{1\le{\mu_1}<{\mu_2}\leq2N} (\gamma_{\mu_1}\gamma_{\mu_2})\otimes(\gamma_{\mu_1}\gamma_{\mu_2}),
\end{align}
which is also relevant for the analytical evaluation of the FAF. The bridge generators for the two-replica system~\cite{Sierant26matchgatecommutant} are defined as \(\Gamma_{\mu_i} = \gamma_{\mu_i} \otimes \gamma_{\mu_i}\). Consequently, we rewrite \(\Upsilon_2^{(2)} =  \binom{2N}{2}^{-1/2}2^{-N}\sum_{{\mu_1} < {\mu_2}} \Gamma_{\mu_1} \Gamma_{\mu_2}\).

Here, the corresponding finite state machine consists of three states yielding an MPO bond dimension \(D=3\). Each state encodes those \(\Gamma_{\mu_i}\) whose local Pauli operators \(\mathcal X\) or \(\mathcal Y\) have contributed so far, while traversing from left to right upto the current site \(j\) over a possible Pauli string generated by the finite state machine. After processing site $j$, the three possible states are: 
\begin{itemize}
    \item \(\mathbf{State \;S_0}:\) Neither \(\mathcal X\) nor \(\mathcal Y\) associated with either \(\Gamma_{\mu_1}\)or \(\Gamma_{\mu_2}\) has yet been encountered in the Pauli string up to site \(j\), corresponding to \(j<  \lceil\mu_i/2\rceil \) (\(i=1,2\)). Consequently, the string consists solely of \(\mathcal Z\) operators upto the current site. 
    \item \(\mathbf{State \;S_1}:\) One \(\mathcal X\) or \(\mathcal Y\) associated only with \(\Gamma_{\mu_1}\) has been encountered upon traversal of the generated Pauli string so far. Equivalently,   $j\geq\lceil\mu_1/2\rceil$ and  $j<\lceil\mu_2/2\rceil$. 
    \item \(\mathbf{State \;S_2}:\) The local operators \(\mathcal X\) or \(\mathcal Y\) associated with both \(\Gamma_{\mu_1}\) and \(\Gamma_{\mu_2}\) has been encountered in the Pauli string up to site \(j\) which equivalently is \(j\geq   \lceil\mu_i/2\rceil \) (\(i=1,2\)).
\end{itemize}

Here, the transition matrix \(\mathbb W_2^{[j]}\) corresponding to site \(j\) is a \(3\times 3\) matrix of two-replica operators acting on the local physical space $\mathbb{C}^2\otimes\mathbb{C}^2$ spanned by \(\{|00\rangle, |01\rangle, |10\rangle, |11\rangle\}\). It governs which two-replica operator will be applied to the current site as we propagate along the chain. At site $j$, at most two terminal operators, $\mathcal X$ and $\mathcal Y$, can be selected which are associated with Majorana
  indices $2j-1$ and $2j$, respectively. The operators in transition matrix at site \(j\) are as follows,
\begin{itemize}
    \item \(\mathbf{S_0 \rightarrow S_0}\), on every site processed so far, both bridge generators contribute
  $\mathcal Z$, so their product is $T_{S_0 \rightarrow S_0} = \mathcal Z^2=\mathcal I$.
    \item \(\mathbf{S_0 \rightarrow S_1}\), such that The local operator \(\mathcal X\) \((\mathcal Y)\) is applied for the term \(\Gamma_{\mu_1}\) for odd (even) \(\mu_1\). Meanwhile, the \(\mathcal Z\) string is continues for \(\Gamma_{\mu_2}\). Hence, the combined contribution of the even and odd case is \(T_{S_0 \rightarrow S_1} = \mathcal X \mathcal Z + \mathcal Y \mathcal Z = -(\mathcal Y + \mathcal X)\).
    \item \(\mathbf{S_0 \rightarrow S_2}\), where the local operators \(\mathcal X\) and \(\mathcal Y\) associated with \(\Gamma_{\mu_1}\) and \(\Gamma_{\mu_2}\) respectively are applied together on the current site, we get \(T_{S_0 \rightarrow S_2} = \mathcal X \mathcal Y = -\mathcal Z\). This transition selects the terminals with $\mu_1=2j-1$ and $\mu_2=2j$.
    \item The transition \(\mathbf{S_1 \rightarrow S_1}\) depicts that \(\mathcal X\) or \(\mathcal Y\) associated with \(\Gamma_{\mu_1}\) has already been applied before in the generated Pauli string and only \(\mathcal I\) is being applied while the string of \(\mathcal Z\) continues for \(\Gamma_{\mu_2}\). Consequently, transition operator \(T_{S_1 \rightarrow S_1} = \mathcal I\mathcal Z = \mathcal Z\).
    \item \(\mathbf{S_1 \rightarrow S_2}\), the operator \(\mathcal X\)(\(\mathcal Y\)) is applied at the current site for odd (even) \(\mu_2\) while only \(\mathcal I\) is applied for \(\Gamma_{\mu_1}\) leading to contribution of \(T_{S_1 \rightarrow S_2} = \mathcal I \mathcal X + \mathcal I \mathcal Y =  \mathcal X +\mathcal Y\).
    \item \(T_{S_2 \rightarrow S_2} = \mathcal I\,\mathcal I = \mathcal I\) corresponding to \({S_2 \rightarrow S_2}\) in which we apply \(\mathcal I\) for both \(\Gamma_{\mu_1}\) and \(\Gamma_{\mu_2}\).
\end{itemize}

The final representation of MPO at site \(j\) is
\begin{align}
\mathbb W_2^{[j]} = \begin{pmatrix}
\mathcal{I} & -(\mathcal{X+Y}) & - \mathcal Z \\
0 & \mathcal{Z} & (\mathcal{X+Y}) \\
0 & 0 & \mathcal{I}
\end{pmatrix}
\end{align}
In presence of the boundary vectors \(v_L = \begin{pmatrix} 1&0&0 \end{pmatrix}\) and \(v_R = \begin{pmatrix} 0&0&1 \end{pmatrix}^T\) the complete MPO representation then becomes,
\begin{align}
    \Upsilon_2^{(2)}= C_2\; v_L \left(\prod_{j=1}^{N} \mathbb W_2^{[j]} \right)v_R.
\end{align}

\subsection{\(\mathit{n=3}\) example}
Having discussed the simple \(\mathit{n=2}\) example, to further elucidate the MPO construction, we consider the representative example of the two-replica Majorana correlator 
\begin{align}
    \Upsilon_3^{(2)} = C_3 \sum_{1\le{\mu_1}<{\mu_2}<{\mu_3} \leq 2N} (\gamma_{\mu_1}\gamma_{\mu_2}\gamma_{\mu_3})\otimes(\gamma_{\mu_1}\gamma_{\mu_2}\gamma_{\mu_3}).
\end{align}
Consequently, using the bridge generators, we rewrite \(\Upsilon_3^{(2)} =  C_3 \sum_{{\mu_1} < {\mu_2}<{\mu_3}} \Gamma_{\mu_1} \Gamma_{\mu_2} \Gamma_{\mu_3}\) such that the corresponding finite state machine consists of four states yielding an MPO bond dimension \(D=4\). After processing site $j$, the four possible FSM states are:
\begin{itemize}
    \item \(\mathbf{State \;S_0}:\) Neither \(\mathcal X\) nor \(\mathcal Y\) associated with either \(\Gamma_{\mu_1}\), \(\Gamma_{\mu_2}\) or \(\Gamma_{\mu_3}\) has yet been encountered in the Pauli string up to site \(j\), corresponding to \(j<\lceil\mu_i/2\rceil\) (\(i=1,2,3\)). Consequently, the string consists solely of \(\mathcal Z\) operators upto the current site.
    
    \item \(\mathbf{State \;S_1}:\) One \(\mathcal X\) or \(\mathcal Y\) associated only with \(\Gamma_{\mu_1}\) has been encountered upon traversal of the generated Pauli string so far. Equivalently, \(j\geq\lceil\mu_1/2\rceil\) and \(j<\lceil\mu_2/2\rceil\).
    
    \item \(\mathbf{State \;S_2}:\) The local operators \(\mathcal X\) or \(\mathcal Y\) associated with both \(\Gamma_{\mu_1}\) and \(\Gamma_{\mu_2}\) has been encountered in the Pauli string up to site \(j\). Although the only the string of \(\mathcal Z\) is encountered for \(\Gamma_{\mu_3}\) which equivalently is \(j\geq\lceil\mu_i/2\rceil\) (\(i=1,2\)) and \(j<\lceil\mu_3/2\rceil\).
    
    \item \(\mathbf{State \;S_3}:\) The local operators \(\mathcal X\) or \(\mathcal Y\) corresponding to all \(\Gamma_{\mu_i}\) have been encountered in the Pauli string up to site \(j\). Mathematically it is expressed as, \(j\geq\lceil\mu_i/2\rceil\) (\(i=1,2,3\)).
\end{itemize}

Moreover, the transition matrix \(\mathbb W_3^{[j]}\) corresponding to site \(j\) is a \(4\times 4\) matrix of two-replica operators acting on the local physical space $\mathbb{C}^2\otimes\mathbb{C}^2$ spanned by \(\{|00\rangle, |01\rangle, |10\rangle, |11\rangle\}\). It governs which two-replica operator will be applied to the current site as we propagate along the chain. Given \(\mu_1 < \mu_2\), at most two consecutive Majorana operators can be supported on the same site, that is, \(\lceil\mu_1/2\rceil \leq \lceil\mu_{2}/2\rceil\). Consequently, the allowed transitions are \(\mathbf{S_k}\!\rightarrow\!\mathbf{S_k}\), \(\mathbf{S_k}\!\rightarrow\!\mathbf{S_{k+1}}\), and \(\mathbf{S_k}\!\rightarrow\!\mathbf{S_{k+2}}\), while the transitions between finite state machine states whose labels differ by more than two are forbidden. Accordingly, the transition matrix looks like,
\begin{align}
\mathbb W_3^{[j]} =
\begin{pmatrix}
T_{00} & T_{01} & T_{02} & 0 \\
0 & T_{11} & T_{12} & T_{13} \\
0 & 0 & T_{22} & T_{23} \\
0 & 0 & 0 & T_{33}
\end{pmatrix}.
\end{align}
The operators in transition matrix at site \(j\) are as follows.

\textbf{Diagonal terms (\(T_{\mathrm{init}}\) and \(\tilde T_{\mathrm{init}}\))}:
\begin{itemize}
\item \(\mathbf{S_0 \rightarrow S_0}.\) Only \(\mathcal Z\) is applied for all \(\Gamma_{\mu_i}\) at the current site resulting in \(T_{S_0 \rightarrow S_0} = \mathcal Z \mathcal Z \mathcal Z  = \mathcal Z\).
\item \(\mathbf{S_1 \rightarrow S_1}.\) This depicts that \(\mathcal X\) or \(\mathcal Y\) associated with \(\Gamma_{\mu_1}\) has already been applied before in the generated Pauli string and now, only \(\mathcal I\) is being applied.  Meanwhile the string of \(\mathcal Z\) continues for \(\Gamma_{\mu_2}\) and \(\Gamma_{\mu_3}\). Consequently, transition operator \(T_{S_1 \rightarrow S_1} = \mathcal I\mathcal Z \mathcal Z = \mathcal I\).
\item \(\mathbf{S_2 \rightarrow S_2}.\) Here, \(T_{S_2 \rightarrow S_2} = \mathcal I\,\mathcal I\,\mathcal Z = \mathcal Z\) corresponding to the application of \(\mathcal I\) from both \(\Gamma_{\mu_1}\) and \(\Gamma_{\mu_2}\) as the local operator \(\mathcal X\) or \(\mathcal Y\) has already been applied, though \(\mathcal Z\) string continues for \(\Gamma_{\mu_3}\).
\item \(\mathbf{S_3 \rightarrow S_3}.\) Finally, all the local operators \(\mathcal X\) or \(\mathcal Y\) associated with all \(\Gamma_{\mu_i}\) have already been encountered and only \(\mathcal I\) is now applied at the current site giving \(T_{S_3 \rightarrow S_3} = \mathcal I\,\mathcal I\,\mathcal I = \mathcal I\)
\end{itemize}

\textbf{Single selection operators (\(T_\mathrm{start}\))}:
\begin{itemize}
    \item \(\mathbf{S_0 \rightarrow S_1}.\) The local operator \(\mathcal X\) \((\mathcal Y)\) is only applied for the term \(\Gamma_{\mu_1}\) at the current site, corresponding to the odd (even) \(\mu_1\) . Meanwhile, the \(\mathcal Z\) string is continues for \(\Gamma_{\mu_2}\) and \(\Gamma_{\mu_3}\). Hence, the combined contribution of the even and odd case is \(T_{S_0 \rightarrow S_1} = \mathcal X \mathcal Z \mathcal Z + \mathcal Y \mathcal Z \mathcal Z = (\mathcal X + \mathcal Y)\).
    \item \(\mathbf{S_1 \rightarrow S_2}.\) the operator \(\mathcal X\) (\(\mathcal Y\)) is applied at the current site associated to \(\Gamma_{\mu_2}\) having odd (even) \(\mu_2\).  Meanwhile, only \(\mathcal I\) is applied for \(\Gamma_{\mu_1}\) and \(\Gamma_{\mu_3}\) contributes a \(\mathcal Z\) leading to contribution of \(T_{S_1 \rightarrow S_2} = \mathcal I \mathcal X \mathcal Z + \mathcal I \mathcal Y \mathcal Z  =  -(\mathcal Y +\mathcal X)\). 
    \item \(\mathbf{S_2 \rightarrow S_3}.\) Here, only the local operator \(\mathcal X\) or \(\mathcal Y\) associated to \(\Gamma_{\mu_3}\) is applied to the current site in conjunction to the \(\mathcal I\) contribution by \(\Gamma_{\mu_1}\) and \(\Gamma_{\mu_2}\). This results in \(T_{S_2 \rightarrow S_3} = \mathcal I \mathcal I \mathcal X + \mathcal I \mathcal I \mathcal Y  =  (\mathcal X +\mathcal Y)\)
\end{itemize}

\textbf{Double selection operators (\(T_\mathrm{local}\) and \(\tilde T_\mathrm{local}\))}:
\begin{itemize}
    \item \(\mathbf{S_0 \rightarrow S_2}.\) In this transition, the local operators \(\mathcal X\) and \(\mathcal Y\) associated with \(\Gamma_{\mu_1}\) and \(\Gamma_{\mu_2}\) respectively are applied together on the current site. Moreover, \(\Gamma_{\mu_3}\) contributes the \(\mathcal Z\) string term, finally resulting in, \(T_{S_0 \rightarrow S_2} = \mathcal X \mathcal Y \mathcal Z = -\mathcal I\). This transition satisfies the relation \(\mu_1 + 1 = \mu_2\).
    \item \(\mathbf{S_1 \rightarrow S_3}.\) Since the \(\mathcal X\) or \(\mathcal Y\) associated with \(\Gamma_{\mu_1}\) has already been encountered before, it only contributes \(\mathcal I\) to the current site. Meanwhile, the terminal operators $\mathcal X$ and $\mathcal Y$ of $\Gamma_{\mu_2}$ and $\Gamma_{\mu_3}$ are selected together, with $\mu_2=2j-1$ and $\mu_3=2j$. Consequently, the final term becomes, \(T_{S_1 \rightarrow S_3} = \mathcal I \mathcal X \mathcal Y = -\mathcal Z\)
\end{itemize}
The final representation of MPO at site \(j\) is
\begin{align}
\mathbb W_3^{[j]} =
\begin{pmatrix}
\mathcal{Z} & (\mathcal{X} + \mathcal{Y}) & -\mathcal{I} & 0 \\
0 & \mathcal{I} & -(\mathcal{X} + \mathcal{Y}) & -\mathcal{Z} \\
0 & 0 & \mathcal{Z} & (\mathcal{X} + \mathcal{Y}) \\
0 & 0 & 0 & \mathcal{I}
\end{pmatrix}.
\end{align}

In presence of the boundary vectors \(v_L = \begin{pmatrix} 1&0&0&0 \end{pmatrix}\) and \(v_R = \begin{pmatrix} 0&0&0&1 \end{pmatrix}^T\) the complete MPO representation then becomes,
\begin{align}
    \Upsilon_3^{(2)} = C_3 \; v_L \left(\prod_{j=1}^{N} \mathbb W_3^{[j]} \right)v_R.
\end{align}

\section{Asymptotic scaling of FAF}
The first-order fermionic antiflatness $\mathcal F_1$ is determined by two-body Majorana correlators. For a pure state $\ket{\psi}$, the covariance matrix has entries
\begin{align}
M_{\mu\nu}
=
-\frac{i}{2}
\langle\psi|
[\gamma_\mu,\gamma_\nu]
|\psi\rangle
=
-i\langle\psi|
\gamma_\mu\gamma_\nu
|\psi\rangle,
\qquad
\mu\neq\nu.
\end{align}
Using the antisymmetry of $M$, the FAF can be expressed as
\begin{align}
\mathcal F_1(\ket{\psi})
&=
N-\frac{1}{2}\Tr(M^{\mathsf T}M)
\nonumber\\
&=
N+
\sum_{1\leq\mu<\nu\leq2N}
\langle\psi|
\gamma_\mu\gamma_\nu
|\psi\rangle^2
\nonumber\\
&=
N+
\sum_{1\leq\mu<\nu\leq2N}
\Tr\left[
\ket{\psi}\!\bra{\psi}^{\otimes2}
(\gamma_\mu\gamma_\nu)^{\otimes2}
\right].
\label{eq:faf_two_replica}
\end{align} 
The quadratic dependence on Majorana correlators therefore permits an exact evaluation in the two-replica formalism. By definition, the normalized pairing tensor in the two-Majorana sector satisfies
\begin{align}
\sum_{1\leq\mu<\nu\leq2N}
(\gamma_\mu\gamma_\nu)^{\otimes2} = C_2^{-1}\Upsilon_2^{(2)}.
\label{eq:Upsilon2}
\end{align}
For the GAMPS ensemble, $\ket{\Psi_{\mathcal G}}=U_G\ket{\Phi_\chi}$, Eq.~\eqref{eq:faf_two_replica} gives
\begin{align}
\left\langle\mathcal F_1\right\rangle_{\mu_{\mathcal G}}
&=
\sum_{1\leq\mu<\nu\leq2N}
\Tr\left[(N\mathbb I^{\otimes N} + (\gamma_\mu\gamma_\nu)^{\otimes2})
\mathbb E_{\mu_{\mathcal G}}
\left[
\rho_{\mathcal G}^{\otimes2}
\right]
\right]\nonumber\\ 
&= \Tr\left[(N\mathbb I^{\otimes N} + C_2^{-1}\Upsilon_2^{(2)})
\mathbb E_{\mu_{\mathcal G}}
\left[
\rho_{\mathcal G}^{\otimes2}
\right]
\right] = \Tr\left[B_{\mathcal F_1}
\mathbb E_{\mu_{\mathcal G}}
\left[
\rho_{\mathcal G}^{\otimes2}
\right]
\right]
\label{eq:faf_gamps_average}
\end{align}
Hence by invoking this PT-operators \(\Upsilon_2^{(2)}\) and \(\Upsilon_0^{(2)}\), Eq.~(\ref{eq:faf_gamps_average}) can be reiterated in form of Eq.~(\ref{eq:quantifier_pt_expansion_supp}) as
\begin{align}
     \langle \mathcal F_1 \rangle_{\mu_\mathcal G} 
     &= \Tr\left[(NC_0^{-1}\Upsilon_0^{(2)} + C_2^{-1}\Upsilon_2^{(2)})
    \mathbb E_{\mu_{\mathcal G}}\left[\rho_{\mathcal G}^{\otimes2}\right]\right]\nonumber\\
    &=\sum_{n=0}^{2N} \Tr\left[(NC_0^{-1}\Upsilon_n^{(2)}\delta_{n,0} + C_2^{-1}\Upsilon_n^{(2)}\delta_{n,2})
    \mathbb E_{\mu_{\mathcal G}}\left[\rho_{\mathcal G}^{\otimes2}\right]\right]\nonumber\\
    &=\sum_{n=0}^{2N} \Tr\left[(NC_0^{-1}\Upsilon_n^{(2)}\Tr[\Upsilon_0^{(2)}\Upsilon_n^{(2)}] + C_2^{-1}\Upsilon_n^{(2)}\Tr[\Upsilon_2^{(2)}\Upsilon_n^{(2)}])
    \mathbb E_{\mu_{\mathcal G}}\left[\rho_{\mathcal G}^{\otimes2}\right]\right]\nonumber\\
    &=\sum_{n=0}^{2N} \Tr[(NC_0^{-1}\Upsilon_0^{(2)}\Upsilon_n^{(2)}) + (C_2^{-1}\Upsilon_2^{(2)}\Upsilon_n^{(2)})]\Tr\left[\Upsilon_n^{(2)}
    \mathbb E_{\mu_{\mathcal G}}\left[\rho_{\mathcal G}^{\otimes2}\right]\right]\nonumber\\
    &=\sum_{n=0}^{2N} \Tr[(N\mathbb I^{\otimes N} + C_2^{-1}\Upsilon_2^{(2)})\Upsilon_n^{(2)}]\Tr\left[\Upsilon_n^{(2)}
    \mathbb E_{\mu_{\mathcal G}}\left[\rho_{\mathcal G}^{\otimes2}\right]\right]\nonumber\\
    &=\sum_{n=0}^{2N} \Tr[B_{\mathcal F_1}\Upsilon_n^{(2)}]\Tr\left[\Upsilon_n^{(2)}
    \mathbb E_{\mu_{\mathcal G}}\left[\rho_{\mathcal G}^{\otimes2}\right]\right]
\end{align}
Meanwhile, simply utilizing Eq.~(\ref{eq:Upsilon2}) in Eq.~(\ref{eq:faf_gamps_average}), we get
\begin{align}
    = N + C_2^{-1} \,\Tr\big[\Upsilon_2^{(2)}\,\mathbb E_{\mu_\chi}[\rho_{\Phi_\chi}^{\otimes2}]\big] = N +C_2^{-1} \, \alpha_2(\varrho^{(2)}_{\mu_\chi}).
\end{align}
\subsection{Transfer matrix analysis for FAF}
The MPO contraction representing $C_2^{-1}\alpha_2(\varrho^{(2)}_{\mu_\chi})$ consists of a translationally invariant bulk tensor and two boundary tensors
\begin{align}
\left\langle\mathcal F_1\right\rangle_{\mu_{\mathcal G}}
&=
N+
\adjustbox{valign=c}{
\includegraphics[height=5em]{ 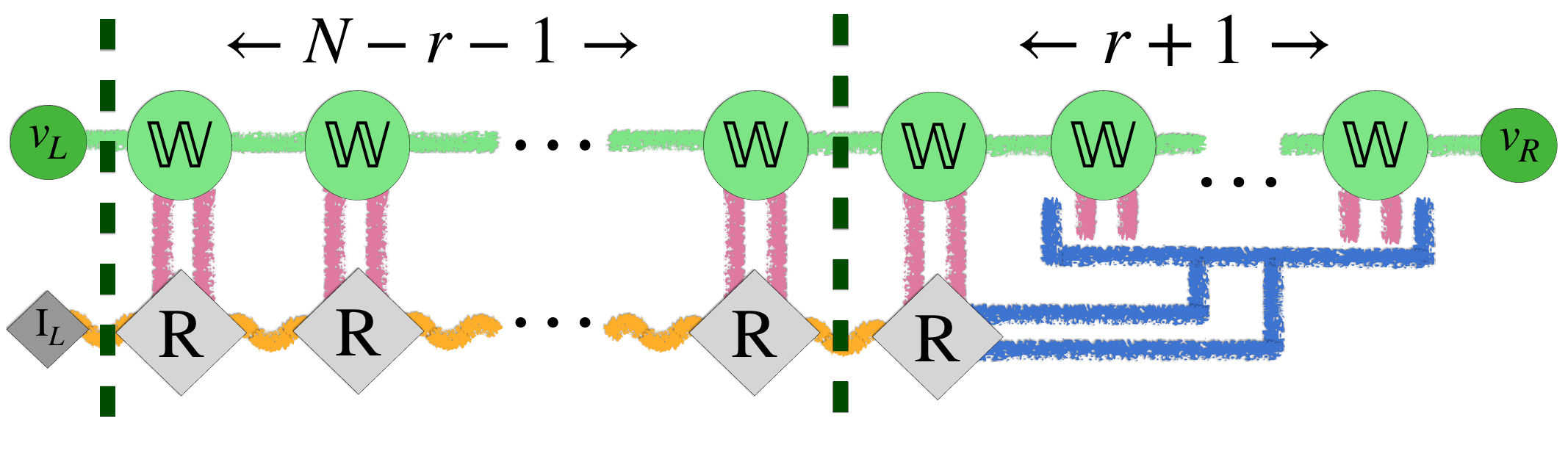}
}
\nonumber\\
&=
N+
\mathcal L\,
T^{N-r-1}\,
\mathcal R,
\label{eq:FAF_mpo_transfermat}
\end{align}
where $\mathcal L$ and $\mathcal R$ are the boundary vectors and $T$ is the site-independent transfer matrix. Combining the two-valued permutation index with the $(n+1)$-dimensional pairing-tensor bond gives a transfer matrix of size $2(n+1)\times2(n+1)$. For FAF, $n=2$, so $T\in\mathbb C^{6\times6}$.

The local RMPS tensor at site \(j\) is \(R_{\tau_j,\tau_{j+1}} = \sum_{\sigma} W_{\tau_j,\sigma}^{(d\chi)} \, P_\sigma^{(d)}\, G_{\sigma,\tau_{j+1}}^{(\chi)}\) where \(W^{(d\chi)}\) and \(G^{(\chi)}\) are the Weingarten and Gram matrices, respectively, while \(P_\sigma^{(d)}\) acts on the two-replica physical space with \(d=2\). Contracting \(P_\sigma^{(d)}\) with the local MPO tensor of \(\Upsilon_2^{(2)}\) defines
\begin{align}
\mathbb{M}^{\sigma}_{a_j,a_{j+1}}=\Tr_d\!\left[P_{\sigma}^{(d)}\,\left(\mathbb{W}_{n}^{[j]}\right)_{a_j,a_{j+1}}\right]
\end{align}
where the trace is taken over the physical indices and \(a_j \in \{0,1,2\}\) is the auxiliary index of the pairing-tensor MPO. The resulting transfer matrix is
\begin{align}
T_{\tau_j,\tau_{j+1}}^{a_j,a_{j+1}} &=\adjustbox{valign=c}{\includegraphics[height=3em]{ transfer_mat.pdf}}
= \sum_{\sigma=\{\mathbb I,\mathbb F\}} W_{\tau_j,\sigma}^{(2\chi)}\, \Tr_d\!\left[P_{\sigma}^{(d)}\,\left(\mathbb{W}_{n}^{[j]}\right)_{a_j,a_{j+1}}\right] \, G_{\sigma,\tau_{j+1}}^{(\chi)} \nonumber \\
&= \sum_{\sigma=\{\mathbb I,\mathbb F\}} W_{\tau_j,\sigma}^{(2\chi)}\, \mathbb M_{a_j,a_{j+1}}^\sigma \, G_{\sigma,\tau_{j+1}}^{(\chi)}.
\end{align} 
The left boundary follows directly from the RMPS boundary tensor and takes the form
\begin{align}
\mathcal L
=
v_L\otimes I_L
=
\begin{pmatrix}
1&0&0
\end{pmatrix}
\otimes
\begin{pmatrix}
1&1
\end{pmatrix}
=
\begin{pmatrix}
1&1&0&0&0&0
\end{pmatrix}.
\end{align}
The right boundary incorporates the normalization imposed by the canonical RMPS construction~\cite{Haag2023TypicalTNcorrlength,Schon2005SeqGeneration} and is given by
\begin{align}
\mathcal R = 
\begin{pmatrix}
W_{\mathbb I,\mathbb I} (\mathbb M^\mathbb I)^{r+1} v_R + W_{\mathbb I,\mathbb F} (\mathbb M^\mathbb F)^{r+1} v_R \\
W_{\mathbb F,\mathbb I} (\mathbb M^\mathbb I)^{r+1} v_R + W_{\mathbb F,\mathbb F} (\mathbb M^\mathbb F)^{r+1} v_R
\end{pmatrix},
\end{align}
where \(\mathbb M^\sigma\) denotes the \(3\times3\) matrix.

The transfer matrix is not diagonalizable and has defective eigenvectors, hence we perform the Jordan decomposition \(T=VJV^{-1}\). Its eigenvalues are \(\left\{0,\; 1,\; \lambda_3(\chi),\; \lambda_2(\chi)\right\},\)
with \(\lambda_2(\chi) = \frac{2(\chi^2 - 1)}{4\chi^2 - 1}\) and \(\lambda_3(\chi) = \frac{2\chi^2 - 1}{4\chi^2 - 1}\). The Jordan normal form is
\begin{align}
J=
\begin{pmatrix}
0&0&0&0&0&0\\
0&1&1&0&0&0\\
0&0&1&0&0&0\\
0&0&0&\lambda_3(\chi)&0&0\\
0&0&0&0&\lambda_2(\chi)&1\\
0&0&0&0&0&\lambda_2(\chi)
\end{pmatrix},
\label{eq:FAF_Jordan_form}
\end{align}
with \(2\times2\) Jordan blocks associated with \(1\) and \(\lambda_2(\chi)\). The explicit similarity matrix \(V\) is given by
\begin{align}
V =
\begin{pmatrix}
0 & \frac{2 - 5 \chi^{2}}{2 \chi^{4} + \chi^{2}} & 0 & 6 \chi - \frac{4}{\chi} + \frac{1}{2 \chi^{3}} & \frac{\chi^{3} \left(- 96 \chi^{4} + 156 \chi^{2} - 60\right)}{16 \chi^{6} + 12 \chi^{4} - 1} & \frac{- 64 \chi^{7} + 120 \chi^{5} - 66 \chi^{3} + 10 \chi}{8 \chi^{6} + 12 \chi^{4} + 6 \chi^{2} + 1}
\\
0 & 0 & \frac{4 - 10 \chi^{2}}{2 \chi^{3} + \chi} & 2 - 4 \chi^{2} & \frac{\chi^{2} \left(32 \chi^{4} - 52 \chi^{2} + 20\right)}{8 \chi^{4} + 2 \chi^{2} - 1} & 0
\\
-\chi & 0 & -\frac{1}{\chi^{2}} & -\frac{1}{2\chi} & 0 & \frac{4\chi(\chi^{2}-1)}{2\chi^{2}+1}
\\
1 & 0 & \frac{2}{\chi} & 1 & 0 & \frac{8\chi^{2}(1-\chi^{2})}{2\chi^{2}+1}
\\
0 & 0 & 1 & 0 & 0 & -\frac{3\chi}{2\chi^{2}+1}
\\
0 & 0 & 0 & 0 & 0 & 1
\end{pmatrix}.
\end{align}
In the limit of \(N \gg r\), we obtain \(T^{N-r-1} \simeq VJ^{N}V^{-1}\). Furthermore, \(|\lambda_2(\chi)|, |\lambda_3(\chi)|<1\) suggests that these eigenvalue contributions vanish exponentially with \(N\) resulting in only the survival of the Jordan block corresponding to eigenvalue equal to \(1\),
\begin{equation}
J^N =
\begin{pmatrix}
0 & 0 & 0 & 0 & 0 & 0\\
0 & 1 & N & 0 & 0 & 0\\
0 & 0 & 1 & 0 & 0 & 0\\
0 & 0 & 0 & \lambda_3^N & 0 & 0\\
0 & 0 & 0 & 0 & \lambda_2^N & N \lambda_2^{N-1}\\
0 & 0 & 0 & 0 & 0 & \lambda_2^N
\end{pmatrix} \xrightarrow{N\to\infty}
\begin{pmatrix}
0 & 0 & 0 & 0 & 0 & 0\\
0 & 1 & N & 0 & 0 & 0\\
0 & 0 & 1 & 0 & 0 & 0\\
0 & 0 & 0 & 0 & 0 & 0\\
0 & 0 & 0 & 0 & 0 & 0\\
0 & 0 & 0 & 0 & 0 & 0
\end{pmatrix}.
\end{equation}
By utilizing this Jordan decomposition form in the large \(N\) limit in Eq.~(\ref{eq:FAF_mpo_transfermat}) we obtain 
\begin{align}
    \mathcal{L}T^{N-r-1}\mathcal{R} &= \mathcal L (V J^{N-r-1} V^{-1}) \mathcal R \nonumber\\
    &\simeq \mathcal L (V J^N V^{-1}) \mathcal R \nonumber\\
    &= Nf(\chi)+g(\chi)
\end{align}
with \(f(\chi) \) and \(g(\chi)\) as closed form functions of \(\chi\). Correspondingly, neglecting corrections that are exponentially suppressed with \(N\) the deviation of Haar FAF evaluates to
\begin{align}
    \Delta \langle \mathcal F_1 \rangle_{\mu_{\mathcal G}} &= \langle \mathcal F_1 \rangle_{\mu_{H}}- \langle \mathcal F_1 \rangle_{\mu_{\mathcal G}} = \Bigg(N-\frac{N(2N-1)}{2^N + 1}\Bigg) - \Bigg(N+Nf(\chi)+g(\chi)\Bigg)\\
    &= -Nf(\chi)-g(\chi)\\
    &=\frac{5}{\chi}
    +\frac{5N}{2\chi^2}
    +\frac{\log^2\chi}{\chi^2\log^2 2}
    +\frac{7\log\chi}{2\chi^2\log 2}
    -\frac{9}{\chi^2}
    -\frac{9}{2\chi^3}
    \nonumber + \frac{9N}{4\chi^4}
    -\frac{\log^2\chi}{2\chi^4\log^2 2}
    -\frac{11\log\chi}{4\chi^4\log 2}
    +\frac{53}{4\chi^4}.
    \label{eq:FAF_asymptotic_expansion}
\end{align}
Consequently,
\begin{align}
\frac{
\Delta\langle\mathcal F_1\rangle_{\mu_{\mathcal G}}
}{N}
=
\frac{5}{N\chi}
+
\frac{5}{2\chi^2}
+
\mathcal O\!\left(
\frac{\log^2\chi}{N\chi^2}
\right)
+
\mathcal O\!\left(\chi^{-4}\right).
\end{align}
At fixed \(N\), the leading finite-size correction scales as \(\chi^{-1}\). In the thermodynamic regime \(N\gg\log_2\chi\), this contribution becomes subleading and the FAF density approaches its Haar value as \(\chi^{-2}\), matching the asymptotic behavior of the underlying RMPS ensemble.

\subsection{FAF dominance in IPR}
Having determined the deviation in FAF, the deviation in inverse participation ratio depicted as can be rewritten as \(\Delta \langle I_{2}\rangle_{\mu_{\mathcal G}} = \frac{1}{{\langle I_{2} \rangle_{\mu_\mathrm H}}}\sum_{n=0}^{2N}\beta_n(I_{2})\,\Delta \alpha_n(\varrho^{(2)}_{\mu_\chi})\) where \(\Delta\alpha_n(\varrho^{(2)}_{\mu_\chi}) = \alpha_n(\varrho^{(2)}_{\mu_\chi})- \alpha_n^{\mathrm{Haar}}\) and \(\alpha^{\mathrm{Haar}}_n\) corresponds to the Haar coefficient. By definition the deviation of FAF is equivalent to 
\(\Delta\alpha_2(\varrho^{(2)}_{\mu_\chi})= C_2 (-\Delta\langle \mathcal F_1 \rangle_{\mu_{\mathcal G}})\).
Consequently, by asymptotically analyzing the functions \(\beta_2(I_{2})\,\Delta\alpha_2(\varrho^{(2)}_{\mu_\chi})\), we observe that this FAF term corresponding to \(n=2\) dominates the sum in \(\Delta \langle I_{2}\rangle_{\mu_{\mathcal G}}\) which in the large \(N-\)limit can be rewritten as
\begin{align}
    \Delta \langle I_{2}\rangle_{\mu_{\mathcal G}} &= \frac{(2^N+1)}{2}\beta_2(I_{2})\,\Delta\alpha_2(\varrho^{(2)}_{\mu_\chi}) =  \frac{(2^N+1)}{2}  \bigg[(-1)\binom{2N}{2}^{-1/2}\binom{N}{1}\bigg] \bigg[2^{-N}\binom{2N}{2}^{-1/2} (-\Delta\langle \mathcal F_1 \rangle_{\mu_{\mathcal G}})\bigg]\nonumber\\ &\approx \frac{\Delta\langle \mathcal F_1 \rangle_{\mu_{\mathcal G}}}{4N}
\end{align} 
where we have utilized the relation \(\beta_{2\ell}(I_{2}) = (-1)^{\ell} \,\binom{2N}{2\ell}^{-1/2}\binom{N}{\ell}\).

\section{Pairing tensor to Gelfand-Tsetlin basis transformation.}\label{sec:pt_to_gt}

The projectors \(\{\mathcal E^{(2)}_p\}\), which diagonalize the matchgate commutant, constitute the Gelfand-Tsetlin (GT) basis~\cite{Sierant26matchgatecommutant,Braccia2026matchgatecommutant}. Accordingly, the second-moment matchgate twirling channel admits the spectral expansions
\begin{align}
\mathcal{T}_G^{(2)}(\rho^{\otimes2})
= \sum_{p=0}^{2N}\kappa_p(\rho)\,\mathcal E^{(2)}_p
=
\sum_{n=0}^{2N}\alpha_n(\rho)\,\Upsilon_n^{(2)}
\label{eq:twirl_matchgate}
\end{align}
with \(\kappa_p(\rho)\)
as the coefficients in the GT-basis.
In this section, we derive the transformation relating the pairing tensor bases and GT bases having a similar approach to~\cite{Braccia2026matchgatecommutant}.

Let \(\Gamma_\mu=\gamma_\mu\otimes\gamma_\mu\) denote the bridge generators. These Hermitian operators mutually commute, \([\Gamma_\mu,\Gamma_\nu]=0\), and therefore admit a simultaneous eigenbasis \(\ket{s_1\cdots s_{2N}}\), with \(s_\mu\in\{+1,-1\}\), satisfying \(\Gamma_\mu\ket{s_1\cdots s_{2N}}=s_\mu\ket{s_1\cdots s_{2N}}\). We define the bridge operator \(\Lambda=\sum_{\mu=1}^{2N}\Gamma_\mu\), whose action on this basis is
\begin{align}
\Lambda\ket{s_1\cdots s_{2N}}
=
\left(\sum_{\mu=1}^{2N}s_\mu\right)
\ket{s_1\cdots s_{2N}}.
\end{align}
If exactly \(p\) of the \(2N\) eigenvalues are positive, the remaining \(2N-p\) are negative, yielding \(\sum_{\mu=1}^{2N}s_\mu=p-(2N-p)=2p-2N\). The spectrum of \(\Lambda\) is consequently \(\operatorname{spec}(\Lambda)=\{-2N,-2N+2,\ldots,2N\}\).

The Gelfand-Tsetlin projectors \(\mathcal E_p^{(2)}\) project onto the eigenspace of \(\Lambda\) with eigenvalue \(2p-2N\), spanned by simultaneous eigenstates containing exactly \(p\) positive eigenvalues. The bridge operator therefore admits the spectral decomposition \(\Lambda=\sum_{p=0}^{2N}(2p-2N)\mathcal E_p^{(2)}\), where \(\dim(\mathcal E_p^{(2)})=\binom{2N}{p}\). By construction, the projectors are mutually orthogonal, diagonal in the simultaneous eigenbasis, and commute with every bridge generator \(\Gamma_\mu\).

For an ordered subset of Majorana indices \(S=\{\mu_1<\mu_2<\cdots<\mu_n\}\), the corresponding two-replica Majorana string can be written in terms of the bridge generators as \(\gamma_S\otimes\gamma_S=\prod_{\mu\in S}\Gamma_\mu\). The pairing-tensor operators therefore take the form
\begin{align}
\Upsilon_n^{(2)}
=
C_n\sum_{|S|=n}\gamma_S\otimes\gamma_S
=
C_n\sum_{|S|=n}\prod_{\mu\in S}\Gamma_\mu,
\label{eq:Upsilon_spec}
\end{align}
where \(C_n\) denotes the normalization factor. The sum over products of bridge generators is precisely the \(n\)-th elementary symmetric polynomial \(e_n(\Gamma_1,\ldots,\Gamma_{2N})=\sum_{|S|=n}\prod_{\mu\in S}\Gamma_\mu\).

Upon projecting these polynomials onto a sector corresponding to \(p\) number of \(\Gamma_\mu\) having the positive eigenvalues \(+1\), the generating polynomials of the elementary symmetric polynomials of these bridge generators become identical to the generating function of the Krawtchouk polynomials,
\begin{align}
\sum_{n=0}^{2N}
\left[
e_n(\Gamma_1,\ldots,\Gamma_{2N})
\mathcal E_p^{(2)}
\right]t^n
&=
\prod_{\mu=1}^{2N}
(1+t\Gamma_\mu)\mathcal E_p^{(2)}
\nonumber\\
&=
(1+t)^p(1-t)^{2N-p}
\mathcal E_p^{(2)}
\nonumber\\
&=
\sum_{n=0}^{2N}
\left[K_n(p;2N)\mathcal E_p^{(2)}\right]t^n.
\label{eq:Krawtchouk_generating_function}
\end{align}
Here, \(K_n(p;2N)=\sum_{j=0}^{n}(-1)^j\binom{p}{n-j}\binom{2N-p}{j}\) denotes the binary Krawtchouk polynomial in the convention fixed by Eq.~\eqref{eq:Krawtchouk_generating_function} \cite{Krawtchouk1929}. This generating-function construction is possible because the bridge generators mutually commute.

By equating coefficients of \(t^n\) in conjunction with the completeness relation \(\sum_{p=0}^{2N}\mathcal E_p^{(2)}=\mathbb I\), the elementary symmetric polynomials admit the spectral decomposition
\begin{align}
e_n(\Gamma_1,\ldots,\Gamma_{2N})
=
\sum_{p=0}^{2N}
K_n(p;2N)\mathcal E_p^{(2)}.
\label{eq:elementary_symmetric_spectral}
\end{align}
Consequently, the pairing-tensor operators are diagonal in the Gelfand-Tsetlin basis, with eigenvalues proportional to the Krawtchouk polynomials.
Substituting the spectral decomposition from Eq.~(\ref{eq:Upsilon_spec}) into the expression for \(\Upsilon_n^{(2)}\) gives the coefficients of the matchgate commutant in GT-basis as 
\begin{equation}
\Upsilon_n^{(2)} = C_n  \sum_{p=0}^{2N} K_n(p;2N)\mathcal E^{(2)}_p
\quad \text{and} \quad \kappa_p(\rho) = \sum_{n=0}^{2N} C_n K_n(p;2N) \, \alpha_n(\rho) 
\label{eq:matchgate_coeff}
\end{equation}

\section{Second Moment Operator for Haar Random States}
\subsection{Expansion in Pairing tensor basis}

Any operator \(O\) acting on \(\mathbb C^{2^N}\otimes\mathbb C^{2^N}\) can be expanded in the two-replica Majorana-string basis as
\begin{align}
O
=
\sum_{S,T}c_{ST}\,
\gamma_S\otimes\gamma_T,
\qquad
c_{ST}
=
\frac{
\Tr\!\left[
(\gamma_S\otimes\gamma_T)^\dagger O
\right]
}{2^{2N}}.
\end{align}
The single-replica Majorana strings satisfy the orthogonality relation \(\Tr[\gamma_S^\dagger\gamma_T]=2^N\delta_{S,T}\). Their tensor products therefore obey
\begin{align}
\Tr\!\left[
(\gamma_S\otimes\gamma_T)^\dagger
(\gamma_{S'}\otimes\gamma_{T'})
\right]
&=
\Tr[\gamma_S^\dagger\gamma_{S'}]\,
\Tr[\gamma_T^\dagger\gamma_{T'}]
\nonumber\\
&=
2^{2N}\delta_{S,S'}\delta_{T,T'}.
\label{eq:majorana_stringbasis}
\end{align}

Schur--Weyl duality expresses the Haar-averaged two-replica state in terms of the identity and SWAP operators as
\begin{align}
\mathcal T_{\mathrm{Haar}}^{(2)}
=
\frac{\mathbb I+\mathbb F}
{2^N(2^N+1)}.
\label{eq:Haar_permbasis}
\end{align}
To expand \(\mathbb F\) in the Majorana-string basis, we use \(\Tr[(A\otimes B)\mathbb F]=\Tr[AB]\), which gives
\begin{align}
c_{ST}^{\mathbb F}
&=
\frac{
\Tr\!\left[
(\gamma_S\otimes\gamma_T)^\dagger
\mathbb F
\right]
}{2^{2N}}
\nonumber\\
&=
\frac{\Tr[\gamma_S^\dagger\gamma_T^\dagger]}{2^{2N}}
=
\frac{
\delta_{S,T}
(-1)^{|S|(|S|-1)/2}
}{2^N}.
\end{align}
The phase follows from \(\gamma_S^\dagger=(-1)^{|S|(|S|-1)/2}\gamma_S\), which accounts for reversing the order of the \(|S|\) mutually anticommuting Majorana operators. The SWAP operator consequently becomes
\begin{align}
\mathbb F
&=
\frac{1}{2^N}
\sum_{n=0}^{2N}
\sum_{|S|=n}
(-1)^{\frac{n(n-1)}{2}}
\gamma_S\otimes\gamma_S
\nonumber\\
&=
\sum_{n=0}^{2N}
(-1)^{\frac{n(n-1)}{2}}
\binom{2N}{n}^{1/2}
\Upsilon_n^{(2)},
\label{eq:swap_upsilon}
\end{align}
where \(n=|S|\) is the Majorana-string length.

Since the two-replica identity satisfies \(\mathbb I=2^N\Upsilon_0^{(2)}\), Eq.~\eqref{eq:Haar_permbasis} can be expanded in the pairing-tensor basis as
\begin{align}
\mathcal T_{\mathrm{Haar}}^{(2)}
&=
\sum_{n=0}^{2N}
\alpha_n^{\mathrm{Haar}}
\Upsilon_n^{(2)}
\nonumber\\
&=
\frac{1}{2^N(2^N+1)}
\left[
2^N\Upsilon_0^{(2)}
+
\sum_{n=0}^{2N}
(-1)^{\frac{n(n-1)}{2}}
\binom{2N}{n}^{1/2}
\Upsilon_n^{(2)}
\right].
\end{align}
The corresponding Haar coefficients are therefore
\begin{align}
\alpha_n^{\mathrm{Haar}}
=
\frac{1}{2^N(2^N+1)}
\left[
2^N\delta_{n,0}
+
(-1)^{\frac{n(n-1)}{2}}
\binom{2N}{n}^{1/2}
\right].
\label{eq:Haar_pt_coeff}
\end{align}

\subsection{Expansion in Gelfand-Tsetlin basis}

The spectral decomposition of the pairing-tensor basis derived in Sec.~\ref{sec:pt_to_gt} determines the corresponding second-moment coefficients in the Gelfand-Tsetlin basis. Substituting the transformation coefficients from Eq.~\eqref{eq:matchgate_coeff} into the Haar second moment in Eq.~\eqref{eq:Haar_pt_coeff} gives
\begin{align}
\kappa_p^{\mathrm{Haar}}
&=
\frac{K_0(p;2N)}{2^{2N}}
+
\sum_{n=1}^{2N}
\frac{
(-1)^{\frac{n(n-1)}{2}}
\binom{2N}{n}^{1/2}
}{
2^N(2^N+1)
}
\frac{
K_n(p;2N)
}{
2^N\binom{2N}{n}^{1/2}
}
\nonumber\\
&=
\frac{1}{2^{2N}}
\left[
1+
\frac{1}{2^N+1}
\sum_{n=1}^{2N}
(-1)^{\frac{n(n-1)}{2}}
K_n(p;2N)
\right],
\end{align}
where we used \(K_0(p;2N)=1\) for every \(p\).
This sum evaluates in closed form,
\begin{align}
\kappa_p^{\mathrm{Haar}}
=
\begin{cases}
\dfrac{2}{2^N\left(2^N+1\right)}, & (p-N)\bmod 4\in\{0,1\},\\[2mm]
0, & \text{otherwise}.
\end{cases}
\label{eq:kappa_haar_closed}
\end{align}
Half of the GT sectors therefore carry no Haar weight, as required by $\Omega_{\mathrm H}^{(2)}$ being supported on the symmetric subspace: the nonvanishing sectors are precisely those sectors contained in the symmetric subspace~\cite{tarabunga2026fermionic}.

\subsection{Trace distance in GT-basis}

The trace distance between the second-moment operator of an ensemble and its Haar counterpart quantifies the ensemble's \(\epsilon\)-approximate state two-design properties. For GAMPS, both second moments have been expanded in the Gelfand-Tsetlin basis,
\(\mathcal T_{\mathcal G}^{(2)}=\mathcal T_G^{(2)}(\mathbb \varrho_{\mu_\chi}^{(2)})=\sum_{p=0}^{2N}\kappa_p^{\mathcal G}\mathcal E_p^{(2)}\)
and
\(\mathcal T_{\mathrm{Haar}}^{(2)}=\sum_{p=0}^{2N}\kappa_p^{\mathrm{Haar}}\mathcal E_p^{(2)}\).
Their difference is consequently diagonal in the bridge-projector basis,
\begin{align}
\delta
=
\mathcal T_{\mathcal G}^{(2)}
-
\mathcal T_{\mathrm{Haar}}^{(2)}
=
\sum_{p=0}^{2N}
\left(
\kappa_p^{\mathcal G}
-
\kappa_p^{\mathrm{Haar}}
\right)
\mathcal E_p^{(2)}.
\end{align}
The corresponding trace distance is
\begin{align}
\mathcal D_{\mathrm H}(\Psi_{\mathcal G})
=
\frac{1}{2}\|\delta\|_1
=
\frac{1}{2}
\left\|
\mathcal T_{\mathcal G}^{(2)}
-
\mathcal T_{\mathrm{Haar}}^{(2)}
\right\|_1,
\end{align}
where \(\|O\|_1=\Tr\sqrt{O^\dagger O}\) denotes the trace norm. The bridge projectors satisfy \(\mathcal E_p^{(2)}\mathcal E_s^{(2)}=\delta_{ps}\mathcal E_p^{(2)}\) and \(\sum_{p=0}^{2N}\mathcal E_p^{(2)}=\mathbb I\). Their orthogonality therefore gives
\begin{align}
\sqrt{\delta^\dagger\delta}
&=
\sqrt{
\sum_{p=0}^{2N}
\left|
\kappa_p^{\mathcal G}
-
\kappa_p^{\mathrm{Haar}}
\right|^2
\mathcal E_p^{(2)}
}
\nonumber\\
&=
\sum_{p=0}^{2N}
\left|
\kappa_p^{\mathcal G}
-
\kappa_p^{\mathrm{Haar}}
\right|
\mathcal E_p^{(2)}.
\end{align}
Taking the trace and using \(\Tr(\mathcal E_p^{(2)})=\binom{2N}{p}\) yields
\begin{align}
\mathcal D_{\mathrm H}(\Psi_{\mathcal G})
&=
\frac{1}{2}
\sum_{p=0}^{2N}
\left|
\kappa_p^{\mathcal G}
-
\kappa_p^{\mathrm{Haar}}
\right|
\Tr(\mathcal E_p^{(2)})
\nonumber\\
&=
\frac{1}{2}
\sum_{p=0}^{2N}
\binom{2N}{p}
\left|
\kappa_p^{\mathcal G}
-
\kappa_p^{\mathrm{Haar}}
\right|.
\end{align}

\section{Schur--Weyl Duality and Weingarten Calculus in the Presence of a \texorpdfstring{$\mathbb Z_2$}{Z2} Symmetry}
The GAMPS construction naturally admits a formulation that incorporates fermionic parity symmetry. We consider parity-resolved RMPS generated by local unitaries that either preserve or exchange the parity sectors and subsequently apply FGUs, including parity reflections. On the two-replica space, these transformations enlarge the permutation basis to the parity-resolved commutant basis $\mathcal B=\{\Pi_sP_\sigma\}_{s=\pm1,\,\sigma\in S_2}$, where $\Pi_s=[\mathbb I+s(\Pi\otimes\Pi)]/2$.

Let $\mathcal H$ be a $D$-dimensional Hilbert space equipped with the global parity operator
\(\Pi=\prod_{i=1}^{N}Z_i,
\) satisfying \(\Pi^2=\mathbb I\). Its eigenspaces define the even- and odd-parity sectors, yielding the decomposition $\mathcal H=\mathcal H_+\oplus\mathcal H_-$. Here, we consider the subgroup
\begin{align}
\mathcal U_\Pi(D)
=
\left\{
U\in\mathcal U(D)
\;\middle|\;
[U,\Pi]=0
\ \text{or}
\{U,\Pi\}=0
\right\},
\label{eq:parity_unitary_group}
\end{align}
which contains both parity-preserving unitaries and parity reflections that exchange $\mathcal H_+$ and $\mathcal H_-$.

The resource quantifiers require the two-replica representation on $\mathcal H^{\otimes2}$. For the full unitary group, Schur--Weyl duality gives
\(\operatorname{Comm}\!\left[\mathcal U(D)^{\otimes2}\right]
= \operatorname{span}\{\mathbb I,\mathbb F\},
\) where $\mathbb F$ exchanges the two replicas. The presence of parity resolves the two-replica Hilbert space into sectors with equal and opposite replica parities,
\(\mathcal H^{\otimes2}
={}
\left[
(\mathcal H_+\otimes\mathcal H_+)
\oplus
(\mathcal H_-\otimes\mathcal H_-)
\right]
\oplus
\left[
(\mathcal H_+\otimes\mathcal H_-)
\oplus
(\mathcal H_-\otimes\mathcal H_+)
\right].\)
These sectors are selected by the basis operators
\begin{align}
\Pi_s
=
\frac{\mathbb I+s(\Pi\otimes\Pi)}{2},
\qquad
s\in\{-1,+1\}.
\end{align}
and the two-replica commutant is consequently spanned by the parity-resolved permutation operators \(\operatorname{Comm}\!\left[\mathcal U_\Pi(D)^{\otimes2}\right] = \operatorname{span}\mathcal B\).
Similar to the main text, to restructure the MPO chain we evaluate the twirling channel of this unitary group applied on the two-replica state \(\rho_0^{\otimes 2} = |0\rangle\langle0|^{\otimes 2}\). The commutant basis operators \(P^{(D)}_{\sigma,s} = \Pi_s P^{(D)}_\sigma \in \mathcal B\) are utilized in the Schur-Weyl duality resulting in
\begin{align}
\mathcal T^{(2)}_\Pi (\rho_0^{\otimes 2}) &= \int_\Pi U_\Pi^{\otimes 2} \rho_0^{\otimes 2} (U_\Pi^\dagger)^{\otimes 2} \,d\mu_\Pi = \sum_{(\sigma,s),(\tau,t)}W^{(d\chi)}_{(\sigma,s),(\tau,t)}\mathrm{Tr}_d\left[\big(P^{(d\chi)}_{(\sigma,s)}\big)^T(\rho_0^{\otimes 2})\right]P^{(d\chi)}_{(\tau,t)}
\end{align}
where \(d\mu_\Pi\) is the probability measure on this group and \(W^{(d\chi)}_{(\sigma,s),(\tau,t)}\) is the Weingarten matrix obtained from inverting the Gram matrix \(G^{(d\chi)}_{(\sigma,s),(\tau,t)} = \Tr[\big(P_{\sigma,s}^{d\chi}\big)^T P_{\tau,t}^{d\chi}]\). 
The parity resolved Gram matrix in this subspace is correspondingly defined as
\begin{align}
G^{(D)}_{(\sigma,s),(\tau,t)}
=\Tr_D\left[\Pi_s P^{(D)}_\sigma \Pi_{t} P^{(D)}_\tau\right] =\Tr_D\left[\Pi_s P^{(D)}_\sigma P^{(D)}_\tau\right]
\end{align}
where we have used the symmetric nature of the basis operators \([\Pi_s,P^{(D)}_\sigma]=0\), Projector orthogonality \(\Pi_s \Pi_{t} = \delta_{st}\Pi_s\) and the cyclic nature of trace.
Consequently the Gram matrix takes the form,
\begin{equation}
G =
\begin{pmatrix}
G_{+} & 0 \\
0 & G_{-}
\end{pmatrix}
\quad \text{where} \quad
G_{s}=
\begin{pmatrix}
\mathrm{Tr}(\Pi_s) & \mathrm{Tr}(\Pi_sF) \\
\mathrm{Tr}(\Pi_sF) & \mathrm{Tr}(\Pi_s)
\end{pmatrix}.
\end{equation}
Upon simplification of \(\mathrm{Tr}(\Pi_s) = (D^2+s(\mathrm{Tr}\Pi)^2)/2\) and \(\mathrm{Tr}(\Pi_s\mathbb F) = \left(D+s\mathrm{Tr}(\Pi^2)\right)/2\), we obtain 
\begin{equation}
G_{+}
= \begin{pmatrix}
D^2/2 & D\\
D & D^2/2
\end{pmatrix}
\quad \text{and} \quad
G_{-}
=
\begin{pmatrix}
D^2/2 & 0\\
0 & D^2/2
\end{pmatrix}.
\end{equation}
The corresponding symmetry-adapted Weingarten matrix is defined by

\begin{equation}
W =
\begin{pmatrix}
W_+ & 0 \\
0 & W_-
\end{pmatrix} =
\begin{pmatrix}
(G_{+})^{-1} & 0 \\
0 & (G_{-})^{-1}
\end{pmatrix} =
\begin{pmatrix}
\dfrac{D^2/2}{D^4/4-D^2} & \dfrac{-D}{D^4/4-D^2} & 0 & 0 \\[0.3cm]
\dfrac{-D}{D^4/4-D^2} & \dfrac{D^2/2}{D^4/4-D^2} & 0 & 0 \\[0.3cm]
0 & 0 & \dfrac{2}{D^2} & 0 \\[0.3cm]
0 & 0 & 0 & \dfrac{2}{D^2}
\end{pmatrix}.
\end{equation}
The commutant basis operators obey the relation \(P_{\sigma,s}^{(d\chi)} = \sum_{t}P_{\sigma,t}^{(d)}P_{\sigma,ts}^{(\chi)}\) simplifying the twirl as
\begin{align}
    \mathcal T^{(2)}_\Pi (\rho_0^{\otimes 2}) &= \sum_{(\sigma,s),(\tau,t)}W^{(d\chi)}_{(\sigma,s),(\tau,t)}\mathrm{Tr}_d\left[\sum_{u=\pm 1}\big(P_{\sigma,u}^{(d)}P_{\sigma,us}^{(\chi)})^T(\rho_0^{\otimes 2})\right]P^{(d\chi)}_{(\tau,t)} \nonumber \\
    &= \sum_{(\sigma,s),(\tau,t)}W^{(d\chi)}_{(\sigma,s),(\tau,t)} \sum_{u=\pm 1}\big(P_{\sigma,us}^{(\chi)}\big)^T \mathrm{Tr}_d\left[\big(P_{\sigma,u}^{(d)})^T(\rho_0^{\otimes 2})\right]P^{(d\chi)}_{(\tau,t)} \nonumber \\
    &= \sum_{(\sigma,s),(\tau,t)}W^{(d\chi)}_{(\sigma,s),(\tau,t)} \big(P_{\sigma,s}^{(\chi)}\big)^T P^{(d\chi)}_{(\tau,t)},
\end{align}
where in the last line we invoke the even parity of \(\rho_0^{\otimes 2} \in \mathcal H_+ \otimes \mathcal H_+\). 

Using the decomposition
\(P_{\tau,t}^{(d\chi)}
=\sum_{u=\pm1}P_{\tau,u}^{(d)}P_{\tau,ut}^{(\chi)}\),
we obtain
\begin{align}
\mathcal T^{(2)}_\Pi (|0\rangle\langle0|^{\otimes 2}) 
&= \sum_{(\sigma,s),(\tau,t)}
W^{(d\chi)}_{(\sigma,s),(\tau,t)}
\left(P_{\sigma,s}^{(\chi)}\right)^T
P_{\tau,t}^{(d\chi)}
\nonumber\\
&=
\sum_{\sigma,\tau}
\sum_{s,t,u}
W^{(d\chi)}_{(\sigma,s),(\tau,t)}
\left(P_{\sigma,s}^{(\chi)}\right)^T
P_{\tau,u}^{(d)}
P_{\tau,ut}^{(\chi)}
\nonumber\\
&=
\sum_{\sigma,\tau}
\sum_{s,u}
W^{(d\chi)}_{(\sigma,s),(\tau,s)}
\left(P_{\sigma,s}^{(\chi)}\right)^T
P_{\tau,u}^{(d)}
P_{\tau,us}^{(\chi)}
\nonumber\\
&=
\sum_{\sigma,\tau}
\sum_{s,t}
\left(P_{\sigma,s}^{(\chi)}\right)^T
W^{(d\chi)}_{(\sigma,s),(\tau,s)}
P_{\tau,t}^{(d)}
P_{\tau,ts}^{(\chi)}.
\end{align}
Here, the second equality follows from the block-diagonal structure
\(W^{(d\chi)}_{(\sigma,s),(\tau,t)}
=\delta_{s,t}W^{(d\chi)}_{(\sigma,s),(\tau,s)}\),
while the final equality follows by relabeling the dummy index \(u\) as \(t\).

Now that we have the RMPS structure in terms of the permutation operators, we apply the MPO structure corresponding to the FGU. Denoting the arbitrary parity-resolved MPS as \(\rho_{\Phi^{\Pi}_\chi}\) and the corresponding the parity-resolved RMPS measure by $\mu_\chi^\Pi$, we evaluate the averaged coefficients $\alpha_n(\varrho^{(2)}_{\mu^{\Pi}_\chi}) = \mathbb E_{\mu_\chi^\Pi}[\alpha_n(\rho_{\Phi^{\Pi}_\chi}^{\otimes 2})]$ using the transfer matrix at site $j$ as,
 \begin{align}
T^{a_j,a_{j+1}}_{\bm{\sigma_j},\bm{\sigma_{j+1}}}
&= \sum_{\tau}
\sum_{t}
W^{(d\chi)}_{\bm{\sigma_j},(\tau,s_j)}
\Tr_d\left[P_{\tau,t}^{(d)}\left(\mathbb{W}_{n}^{[j]}\right)_{a_j,a_{j+1}}\right] \mathrm{Tr}_\chi\left[P_{\tau,(ts_j)}^{(\chi)} \left(P_{\bm{\sigma_{j+1}}} ^{(\chi)}\right)^T\right] \nonumber\\
&= \sum_{\tau,t}
W^{(d\chi)}_{\bm{\sigma_j},(\tau,s_j)} \mathbb M^{\tau,t}_{a_j,a_{j+1}}G^{(\chi)}_{\bm{\sigma_{j+1}},(\tau,ts_{j})} 
\end{align}
such that \(\bm{\sigma_j} = (\sigma_j, s_j)\). Similar to the previous scenario, this Transfer matrix also is site independent as the all the site dependent terms get contracted. The combined auxiliary index $(a,\tau,t)$ has $4(n+1)$ values, so the transfer matrix has size $4(n+1)\times4(n+1)$ before any block reduction.
    
To evaluate the left boundary tensor, we again utilize the fact that \(|0\rangle\langle 0 |^{\otimes2}\) is an even parity state and it enforces the condition of \(s=+\) to simplify the calculation further upon contraction,
\begin{align}
\mathcal L^{a_{1}}_{\bm{\sigma_{1}}}
&= \sum_{\eta,\tau}
\sum_{s,t}
\mathrm{Tr}_\chi\left[\left(P_{\eta,s}^{(\chi)}\right)^T |0\rangle\langle 0 |^{\otimes2} \right] W^{(d\chi)}_{(\eta,s),(\tau,s)}\,\Tr_d\left[P_{\tau,t}^{(d)} \big(v_L\mathbb W_n^{[1]}\big)_{a_1}\right]
\mathrm{Tr}_\chi\left[P_{\tau,(ts)}^{(\chi)}(P_{\bm{\sigma_1}}^{(\chi)})^T \right] \nonumber\\
&=
\sum_{\eta,\tau}
\sum_{t}W^{(d\chi)}_{(\eta,+),(\tau,+)} \mathbb M_{a_{j}}^{\tau,t} G^{(\chi)}_{\bm{\sigma_1},(\tau,t)}
\end{align}

Due to the normalization condition in the RMPS structure, the calculation right boundary tensor is more involved.
The physical factor $P_{\sigma,s}^{(d)}$ contracts with one
local pairing-tensor MPO tensor, while as discussed in Sec.~\ref{sec:rmps}, $P_{\sigma,s}^{(\chi)}$
contracts with $r$ further tensors because $\chi=d^r$. Starting from the right boundary of the RMPS 

\begin{align}
\tilde{\mathcal R}_{\bm{\sigma_{R}}} = \sum_{\tau,s_0} W_{\bm{\sigma_{R}},(\tau,s_{R})} P_{\tau,s_0}^{(d)} P_{\tau, s_0 s_R}^{(\chi)},
\end{align}

where \(R=N-r-1\). By applying the decomposition \(P_{\sigma,s}^{(d\chi)} = \sum_{t} P_{\sigma,t}^{(d)} P_{\sigma,ts}^{(\chi)}\) on \(\chi=d^{r}\) discussed earlier, we obtain the expression
\begin{align}
\tilde{\mathcal R}_{\bm{\sigma_{R}}}
=
\sum_{\tau,s_0,s_1} W_{\bm{\sigma_R},(\tau,s_R)}
P_{\tau,s_0}^{(d)}
P_{\tau,s_1}^{(d)}
P_{\tau,s_1s_0s_R}^{(d^{r-1})}.
\end{align}
Further recursively applying this decomposition on the rest of the higher-dimensional permutation operators, the original operator acting on \(d^r\)-dimension is completely decomposed into (r+1) parity-resolved single-site operators finally yielding,

\begin{align}
\tilde{\mathcal R}_{\bm{\sigma_{R}}}
= 
\sum_{\tau}
\sum_{s_0,s_1,\ldots,s_{r-1}}
W_{\bm{\sigma_R},(\tau,s_R)} P_{\tau,s_0}^{(d)} P_{\tau,s_1}^{(d)}
\cdots P_{\tau,s_{r-1}}^{(d)} P_{\tau,s_0s_1\cdots s_{r-1}s_R}^{(d)}.
\end{align}
Consequently, the right boundary tensor in the transfer matrix formalism utilizes the contraction \(\mathbb M^{\tau,s}_{a_j, a_{j+1}} = \mathrm{Tr}_d \left[ P_{\tau,s}^{(d)} \mathbb W^{[j]}_{a_j a_{j+1}} \right],\) and takes the form,
\begin{align}
\mathcal R_{\bm{\sigma_{R}}}^{a_R}
=
\sum_{\tau}
W_{\bm{\sigma_R},(\tau,s_R)}
\sum_{s_0,\ldots,s_{r-1}}
\sum_{a_1,\ldots,a_r}
\mathbb M_{\tau,s_0}^{a_{R},a_1}
\mathbb M_{\tau,s_1}^{a_1,a_2}
\; \cdots \;
\mathbb M_{\tau,s_{r-1}}^{a_{r-1},a_r}
\mathbb M_{\tau,s_0s_1\cdots s_{r-1}s_R}^{a_r}.
\end{align}
The final parity label \(s_0s_1\cdots s_{r-1}s_R\) non-trivially contributes to the sum by having dependency on the cumulative product. Therefore, we introduce cumulative parity variables \(p_0 = s_0\) and \(p_i = s_0s_1\cdots s_i, \forall i\ge 1\) to reconstruct the local parity labels as \(s_i=p_{i-1}p_i\). Substituting into the right-boundary expression yields
\begin{align}
\mathcal R_{\bm{\sigma_{R}}}^{a_R}
=
\sum_{\tau}
W_{\bm{\sigma_R},(\tau,s_R)}
\sum_{p_0,\ldots,p_{r-1}}
\sum_{a_1,\ldots,a_r}
\mathbb M_{\tau,p_0}^{a_{R},a_1}
\mathbb M_{\tau,p_0p_1}^{a_1,a_2}
\mathbb M_{\tau,p_1p_2}^{a_2,a_3}
\; \cdots \;
\mathbb M_{\tau,p_{r-2}p_{r-1}}^{a_{r-1},a_r}
\mathbb M_{\tau,p_{r-1}s_R}^{a_r}.
\end{align}

Now, in order to evaluate \(\sum_{p_0,\ldots,p_{r-1}} \sum_{a_1,\ldots,a_r} \mathbb M_{\tau,p_0}^{a_{R},a_1} \mathbb M_{\tau,p_0p_1}^{a_1,a_2} \mathbb M_{\tau,p_1p_2}^{a_2,a_3} \; \cdots \; \mathbb M_{\tau,p_{r-2}p_{r-1}}^{a_{r-1},a_r} \mathbb M_{\tau,p_{r-1}s_R}^{a_r}\), for fixed \(\tau\) we again utilize the transfer matrix formalism by considering

\begin{align}
\ell_\tau^{a_{R}} = \mathbb M_{\tau,p_0}^{a_{R},a_1}, \quad \mathbb{T}_\tau = \mathbb M_{\tau,(p_{i-1},p_{i})}^{(a_i,a_{i+1})} \quad \text{and} \quad r_{\tau,s_R} = \mathbb M_{\tau,p_{r-1} s_R}^{a_r}.
\end{align}
This representation replaces the explicit sum over parity configurations by powers of $\mathbb T_\tau$, which acts on
the combined parity and MPO index $(p,a)$ and has size $2(n+1)\times2(n+1)$.
    
With these definitions, all sums over the intermediate MPO indices
\((a_1,\ldots,a_r)\) and parity variables \((p_0,\ldots,p_{r-1})\) are absorbed into matrix multiplication, yielding
\begin{align}
\mathcal R_{\bm{\sigma_{R}}}^{a_R} = \sum_{\tau}
W_{\bm{\sigma_R},(\tau,s_R)}\,
\ell_\tau^{a_{R}}\,
\mathbb{T}_\tau^{\,r-1}\,
r_{\tau,s_R}.
\end{align}
This representation replaces the explicit sum over \(d^r\) parity
configurations by powers of a transfer matrix acting on a parity bond
of dimension \(d=2\). To determine the power of the bulk transfer
matrix, we explicitly count the physical sites absorbed into the
boundary tensors. The left boundary tensor \(\mathcal L\) includes the
contraction of the first physical site. Moreover, the right boundary
tensor
\(\mathcal R\) contains \(r+1\) local tensors, that is, one in \(\ell_\tau\), \(r-1\) in \(\mathbb T_\tau^{\,r-1}\), and one in \(r_{\tau,s_R}\). The remaining number of bulk sites is therefore \(N-1-(r+1)=N-r-2\). Hence,
\begin{align}
    \alpha_n(\varrho^{(2)}_{\mu^{\Pi}_\chi}) = C_n \sum_{\bm{\sigma_{1}},\bm{\sigma_{R}}}\sum_{a_{1},a_{R}} (\mathcal L^{a_{1}}_{\bm{\sigma_{1}}}) (T^{(N-r-2)})^{a_1,a_{R}}_{\bm{\sigma_1},\bm{\sigma_{R}}} (\mathcal R_{\bm{\sigma_{R}}}^{a_R})
\end{align}
This site-counting convention differs from that used in the
non-parity-resolved calculation. There, the left boundary
\(\mathcal L=v_L\otimes I_L\) acts only on the incoming auxiliary
indices and does not absorb a physical site. With the same \(r+1\)
sites included in the right boundary, the number of remaining bulk
sites is \(N-(r+1)=N-r-1\), giving
\(\mathcal L T^{N-r-1}\mathcal R\). The difference between the two
exponents therefore results solely from whether the first physical
site is included in the left boundary tensor.

Upon performing a numerical analysis by utilizing the above evaluation of $\alpha_n(\varrho^{(2)}_{\mu^{\Pi}_\chi})$ in the resource-quantifier definitions given in the main text, we find that both the entanglement and the IPR deviation exhibit the same behaviour as in the non-$\mathbb{Z}_2$ case. Hence, the $\mathrm{GAMPS}$ retain the same qualitative behaviour under the $\mathbb{Z}_2$ symmetry, indicating that the symmetry restriction does not alter their physical resource content or their localization properties.
\begin{figure}
    \centering
    \includegraphics[width=\linewidth]{ 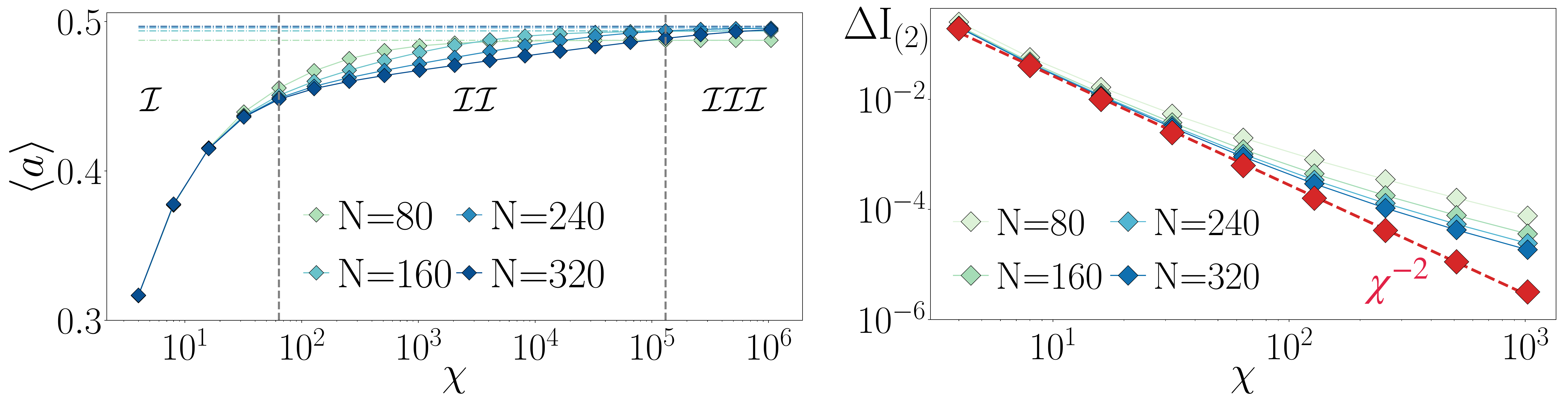}
    \caption{\textbf{Entanglement and anticoncentration of parity-resolved GAMPS.}
Entanglement and anticoncentration as functions of the bond dimension $\chi$ for system sizes $N=80,160,240,$ and $320$. For Haar-random states within either fixed-parity sector, the effective Hilbert-space dimension is $D_{\Pi}=2^{N-1}$. \textbf{(Left)} The annealed half-chain R\'enyi-2 entropy density, $a_{\mathrm{ann}}=-N^{-1}\log_{2}\langle\operatorname{Tr}(\rho_{N/2}^{2})\rangle$, exhibits rapid entanglement growth (I), an intermediate regime with logarithmic corrections (II), and convergence toward the parity-resolved Haar value (III), analogously to the case without an imposed $\mathbb{Z}_{2}$ symmetry. For a Haar-random state in a fixed-parity sector, $\langle\operatorname{Tr}(\rho_m^{2})\rangle_{\mathrm{Haar},\Pi}=(2^m+2^{N-m})/[2(2^{N-1}+1)]$; hence, at $m=N/2$, $a_{\mathrm{Haar},\Pi}=-N^{-1}\log_{2}[2^{N/2}/(2^{N-1}+1)]$, as indicated by the horizontal dashed lines. \textbf{(Right)} Relative deviation of the second-order inverse participation ratio from its parity-resolved Haar value, $\Delta I_{2}=\langle I_{2}\rangle_{\mu_G}/\langle I_{2}\rangle_{\mathrm{Haar},\Pi}-1$, where $\langle I_{2}\rangle_{\mathrm{Haar},\Pi}=2/(D_{\Pi}+1)=2/(2^{N-1}+1)$. In the large-$N$ limit, the deviation approaches the asymptotic $\chi^{-2}$ scaling observed in the case without an imposed $\mathbb{Z}_{2}$ symmetry.}
    \label{fig:placeholder}
\end{figure}

\putbib[bib.bib]
\end{bibunit}

\end{document}